\documentclass[a4paper,11pt]{article}
 \pdfoutput=1
 \usepackage{jheppub} 
 \usepackage{graphicx}
 \usepackage{amssymb}
\usepackage{dcolumn} 
\usepackage{bm} 
\usepackage{color}
\usepackage{multirow}

\usepackage{fancyvrb}
 
\newcommand{\met}{{/\!\!\! E_T}} 
\newcommand{\mpt}{{/\!\!\!\! \vec{P}_T}} 
\allowdisplaybreaks

\usepackage{empheq}
\usepackage{hyperref}
 
\makeatletter
\newcommand\footnoteref[1]{\protected@xdef\@thefnmark{\ref{#1}}\@footnotemark}
\makeatother
 
\newlength\dlf  

\newcommand{\lsim}{{\;\raise0.3ex\hbox{$<$\kern-0.75em\raise-1.1ex\hbox{$\sim$}}\;}}
\newcommand{\gsim}{{\;\raise0.3ex\hbox{$>$\kern-0.75em\raise-1.1ex\hbox{$\sim$}}\;}}
\newcommand{\beq}{\begin{equation}}
\newcommand{\eeq}{\end{equation}}
\newcommand{\bea}{\begin{eqnarray}}
\newcommand{\eea}{\end{eqnarray}}

\def\ra{\rightarrow}
\def\baa{\begin{array}}
\def\eaa{\end{array}}

\def\met{E_T\hspace{-0.45cm}/\hspace{0.25cm}}
\mathchardef\minus="002D

\newcommand{\AugLag}{\tilde{\mathcal{L}}}

\title{OPTIMASS: A Package for the Minimization of Kinematic Mass Functions with Constraints} 
 
\author[a]{Won Sang Cho,\footnote{\label{note1}Corresponding author.}}
\author[b]{James S.~Gainer,}
\author[b]{Doojin Kim,}
\author[a,c]{Sung Hak Lim,\footnoteref{note1}}
\author[b]{Konstantin~T.~Matchev,}
\author[d]{Filip Moortgat,}
\author[d]{Luc Pape,}
\author[e,f,g]{Myeonghun Park}

\affiliation[a]{Center for Theoretical Physics of the Universe, Institute for Basic Science (IBS), Daejeon 305-811, Korea}
\affiliation[b]{Department of Physics, University of Florida, Gainesville, FL 32611, USA}
\affiliation[c]{Department of Physics, Korea Advanced Institute of Science and Technology, 335 Gwahak-ro, Yuseong-gu, Daejeon 305-701, Korea}
\affiliation[d]{CERN, Physics Department, CH-1211 Geneva 23, Switzerland}
\affiliation[e]{Asia Pacific Center for Theoretical Physics, San 31, Hyoja-dong, Nam-gu, Pohang 790-784, Korea} 
\affiliation[f]{Department of Physics, Postech, Pohang 790-784, Korea} 
\affiliation[g]{Kavli IPMU (WPI), The University of Tokyo, Kashiwa, Chiba 277-8583, Japan} 
\emailAdd{hepkosmos@ibs.re.kr, immworry@ufl.edu, sunghak.lim@kaist.ac.kr, matchev@phys.ufl.edu, filip.moortgat@cern.ch, luc.pape@cern.ch, parc@ibs.re.kr}

\abstract{ 
Reconstructed mass variables, such as $M_2$, $M_{2C}$, $M_T^\star$, and $M_{T2}^W$, play an essential role in searches for new physics at hadron
colliders.  The calculation of these variables generally involves constrained minimization in a large parameter space, which is numerically
challenging.  We provide a {\sc  C++} code, {\sc Optimass},
which interfaces with the {\sc  Minuit} library to perform this constrained minimization using the Augmented Lagrangian Method.
The code can be applied to arbitrarily general event topologies, thus allowing the user to significantly extend the existing set 
of kinematic variables. We describe this code, explain its physics motivation, and demonstrate its use in the analysis of the fully leptonic 
decay of pair-produced top quarks using $M_2$ variables. }

\preprint{
\begin{flushright}
CTPU-15-09\\
APCTP-PRE2015-020\\
IPMU15-0115
 \end{flushright}
}
\date{October 22, 2015}
\begin{document} 
\maketitle
\flushbottom

\section{Introduction \label{sec:intro}}
The CERN Large Hadron Collider (LHC) has successfully completed the
decades-long quest to discover the particles of the Standard Model (SM) by 
finding the Higgs Boson~\cite{Aad:2012tfa, Chatrchyan:2012ufa}.
The paramount question in the current Run 2 of the LHC is whether 
the LHC can reach the relevant energy scale to discover physics beyond the standard model (BSM).
Popular frameworks for new physics such as Supersymmetry (SUSY)~\cite{Martin:1997ns, Drees:2004jm, Baer:2006rs, Dine:2007zp} 
and Universal Extra Dimensions (UED)~\cite{Appelquist:2000nn, Cheng:2002ab} predict:
\begin{enumerate}
\item The presence of particles, such as neutralinos or KK-photons, that are not reconstructed in the detector, and are hence termed 
``invisible".  In general, the production of these particles will lead to 
``missing transverse energy"\footnote{Or more precisely, missing transverse momentum.} (MET) in an event.
\item Relatively complex decay topologies, in which pair-produced, generally colored, particles undergo several subsequent decays.  Each 
``decay chain" thus produces one or more visible particles, as well as at least one invisible particle.
\end{enumerate}
Searches for new particles that produce such long decay chains in combination with a MET signature 
are complicated by large backgrounds from $t\bar{t}$, $W$, and $Z$ production, often with additional 
jets from initial state radiation (ISR).  
The severity of these backgrounds in multijet and multilepton channels increases with the collider energy.  
Even if a signal of such new physics is seen, the corresponding measurements of particle properties 
such as masses, couplings, and spins, are highly nontrivial \cite{Barr:2010zj,Wang:2008sw,Burns:2008cp}.
Therefore, sophisticated procedures must be used to separate signal from background and to extract the quantum 
numbers of the new particles.

In a given event, one observes some number of ``physics objects" 
which correspond to physical particles that have produced appropriate energy deposits in
the tracker and calorimeters of the detector; we will refer to these objects as ``visible particles". 
We shall denote their measured four-momenta by $p^\mu_j$, where $j$ is the visible particle label. 
At the same time, the existence of a non-vanishing MET in the event indicates the presence of 
some number of additional, invisible, particles with four-momenta $q^\mu_k$, where $k$ now labels the invisible particles.
The individual momenta $q^\mu_k$ are not measured, and the only available piece of information is the 
missing transverse momentum
\beq
\mpt \equiv \sum_{k} \vec{q}_{Tk} = - \sum_{j} \vec{p}_{Tj}.
\label{eq:mpt}
\eeq
The MET, $\met$, is then simply the magnitude of the missing transverse momentum vector:
\beq
\met \equiv |\, \mpt |.
\eeq
The essential question is how to take these sets of measured four-momenta, $\{p^\mu_j\}_a$,
(one for each event $a$) and determine whether the events
are produced purely by SM processes or whether new physics is at work.  
The standard procedure is to construct some kinematic variable, $v$, and compare the 
$v$ distribution predicted by the SM to the data. In choosing a suitable variable, $v$, 
one typically follows one of these approaches:
\begin{enumerate}
\item {\em The variable, $v$, is an analytic function of some, but not all, of the measured momentum degrees of freedom.}
This is the preferred approach in inclusive analyses, where one targets a specific subset of
the event. For example, in an inclusive search for a visibly decaying resonance, $X$, 
one would select only the momenta of the hypothesized decay products $j_1, j_2, ... , j_n$, 
and form the invariant mass of the resonance $X$,
\beq
M_X \equiv \sqrt{\left( p_{j_1}+p_{j_2}+...+p_{j_n} \right)^2},
\label{MRdef}
\eeq
leaving out all other objects in the event.\footnote{In forming the variable 
(\ref{MRdef}), one also ignores, e.g., additional angular information.} 
In the case of missing energy events, the situation is 
much more complicated --- we cannot reconstruct the mass of the resonance as in eq.~(\ref{MRdef}),
since we have not measured the momenta, $q^\mu_k$, of the invisible decay products.
Then, one typically tries to form a variable which correlates with the scale of $M_X$.
Various candidates have been tried, including the transverse momentum of the hardest
object of a given type (lepton, jet, etc.) \cite{Abbott:2000gx,Aaltonen:2011wt}, 
the scalar $p_T$ sum of the four hardest jets (or of all jets) \cite{Tovey:2000wk}, 
the jet multiplicity \cite{Bramante:2011xd,Hedri:2013pvl}, the ``fat" jet mass \cite{Hook:2012fd},
the ``contransverse mass", $M_{CT}$~\cite{Tovey:2008ui,Polesello:2009rn, Matchev:2009ad},
the lepton energy \cite{Agashe:2012bn,Agashe:2012fs,Agashe:2013eba} or
lepton energy ratios \cite{Nojiri:2000wq,Cheng:2011ya}, and many more. 
The advantage of such techniques is their simplicity and robustness --- 
they do not involve too many theoretical assumptions, making them ideal for
model-independent searches for new physics. At the same time, they appear to be
suboptimal, since they do not utilize the full set of measured degrees of freedom, leading to
a certain loss of information. It is also rather challenging to assign a proper physical meaning
to a kinematic variable which only uses such partial information (for more
detailed discussion, see refs.~\cite{Tovey:2000wk,Barr:2011xt,CL_tasi}).
\item {\it The variable, $v$, is an analytic function of some measured momentum degrees of freedom 
and the measured $\met$.} The explicit inclusion of the measured $\met$ in the definition
of $v$ was the next attempt to design a better performing class of variables. 
Perhaps the best known example is the $W$ transverse mass
\cite{Smith:1983aa,Barger:1983wf}, where one identifies the transverse 
momentum of the missing neutrino with the measured $\mpt$. Other possibilities include the
``effective mass", $M_{eff}$~\cite{Hinchliffe:1996iu,Tovey:2000wk}, the
$\sqrt{\hat{s}}_{min}$ variable~\cite{Konar:2008ei,Konar:2010ma,Robens:2011zm},
and the ``razor" variables \cite{Rogan:2010kb,Buckley:2013kua,CMS:2014ets,Khachatryan:2015pwa}.
The outputs of neural nets, boosted decision trees, and other multivariate analyses~\cite{Bhat:2010zz},
particularly those involving some form of machine learning, are also variables in this class.
Incorporating the measured $\met$, which is often a sensitive variable all by itself \cite{Hubisz:2008gg,Alves:2012ft},
into the definition of the kinematic variable, $v$, is certainly a step in the right direction. 
%
\item {\em The definition of the variable, $v$, involves all measured momentum degrees of freedom 
and the individual invisible momenta, $q^\mu_k$.}
Finally, one may construct the variable, $v$, so that from the onset it has explicit dependence on the 
individual invisible momenta, $q^\mu_k$. The advantage of this approach is that one works with
theoretically motivated quantities with clear physical meaning \cite{Barr:2011xt}. The obvious downside is that the
individual invisible particle momenta, $q^\mu_k$, are unknown, and something must be done to 
fix their values in the calculation. There are two possible alternatives:
\begin{itemize}
\item {\em Integrate over all possible values of the invisible momenta.}
Perhaps the simplest solution is to allow all possible values of the invisible momenta, $q^\mu_k$,
which are consistent with the measured missing transverse momentum (\ref{eq:mpt}),
and compute the variable, $v$, as a suitably weighted average.
A celebrated example of this approach is the Matrix Element Method 
(MEM)~\cite{kondo1, kondo2, Gainer:2013iya} 
which is finding increased use in hadron collider physics 
\cite{kondo3,dalitz,oai:arXiv.org:hep-ex/9808029,vigil,canelli,abazov,Aaltonen:2010yz,
  Volobouev:2011vb, Demilly:2014baa, Khachatryan:2015ila}.  
However, the method is often too computationally challenging, requiring novel ideas and approaches 
\cite{Gainer:2014bta,Alwall:2014hca,Schouten:2014yza}.  
Additionally, it is generally difficult to incorporate ``reducible" backgrounds, which consist of events
where the reconstructed particles are misidentified and/or their momenta significantly mismeasured. 
\item {\em Use a physically motivated ansatz for the invisible momenta.} Alternatively, instead of considering all possible
values of the set of invisible momenta, $\{q_k^\mu\}$, one could fix them by following some prescription specified in advance.
With this approach, one gives up on trying to ``guess" the correct values of the invisible momenta, and instead
focuses on constructing a useful variable, $v$, whose properties can reveal important information about the 
underlying physics.  Examples of such variables include the Cambridge $M_{T2}$ variable~\cite{Lester:1999tx,Barr:2003rg} 
and its variants \cite{Konar:2009wn,Barr:2009jv,Konar:2009qr}, and the variables 
$M_{2C}$ \cite{Ross:2007rm,Barr:2008ba},
$M_{CT2}$ \cite{Cho:2009ve,Cho:2010vz}, 
$M_T^\star$ \cite{Barr:2011ux}, 
$M_{T2}^W$ \cite{Bai:2012gs}, and 
$M_{min}$ \cite{Papaefstathiou:2014oja}.
The variables in this class are often specified, not by analytic formulae, but by the 
algorithm used to calculate them\footnote{For some of the variables, analytical formulas may exist in certain 
special cases
\cite{Cho:2007qv,Gripaios:2007is,Barr:2007hy,Cho:2007dh,Burns:2008va,Lester:2011nj,Lally:2012uj}, 
but not in general.}.
\end{itemize}
\end{enumerate}

Our focus in this paper will be on the algorithmically specified variables from the very last category,
which are known to possess several attractive features:
\begin{enumerate}
\item They are ``maximally constraining" \cite{Barr:2011xt,Mahbubani:2012kx} in the sense that, on an event per event basis,
they provide the best possible lower bound on an invariant mass quantity of interest, 
such as the parent masses or the center-of-mass energy, $\sqrt{\hat{s}}$. This is particularly useful 
in cases where it is not possible to determine the actual values of that quantity due to incomplete event information.
\item Their kinematic distributions exhibit sharper endpoints which are easier to measure over the SM backgrounds 
\cite{Cho:2014naa}, leading to a more precise determination of the new physics mass spectrum.
\item Certain measurements of their properties can be used as a self-consistency check on the assumed signal event topology
\cite{Cho:2014naa,Cho:2014yma}. If the check fails, our conjecture about the event topology is falsified, 
thus narrowing down the allowed set of possibilities.
\end{enumerate}

At the same time, such algorithmically defined variables are notoriously difficult to compute.
The algorithmic procedure typically involves calculating a mass for a set of hypothesized particles 
in an event (possibly after projecting their momenta onto the transverse plane), and minimizing
the value of that mass with respect to {\em all} invisible momenta, $q^\mu_k$. What makes the problem 
particularly challenging, however, is the presence of additional non-linear
mass-shell constraints. In essence, we are faced with a multidimensional constrained optimization problem,
where our objective function is an energy function which is to be minimized. 
Given the importance of the maximally constraining invariant mass variables 
for new physics searches and measurements \cite{Barr:2011xt,Mahbubani:2012kx,Cho:2014naa,Cho:2014yma},
it is important to have publicly-available software for performing constrained minimizations 
to calculate kinematic variables from high energy collision data 
with sufficient generality and efficiency. The existing packages described in the literature
are typically only applicable to a specific variable, e.g., 
$M_{T2}$ \cite{Barr:2003rg,Cheng:2008hk,Walker:2013uxa,Lester:2014yga}, or
$M_{T2}^W$ \cite{Bai:2012gs} and cannot be readily generalized to the whole class of
on-shell constrained variables \cite{Mahbubani:2012kx,Cho:2014naa}. 
The standard approach is to try to solve the constraining equations, 
thus reducing the unknown number of degrees of freedom (d.o.f.), then 
implement an unconstrained minimization over the remaining d.o.f. 
While this approach generally provides the most efficient algorithm for a specific event topology,
it is not extendable to more general event topologies, 
where not all constraints can be simultaneously solved analytically. 

In this paper, we describe an alternative approach that is sufficiently universal and flexible,
and can be applied to arbitrarily general event topologies.
The main idea is to use the Augmented Lagrangian Method (ALM) \cite{ALM1, ALM2}, 
briefly described below in section~\ref{sec:ALMmethod}.
In this approach, the feasibility (i.e., the validity of the constraints) is ensured by penalizing 
infeasibility by adding ``penalty terms" to the objective function,
rather than by directly solving the constraining equations. 
The fact that the method does not require the solving of any constraints beforehand
makes it very flexible and applicable to a very general class of event topologies. 
Of course, we still have to perform a standard unconstrained minimization, for which we can
take advantage of any one of the many publicly available packages --- we have chosen to use
{\sc  Minuit} \cite{minuit}, which is widely popular in high energy physics.
We also supply a comprehensive software package, {\sc  Optimass}\footnote{For OPTImization of 
MASS variables.}, in the form of a library, 
which interfaces with {\sc  Minuit} to perform the constrained minimization
of a user-specified kinematic function using the ALM.
Appendix~\ref{app:optimass} contains instructions on the installation and usage of {\sc  Optimass}.

The paper is organized as follows.
In section \ref{sec:overview}, we review the general problem of constrained optimization 
with special emphasis on the motivation and the techniques used by {\sc  Optimass}, 
in particular the ALM. The relevant {\sc  Minuit} routines with which it interfaces are 
described in Appendix~\ref{sec:minuit}.
Section~\ref{sec:algorithm} describes in detail the algorithm behind the {\sc  Optimass} package
and presents several toy examples for its validation.
In section \ref{sec:application} we briefly review the $M_2$ variables \cite{Barr:2011xt,Mahbubani:2012kx,Cho:2014naa}, 
and provide examples of their use, in the study of the fully leptonic decays of pair-produced top quarks.
We use this as an opportunity to compare the results from {\sc  Optimass} with previous studies and known 
analytical calculations.
Finally, section~\ref{sec:conclusions} is reserved for our conclusions and a brief discussion of the future of {\sc  Optimass}.

\section{Review of constrained minimization}
\label{sec:overview}

While many excellent textbooks and references discuss the optimization
techniques that we will utilize (see, e.g., refs.~\cite{nocedal,alm,Press:1992zz} 
and references therein), we feel that a brief review of the elements
of optimization theory relevant to the operation of {\sc Optimass} will prove useful.
We first note that while we will sometimes speak of ``optimization"
rather than ``minimization": (i) in the calculation of kinematic
variables we will be interested only in minimization, and (ii) the methods
used to find a maximum are, up to obvious changes in the sign of
certain parameters, identical to those used to find a minimum.  So in what follows 
we will not worry much about this distinction.

The first issue that \emph{will} concern us is the important division 
of minimization problems into two types:
(i) constrained and (ii) unconstrained. 
In constrained minimization, we want to minimize an {\bf objective function},
\begin{equation}
f(\vec{x}),
\label{eq:objective0} 
\end{equation}
subject to a set of $m$ {\bf constraints} 
\begin{equation}
c_{a}(\vec{x})=0, \quad a=1,...,m,
\label{eq:constraints0}
\end{equation}
where $\vec{x}$, in general, refers to a point in $\mathbb{R}^n$ formed from
some unknown momentum degrees of freedom $x_1,x_2,\ldots, x_n$.
In what follows, we shall assume that the number of constraints, $m$, is 
always less than the number of independent degrees of freedom, $n$, 
so that we are dealing with a true minimization problem.  

If the constraints in eq.~(\ref{eq:constraints0}) are all
independent, then the parameter space is effectively reduced from
$n$ dimensions to $n - m$ dimensions.  Sometimes this reduction can be
performed analytically. For example, consider an optimization problem in $\mathbb{R}^3$, 
subject to the constraint $x^2 + y^2 + z^2 = 1$ --- then we should
parameterize the two dimensional subspace that satisfies the
constraints, i.e., the surface of the unit sphere, $S_2$, by the angles 
$\theta$ and $\varphi$ in the standard way. 

However, this reduction of dimensionality cannot always be performed analytically.  
A useful alternative procedure, therefore, is to turn the constrained minimization
problem into the problem of an unconstrained minimization\footnote{Or possibly, 
a {\em series of} unconstrained minimization problems.}
of {\bf a modified objective function}, $\tilde{f}(\vec{x})$, over the
full, unconstrained, parameter space, $\mathbb{R}^n$.
This new problem can then be solved by setting the gradient of some function equal to zero, 
or by searching for a local/global minimum using one of the many
standard numerical algorithms conventionally used for this purpose. 
   
When performing this unconstrained minimization iteratively, one
develops an algorithm for finding the location of the minimum of $\tilde{f}(\vec{x})$, $\vec{x}^\ast$, 
which is also referred to as the {\bf minimizer}. At each iteration, one starts 
with some initial estimate, $\vec{x}_k$, (typically taken to be the minimizer of 
the previous, $k-1^{\rm st}$, iteration), then refining this estimate by obtaining a new
minimizer, $\vec{x}_{k+1}$, in some prescribed way, until certain convergence 
criteria are met. Since we have not analytically solved the constraints (\ref{eq:constraints0}), 
the estimates, $\vec{x}_k$, {\em will not}, in general, satisfy the constraints exactly.
Following the standard mathematical terminology, we shall refer to values of $\vec{x}$ 
that satisfy the constraints in eq.~(\ref{eq:constraints0}) as {\bf feasible}.
The absolute value of $c_a(\vec{x})$ is then a measure of the {\bf feasibility}\footnote{A point $\vec{x}$ in 
the unknown momentum space $\mathbb{R}^n$  is feasible if it is 
an element of the feasible set $\Omega$ defined by
\beq
\Omega = \left\{ \, \vec{x} \,|\, c_a(\vec{x})\,=\,0, \, a=1,..,m\right\}.\nonumber
\eeq 
}.
Even though feasibility is not strictly guaranteed, the ultimate solution found by the method should
nevertheless be such that the constraints are satisfied to within a required degree of numerical precision.  

In the remaining three subsections of this section, we discuss three possible ways 
to transform a constrained minimization problem into an unconstrained minimization
problem, namely (i) the method of Lagrange multipliers (in section~\ref{sec:lag}), 
(ii) penalty methods (in section~\ref{sec:pen}), and (iii) the Augmented Lagrangian Method
(in section~\ref{sec:ALMmethod}).  We shall see how
penalty methods solve some of the problems associated with the use of
Lagrange multipliers, while the ALM, in turn, resolves certain numerical 
issues related to the use of penalty methods.   
Of course we also must be able to solve the resulting unconstrained
optimization problem; we discuss approaches to this challenge
in Appendix~\ref{sec:minuit}.

\subsection{The method of Lagrange multipliers}
\label{sec:lag}

As noted above, in a generic constrained minimization problem 
we are looking for the minimum value of the target function, $f(\vec{x})$, 
subject to the constraints (\ref{eq:constraints0}):
\bea
\boxed{\qquad
f(\vec{x}^{\,\ast}) \equiv \min_{x\in\mathbb{R}^n}\,f(\vec{x})\quad \text{such that}\quad c_{a=1,...,m}(\vec{x}^{\,\ast})=0.
\qquad}
\label{eq:problem}
\eea
Alternatively, one is trying to find the location of the minimizer, $\vec{x}^\ast$, in $\mathbb{R}^{n}$.
We note that in practical applications of the maximally constraining invariant mass variables
both the minimum value of the function, $f(\vec{x}^\ast)$, and the minimizer, $\vec{x}^\ast$,
itself can serve a useful purpose. For example, in the $M_{T2}$-assisted on-shell (MAOS)
reconstruction method, the minimizer is used to provide an ansatz for certain transverse components 
of the invisible momenta \cite{Cho:2008tj,Park:2011uz}. 
Both  the function, $f(\vec{x})$, and the constraints, $c_a(\vec{x})$,
are assumed to be smooth\footnote{In some cases, the objective function may depart
from smoothness in a specific way; we discuss this point and related issues in 
sections~\ref{sec:example-2} and~\ref{m2-intro}, see also ref.~\cite{Cho:2014naa}.} 
real-valued functions in $\mathbb{R}^n$.

The method of Lagrange multipliers provides necessary and sufficient 
conditions for finding local solutions of the minimization problem (\ref{eq:problem}) above. 
In this method, we define a corresponding Lagrangian, henceforth 
denoted by $\mathcal{L}$, for an objective function, $f(\vec{x})$, and constraints, 
$c_a(\vec{x})$, by
\bea
\mathcal{L}(\vec{x},{\boldsymbol \lambda})=f(\vec{x})\, - \, \sum_{a=1}^{m}\, \lambda_{a}\, c_{a}(\vec{x}),
\label{eq:lag function}
\eea
where ${\boldsymbol \lambda}\equiv (\lambda_1, \lambda_2,\ldots,\lambda_m)$ 
is an $m$-component vector\footnote{Throughout this paper we shall use the notation 
$\vec{v}$ for $n$-dimensional vectors in the space of independent variables, $\mathbb{R}^n$,
and ${\boldsymbol v}$ to denote $m$-dimensional vectors in the space of constraints, $\mathbb{R}^m$.} 
of Lagrange multipliers, $\lambda_a$.  
We now describe the conditions that must
be satisfied in this method in order to establish the existence of 
a local minimum at the proposed minimizer, $\vec{x}^\ast$.

\paragraph{First order condition (FOC).} 

In unconstrained optimization, a \textit{necessary} condition for the existence
of a local minimum of $f$ at $\vec{x}^\ast$ is that the gradient vector, $\nabla_x f(\vec{x}^{\ast})$, 
vanishes.  As this condition involves the gradient, we may term it a 
{\bf first order condition}. 
In the method of Lagrange multipliers, an analogous condition holds for 
the existence of a local minimum.  Here it is only necessary that the 
gradient of the objective function, $f$, at $\vec{x}^\ast$ is orthogonal to the 
surface defined by the constraints.  This condition will hold if the gradient
 of $f$, $\nabla_x f(\vec{x}^{*})$, is an element of the vector space spanned by 
 the gradient vectors of $c_a$, $\nabla_x\, c_a(\vec{x}^{*})$.  Thus the FOC is that 
 there exists a Lagrange multiplier vector, ${\boldsymbol \lambda}^\ast$, in $\mathbb{R}^m$, such that
 at the point $(\vec{x}^{*},{\boldsymbol \lambda}^\ast)\in \mathbb{R}^n\otimes\mathbb{R}^m$,
 the following conditions hold
\bea
\label{eq;foc}
\left\{
  \begin{array}{llr}
    \nabla_\lambda \,\mathcal{L}(\vec{x}^*,{\boldsymbol \lambda}^\ast) = c_{a}(\vec{x}^\ast)=0,\\ [2mm]
    \nabla_x \,\mathcal{L}(\vec{x}^*,{\boldsymbol \lambda}^\ast)
	=\nabla_x \,f(\vec{x}^*) - \sum_{a} \lambda_a^* \nabla_x\, c_a(\vec{x}^*)  = 0.
  \end{array}
\right.\label{eq:1stcond}
\eea
Therefore, at least at the level of the FOC, the problem of constrained
optimization has been reformulated as an unconstrained
optimization problem.

\paragraph{Second order condition (SOC).} 
As is well-known, the condition that a first derivative vanishes is not sufficient
to establish that there is an extremum at that point --- one must also verify 
that the second derivative is, in the case of a minimum, positive.
The extension of this idea to unconstrained optimization in many dimensions 
is to require that the {\bf Hessian matrix}, defined for the objective function, $f$, by
\begin{equation}
H_{jk}(\vec{x}^\ast) = \frac{\partial^2 f(\vec{x}^\ast)}{\partial x_j \partial x_k},
\label{eq:hess}
\end{equation}
and evaluated at the prospective minimizer, $\vec{x}^\ast$,
be positive definite\footnote{In linear algebra, a symmetric $n \times n$ 
real matrix, $M$, is said to be positive definite if $\vec{z}^{\, T} M \vec{z}$ 
is positive for every non-zero column vector, $\vec{z}$, of $n$ real numbers.}.  
This is the ``second order condition" (SOC); 
that both the FOC and the SOC are satisfied is sufficient 
for the existence of a local minimum.  As long as we are not at a boundary of the
parameter space, these conditions are both necessary and sufficient.

We now consider the corresponding SOC for the Lagrange multiplier method, hoping
that, as in the case of the FOC, we can obtain a condition analogous to the case
of unconstrained minimization, i.e., a constraint involving the positive definiteness of some Hessian.
We therefore evaluate the Hessian of the Lagrange function (\ref{eq:lag function}) 
with respect to $\vec{x}$: 
\bea
\label{eq:hess-lag}
(H_{\mathcal{L}})_{jk}(\vec{x}^\ast,{\boldsymbol \lambda}^\ast) \equiv 
\frac{\partial^2 \mathcal{L}(\vec{x}^\ast,{\boldsymbol \lambda}^\ast)}{\partial
  x_j \partial x_k} = 
\frac{\partial^2 f(\vec{x}^\ast)}{\partial x_j \partial x_k} - 
\sum_{a} \lambda_{a}^\ast \frac{\partial^2 c_a(\vec{x}^\ast)}{\partial
  x_j \partial x_k}.
\eea
To determine the SOC, we note that for $\vec{x}^\ast$ to be a minimizer 
(with Lagrange multiplier vector, ${\boldsymbol \lambda}^\ast$), we must have
\bea
\vec{d}^{\,T}\, H_{\mathcal{L}}(\vec{x}^\ast,{\boldsymbol \lambda}^\ast)\, \vec{d}\, >\, 0,
\label{eq:2ndcond}
\eea
for any infinitesimal displacement, $\vec{d}$, from the proposed minimizer,
$\vec{x}^\ast$, in a direction allowed by the constraints.

Thus the relevant condition is \emph{not} the positive definiteness of the Hessian
(\ref{eq:hess-lag}), but the positive definiteness of the restriction
of the Hessian to the space allowed  by the constraints.  The fact that we must still use the constraints to see if a
given stationary point (i.e., a point satisfying the FOC) is a minimum, means that we
have failed in our mission to convert the constrained minimization problem to
an unconstrained minimization problem.
We therefore proceed to look for methods for which the SOC does not
explicitly involve the constraints.

\subsection{Penalty methods}
\label{sec:pen}

A natural approach to our problem of transforming a constrained
optimization problem to an unconstrained optimization problem is to
follow the example of the method of Lagrange multipliers in modifying the
objective function, but to do so in a different way.
Clearly, we would like to modify the function so that infeasibility incurs
a penalty.  One possible way to achieve this is via ``convexification'' of the 
geometry near the desired solution point, i.e., making sure 
that a solution to the constrained minimization problem is a local
minimum of the transformed function, $\tilde f(\vec{x})$, even if it is \emph{not} a local
minimum of $f(\vec{x})$ in the absence of constraints.
We now proceed to give an example of one such
convexification approach.

In the so-called {\bf penalty methods}, the original objective function, $f(\vec{x})$, is
modified by the addition of a {\bf penalty term}, i.e., a functional of
$c_a(x)$ weighted by a positive penalty parameter, $\mu$, so that the term
vanishes when $c_a(\vec{x})=0$, but becomes large if $c_a(\vec{x})\ne 0$ in
the $\mu \to 0$ limit.
While there are various penalty methods, which differ in the form of the 
penalty function, we consider the ``Quadratic Penalty Method" (QPM)
here, because of its connection with the ALM discussed below.
In the QPM, the penalty term is chosen so that the modified function\footnote{From here on, we shall use 
a semicolon, ``;", to separate the arguments of a function into two groups:
independent variables, with respect to which an optimization is to be done, 
and fixed parameters. } under consideration is
\bea
P(\vec{x}\,;\mu) \equiv f(\vec{x}) + \frac{1}{2 \mu}\sum_a  c_a^{2}(\vec{x}).
\label{eq:penaltymethod}
\eea
In the course of the algorithm, the parameter, $\mu$, will be reduced, as the desired
properties of this function hold in the $\mu \to 0$ limit.
The gradient of (\ref{eq:penaltymethod}) is 
\beq
\nabla P(\vec{x}\,;\mu) = \nabla f(\vec{x}) + \sum_a \frac{c_a}{\mu}\,\nabla c_a(\vec{x}),
\eeq 
which reproduces the gradient of the Lagrange function (\ref{eq:1stcond}) if we take
\bea
\lambda_a(\mu) \, &\equiv& -\frac{c_a(\vec{x})}{\mu}
\label{eq:lmpenalty}
\eea
to be (the components of) the Lagrange multiplier vector, ${\boldsymbol \lambda}$.  
This shows that the necessary FOC for a minimum is the same as for the
method of Lagrange multipliers, only the Lagrange multiplier vector is
now determined for us.
Since a minimization using a QPM should give the same value for the
solution, $\vec{x}^\ast$, as the Lagrange multiplier method (which gives
the solution $(\vec{x}^\ast, {\boldsymbol \lambda}^\ast)$), 
it follows that as we approach this solution, we must have
\begin{equation}\label{eq:asymp-QPM}
\lim_{\vec{x}\to\vec{x}^\ast}\left( -\frac{c_a(\vec{x})}{\mu} \right) = {\lambda}_a^\ast,
\end{equation}
thus $|c_a(\vec{x})| \to 0$ as $\mu \to 0$.

In determining the SOC for the QPM, we note that the Hessian of the
function in (\ref{eq:penaltymethod}) is
\begin{equation}\label{eq:pen-hess}
H_P(\vec{x}^\ast)_{jk} = \frac{\partial^2 f(\vec{x}^\ast)}{\partial x_j \partial
  x_k} + \frac{1}{\mu} \sum_a \bigg( (\nabla c_a)_j (\nabla
c_a)_k + c_a \nabla^2_{jk} c_a \bigg) \bigg|_{\vec{x}=\vec{x}^\ast}.
\end{equation} 
If $\vec{x}^\ast$ satisfies the constraints, then this expression simplifies
to 
\begin{equation}\label{eq:pen-hess-simp}
H_P(\vec{x}^\ast)_{jk} = \frac{\partial^2 f(\vec{x}^\ast)}{\partial x_j \partial
  x_k} + \frac{1}{\mu} \sum_a (\nabla c_a)_j (\nabla
c_a)_k \bigg|_{\vec{x}=\vec{x}^\ast}.
\end{equation} 
If we consider an infinitesimal displacement, $\vec{d}$, along the surface
allowed by the constraints, then as
\begin{equation}
\vec{d} \cdot \nabla c_a(\vec{x}^\ast)  = 0, 
\label{eq:jacobian}
\end{equation}
we find that
\begin{equation}
\vec{d}^{\,T}\,H_P(\vec{x}^\ast)\,\vec{d} = \vec{d}^{\, T}\,H_f(\vec{x}^\ast)\,\vec{d},
\end{equation} 
where $H_f(\vec{x})$ is the Hessian (\ref{eq:hess}) for the original objective function,
i.e., $\partial_j \partial_k f(\vec{x})$.

On the other hand, for displacements, $\vec{d}^\prime$, orthogonal to this
surface, we obtain non-zero terms
 from both the objective function and the penalty term, i.e.
\begin{equation}
\vec{d}^{\prime\,T}\,H_P(\vec{x}^\ast)\,\vec{d}^\prime =
\vec{d}^{\prime\,T}\,H_f(\vec{x}^\ast)\,\vec{d}^\prime + \frac{1}{\mu} \sum_a (\nabla
c_a(\vec{x}^\ast) \cdot \vec{d}^\prime)^2,
\end{equation}
where $\nabla c_a(x) \cdot \vec{d}^{\prime}$ is now non-vanishing.  In the $\mu \to
0$ limit, the second term will dominate.  
So if the Hessian is positive definite with respect to allowed
displacements, it is automatically positive definite in general.
Thus, we now have that in
the limit of $\mu \to 0$, the sufficient condition that the stationary
 point, $\vec{x}^\ast$, be a local minimum is simply that the 
Hessian of the function (\ref{eq:penaltymethod}) is positive definite.  
Thus we have succeeded in overcoming the limitation of the method of
Lagrangian multipliers in that both the FOC and SOC are the same as in
unconstrained minimization.

Unfortunately, all is not well when it comes to practical applications of the
QPM.  The basic problem is that in the $\mu \to 0$ limit the algorithm becomes too
sensitive to small departures from feasibility, hence numerical
instabilities may prevent convergence to the solution.  In more formal
language, small values of $\mu\rightarrow 0$ can result in severe {\bf
  ill-conditioning} of the Hessian, since the rightmost term in
(\ref{eq:pen-hess-simp}), which dominates in the $\mu \to 0$ limit, is
not invertible.  Therefore, we will need to further modify the QPM, 
retaining the way in which it maps constrained optimization
problems to unconstrained optimization problems, but reducing the
relative importance given to feasibility in the $\mu \to 0$ limit.
The solution, presented in the next subsection, is the Augmented
Lagrangian Method (ALM).

\subsection{Augmented Lagrangian Method} 
\label{sec:ALMmethod}

We seek a method which preserves the success of the QPM in
generating FOC and SOC that correspond exactly to those obtained in
unconstrained minimization, but which overcomes the difficulties
encountered by the QPM in the $\mu \to 0$ limit.
One approach is to introduce {\bf augmented Lagrange multiplier
  terms}, leading to the following modified objective function:
\bea
\AugLag(\vec{x}\,;{\boldsymbol \lambda},\,\mu) &\equiv& f(\vec{x}) - \sum_a \lambda_a c_a(\vec{x}) +
\frac{1}{2 \mu}\sum_a  c_a^{2}(\vec{x}).
\label{eq:auglag}
\eea
Note that $\boldsymbol \lambda$ is no longer determined numerically by the
optimization procedure, but is instead a fixed vector, just like the penalty parameter, $\mu$. 
As was the case for the QPM, (\ref{eq:auglag}) is used iteratively;
unconstrained minimization of $\AugLag(\vec{x}\,;\boldsymbol \lambda,\,\mu)$ is
performed for the chosen values of $\boldsymbol \lambda$ and $\mu$ in each step of
the procedure.  After optimizing $\AugLag(\vec{x}\,;\boldsymbol \lambda,\,\mu)$ to
within some tolerance, new values of $\boldsymbol \lambda$ and $\mu$ are chosen; 
the process is repeated until the desired levels of
optimality and feasibility are satisfied.  This procedure is the ALM.

We must verify that the ALM indeed avoids the problems of the QPM.
To do this, we note that 
\begin{equation}
\nabla \AugLag(\vec{x}\,;\boldsymbol \lambda,\,\mu) = \nabla f(\vec{x}) - \sum_a \lambda_a
\nabla c_a(\vec{x}) + \sum_a \frac{c_a}{\mu}\nabla c_a(\vec{x}),
\label{eq:alm-gradient}
\end{equation}
while the Hessian is given by
\begin{equation}\label{eq:alm-hessian}
H_{\AugLag}(\vec{x})_{jk} = \frac{\partial^2 f(\vec{x})}{\partial x_j \partial x_k} -
\sum_a \lambda_a \frac{\partial^2 c_a(\vec{x})}{\partial x_j \partial x_k}
+ \frac{1}{\mu} \sum_a \bigg( (\nabla c_a)_j (\nabla c_a)_k + c_a(\vec{x}) \nabla^2_{jk} c_a(\vec{x}) \bigg).
\end{equation}
We note that (\ref{eq:alm-gradient}) recovers the expression for the
method of Lagrange multipliers (\ref{eq:1stcond}) with the substitution
\begin{equation}\label{eq:alm-sub}
\lambda_a \to \lambda_a - \frac{c_a(\vec{x})}{\mu},
\end{equation}
which shows that the FOCs for optimality are the same as for the
problem of unconstrained minimization, just as in the case of the QPM.
It is also possible, albeit more challenging, to show that the SOC for a minimum is
the same as in the unconstrained case; see refs.~\cite{ALM1},
\cite{ALM2}, and~\cite{convexification_alm} for details.

We also conclude from (\ref{eq:alm-sub}) that in the asymptotic limit
\begin{equation}\label{eq:asymp-ALM}
\lambda^\ast_a \to \lambda_a - \frac{c_a(\vec{x}^\ast)}{\mu}.
\end{equation}
in analogy to (\ref{eq:asymp-QPM}).
The point of the ALM is that now that we are free to choose both
$\lambda_a$ and $\mu$, we can enforce (\ref{eq:asymp-ALM}) 
without taking the $\mu \to 0$ limit.  We shall do this by choosing the value 
of $\lambda_a$ in the $(k+1)^{\rm st}$ iteration as follows
\bea
\lambda_a^{k+1} = \lambda_a^{k} - \frac{c_a(\vec{x}_k)}{\mu_k};
\label{eq:lmalm2} 
\eea
As we do not take $\mu_k \to 0$, the Hessian
defined in (\ref{eq:alm-hessian}) should never become ill-conditioned.  
Hence the ALM avoids the major drawback of the QPM, while preserving
its successes.  We therefore implement the ALM in the {\sc Optimass} code,
as described in the following section.

\section{The {\sc Optimass} code}
\label{sec:algorithm}

\begin{figure}[t]
\centering
\includegraphics[width=12cm]{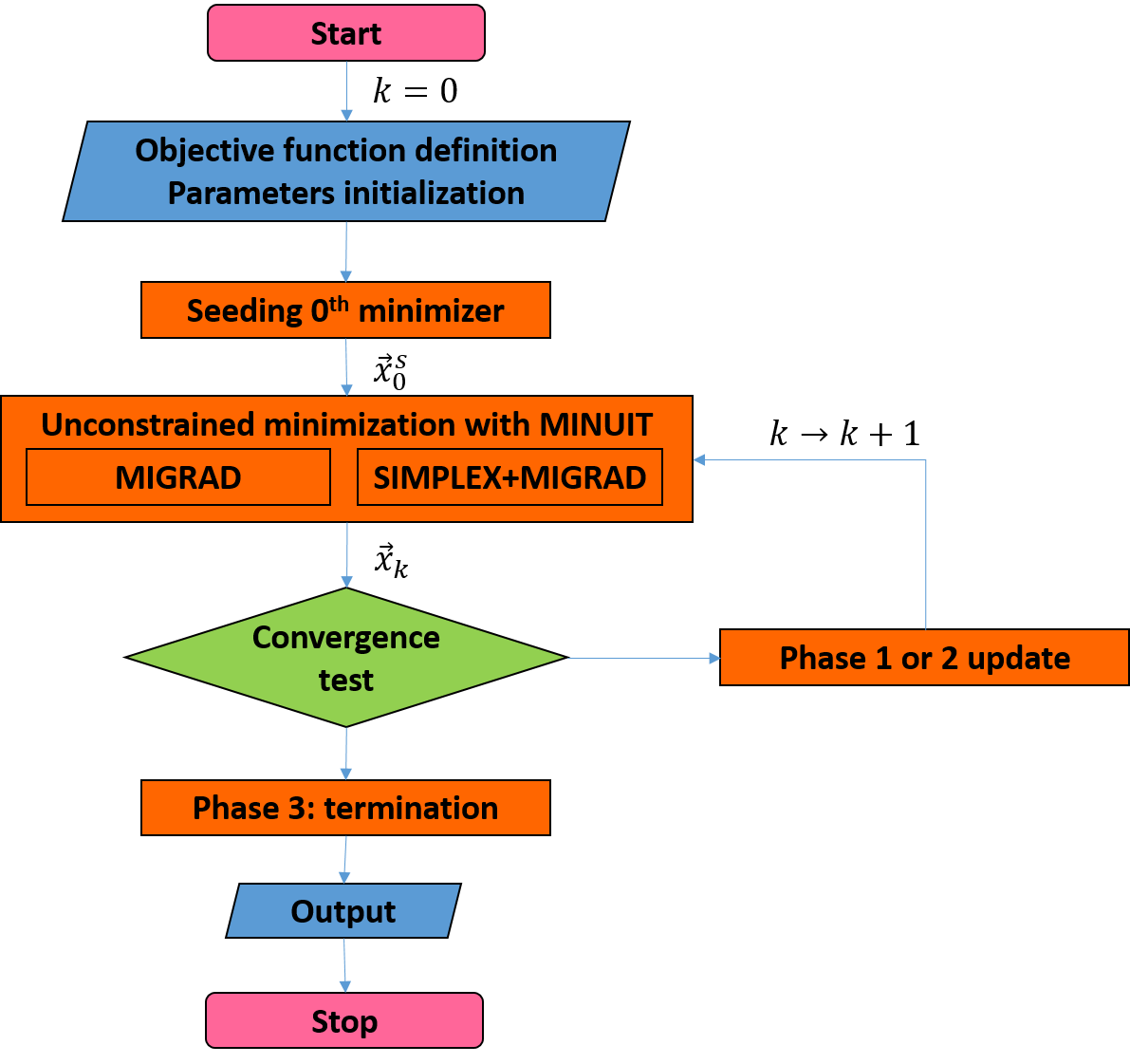}
\caption{Flowchart of the minimization procedure in {\sc Optimass}.}
\label{fig:flowchart}
\end{figure}

Having reviewed  both unconstrained and constrained minimization, 
we now state precisely how {\sc Optimass} accomplishes the task of minimizing a mass function with constraints.  
The basic algorithm is presented in figure~\ref{fig:flowchart} and is the main subject of this section\footnote{Readers 
who are not interested in the details of the code may skip directly to the next section.}.

\subsection{Operational algorithm}

\subsubsection{Step one: initialization}

At the onset, we must specify certain parameters. We first discuss
the parameters associated with the optimality condition, 
followed by those related to the feasibility condition.
Finally, we will discuss the penalty/Lagrange multiplier parameter,
and the starting value for the minimizer.

\paragraph*{Optimality condition.} The test of optimality, i.e.,  
whether the relevant sub-minimization procedure, has reached an
acceptable local minimum in the $k^{\rm th}$ iteration, 
should be performed in terms of some relevant geometric quantity. 
In {\sc Optimass}, we choose the {\sc Migrad} algorithm of {\sc Minuit}
(see Appendix~\ref{sec:Migrad}), where the optimality criterion utilizes 
the so called ``Estimated vertical Distance to the Minimum" (EDM) defined as follows
\beq
\text{EDM} \equiv 
 \frac{1}{2} 
\left( \vec{\nabla} \AugLag(\vec{x}_0) \right)^{T} \cdot
H^{-1}(\vec{x}_0) \cdot
\vec{\nabla} \AugLag(\vec{x}_0) 
= \frac{1}{2}\, \Delta\vec{x}^{\, T}\cdot H(\vec{x}_0)\cdot \Delta \vec{x}
\approx f(\vec{x}) - f(\vec{x}_0),
\label{EDMdef}
\eeq
where 
\beq
\Delta\vec{x}\equiv \vec{x}-\vec{x}_0 = - H(\vec{x}_0)^{-1}\cdot \vec{\nabla} \AugLag(\vec{x}_0). 
\eeq
Using the EDM (\ref{EDMdef}), at each iteration $k$ in {\sc Optimass} 
the optimality convergence test in {\sc Minuit} 
is performed by checking if the EDM is smaller than the optimality tolerance, $\omega_k$: 
 \beq
 \text{EDM}<\omega_k.
 \label{defoptimality}
 \eeq 
We have observed that simply setting $\omega_k$ to a constant,
$\omega^\ast$, suffices. 
By default in {\sc Migrad}, this constant is set by the internal re-parametrization
\beq
\omega^\ast = 0.001\cdot {\text {\sc tolerance}} \cdot {\text {\sc up}}.
\label{def:opt_tolerance}
\eeq
In {\sc Optimass},  the optimality tolerance $\omega^\ast$ is controlled 
and set by the ${\text {\sc tolerance}}$ parameter through the interface 
with {\sc Migrad}\footnote{The specific method to set this quantity is
$$
\label{eq:MnMigrad}
\text{{\sc MnMigrad::operator()(unsigned int maxfcn, double
    tolerance)}},
$$
where {\sc maxfcn} denotes maximum number of function calls after which the {\sc Minuit} minimization
routine will be stopped even if it has not yet obtained satisfactory
convergence to an acceptable minimum.  The default value is $\text{{\sc maxfcn}}=5000$.
}.
Its default value is set to be
\bea
{\text {\sc tolerance}}&=&0.1,
\eea
while the parameter ${\text {\sc up}}$ is not used, it is set internally in {\sc Optimass} to $\text{\sc up}=1$.

\paragraph*{Feasibility condition:} Throughout the whole minimization procedure, we
test for feasibility by computing the quantity\footnote{In eq.~(\ref{deffeasibility}), we assume that 
all constraints $c_a$ have the same mass dimension $dim(c_a) = dim(c) = 1$, 
otherwise each term in the RHS should be raised to the appropriate power of $1/dim(c_a)$.} 
\beq
||c(\vec{x}_k)||^2\equiv \sum_a c_a^2(\vec{x}_k)
\label{deffeasibility}
\eeq
in each iteration. Then, the test for feasibility is 
\bea
|| c(\vec{x}_k) || < \eta_k, 
\label{eq:eta_k}
\eea
where $\eta_k$ denotes the feasibility tolerance in the $k^{\rm th}$
iteration; this criterion evolves iteration-by-iteration,
 unlike the optimality parameter $\omega_k$. Although $\eta_k$ eventually approaches zero as $k\rightarrow\infty$, 
 we set the final convergence criterion as follows: 
\bea
\label{eq:eta_ast}
|| c(\vec{x}_k) || < \eta^\ast,
\eea
where $\eta^\ast$ denotes the terminal feasibility tolerance set to be 
\bea
\eta^\ast=0.001\times M.
\label{eq:eta_star}
\eea
Here $M$ is the appropriate typical mass scale associated with the target mass function,
its value should be chosen depending on the specific physics process at hand.
In the current version of {\sc Optimass}, the user is expected to provide 
the relevant fixed value of the scale $M$.\footnote{One possibility is
$M = \left[ f(\vec{x}_0)\right]^{1/dim(f(\vec{x}))}$,
where $dim(f(\vec{x}))$ is the mass dimension of $f(\vec{x})$.}
The rules for updating $\eta_k$ will be described in section~\ref{sec:step-three}. 
The starting feasibility tolerance is initialized by
\bea
\eta_0= \bar{\eta}\left(\min\left[\mu_0,\bar{\gamma}\right]\right)^{\beta_{\eta}^0}.
\eea
Here the pre-factor $\bar{\eta}$ is given by
\beq
\bar{\eta}=\alpha\,\cdot\,\eta^\ast,
\eeq
with some coefficient $\alpha$ chosen in the range
\beq
10\lsim \alpha \lsim 1000,
\label{eq:feasibility0}
\eeq 
$\mu_0$ is the starting penalty parameter given in (\ref{mu0}) below,
while $\bar{\gamma}\in(0,1)$, and our default for it is 
\bea
\bar{\gamma}=0.2.
\eea
By adjusting the coefficient $\alpha$ within the range (\ref{eq:feasibility0}), 
one can control the relative scale of the initial feasibility tolerance, $\eta_0$, to its terminal value, $\eta^\ast$. 
This in turn determines the relative number of iterations within each of the two regimes (phase-1 and phase-2)
described in section~\ref{sec:step-three} below. If $\alpha$ is too large, the ALM iterations in phase-1 
terminate very quickly, and the majority of the ALM iterations are performed in the regime of phase-2, where
the Lagrange-multipler driven evolution may not be efficient. On the other hand, if $\alpha$ is chosen to be
too small, most of the ALM iterations will be done in the regime of phase-1, which only reduces the penalty parameter 
$\mu_k$. In that case, {\sc Optimass} will not be able to take full advantage of the ALM method in avoiding the
ill-conditioning as explained in section~\ref{sec:pen}. We recommend that users test several different values of 
$\alpha$, until the number of iterations in each phase is adequate and the results are stable.

Finally, the ``tightening'' parameters, $\beta_{\eta}^k\in(0,1)$, for the feasibility constraint are set to be
\bea
\beta_{\eta}^0&=&0.5, \\
\beta_{\eta}^k&=&0.3 \;\;\; (k\geq 1).
\eea
\paragraph*{Penalty and Lagrange multiplier parameters:} The penalty
parameter, $\mu$, and Lagrange multiplier parameter vector,
 ${\boldsymbol \lambda}$, are updated in each iteration (the actual assignment rule will be 
explained below in section~\ref{sec:step-three}). Their starting values are 
\bea
\mu_0 &=& 0.1 \label{mu0}\\
\lambda_{a}^0 &=& 0\;\;\;(a=1,\cdots,m).
\eea
For the reduction of the penalty parameter $\mu$, {\sc Optimass} introduces another parameter, $\tau_{\mu}\in(0,1)$, 
defined in (\ref{tauupdate}) and by default set to 
\bea
\tau_{\mu}=0.5.
\eea
If the value of $\tau_{\mu}$ is too small, the penalty parameter $\mu$ may decrease too quickly, 
causing strong convexification, which can lead to ill-conditioning. Conversely, if the value of 
$\tau_{\mu}$ is too large, one can experience slow convergence and a premature transition 
to phase-2 (see section~\ref{sec:step-three} below).

\paragraph*{Initial minimizer and initial step size for {\sc Minuit}:} The initial guess for the minimizer of the objective function, $\vec{x}_0^s$, 
(referred to as ``seeding'' in the flowchart of figure~\ref{fig:flowchart}),
 is set by the invisible momentum configuration corresponding to $\hat{s}_{\min}$~\cite{Konar:2008ei},
 i.e., the invisible momenta are such that the total invariant mass in the event is minimized. 
In addition, the initial step size, $\Delta \vec{x}_0^s$, from from the $\vec{x}_0^s$ toward the final minimizer, $\vec{x}^*$, 
is another input parameter for the {\sc Minuit} initialization\footnote{The initial input values for $\vec{x}_0^s$ and $\Delta \vec{x}_0^s$
can be set via the method
$$
\label{eq:MnUserParameters}
\text{{\sc MnUserParameters::Add(const char* par-name, double init-point, double init-step-size)}}.
$$
}.

\subsubsection{Step two: unconstrained minimization with {\sc Minuit}}

Once the initial parameters have been chosen, 
we use the ALM mapping of a constrained minimization problem to an unconstrained minimization problem
and then perform the latter minimization with {\sc Minuit} (see Appendix~\ref{sec:minuit}).  
In the process of this minimization, we perform an appropriate adjustment of parameters in each step of the algorithm. 
The code has two different options for minimization with {\sc Minuit} in the $k^{\rm th}$ iteration:
\begin{enumerate}
\item {\sc Migrad}: \\
Using  as a starting value the minimizer, $\vec{x}_{k-1}$, obtained in the previous iteration, 
{\sc Migrad} searches for a minimizer, $\vec{x}_{k,1}$.
\item {\sc Simplex} {\bf and} {\sc Migrad}: \\
Again using as a starting value the minimizer, $\vec{x}_{k-1}$, obtained in the previous iteration, 
{\sc Simplex} first finds a minimizer, $\vec{x}_{k,S}$, and then {\sc Migrad} 
uses $\vec{x}_{k,S}$ as a starting value to find the final minimizer, $\vec{x}_{k,2}$. 
\end{enumerate}
Among the two possible answers, $\vec{x}_{k,1}$ and $\vec{x}_{k,2}$, we 
choose as our minimizer, $\vec{x}_{k}$, the one which gives a lower value for the objective function. 
One could repeat either algorithm 1 or algorithm 2 (or other
minimization procedures) with various starting points, $\vec{x}_0$, to find a global minimum more accurately. 
However, we find that the combination of algorithms 1 and 2 described above is adequate to obtain 
accurate on-shell constrained $M_2$ values, while keeping the computational effort to a minimum.
Interestingly, we find that algorithms 1 and 2 are complementary to each other. For example,
in the $M_2$ calculations of section~\ref{sec:application}, we found that for the maximally 
constrained case of $M_{2CC}$, the final solution, $\vec{x}_{k}$, was given by the answer $\vec{x}_{k,1}$ from 
algorithm 1 ($\vec{x}_{k,2}$ from algorithm 2)  in $83\%$ ($17\%$) of the events.
This trend was reversed in the calculation of the minimally constrained case of $M_{2XX}$,
where algorithm 1 (algorithm 2) supplied the final solution $\vec{x}_{k}$ in $19\%$ ($81\%$) of the events.
This behavior can be understood as follows. {\sc Migrad} relies on gradient information, 
thus it can be fast and accurate if the objective function is smooth and continuous. 
On the other hand, {\sc Simplex} does not require gradient information, and can handle
more complicated functions (including ``folds" and ``creases"), although the ultimate accuracy is not as high. 
In the case of $M_{2CC}$, the objective function is convexified by the penalty terms, 
which makes the relevant geometry near the local minimum smooth and well-defined, 
thus we expect algorithm 1 by {\sc Migrad} to outperform algorithm 2.
However, it is known that the $M_{2XX}$ objective function,
``the maximum of the two invariant masses", develops a crease, 
on which the solution is found \cite{Cho:2014naa}, and
therefore one might expect algorithm 2 by {\sc Simplex} to work better for this case. 
The performance of {\sc Optimass} with regard to the $M_2$ variables will
be discussed in more detail in section~\ref{sec:application}.

\subsubsection{Step three: ALM parameter adjustment}
\label{sec:step-three}

Once the minimization routine has obtained a value of
the minimizer, $\vec{x}_k$, we then evaluate the constraints at this minimizer, i.e., $c_a(\vec{x}_k)$.  
Depending on the value of the feasibility (\ref{deffeasibility}), 
we define three phases: ``Phase 1'', ``Phase 2'', and ``Phase 3''.  
While ``Phase 3'' is nothing but terminating the entire minimization procedure, 
the other two phases basically tighten the feasibility tolerance. 
Our tightening scheme is inspired by the {\sc LANCELOT} package~\cite{lancelot}. 
As mentioned earlier, the tolerance, $\omega_k$, for the optimality condition (\ref{defoptimality}) is {\it not} evolved:
$\omega_k=\omega^\ast$. 

\paragraph*{Phase 1.} The feasibility condition is far from satisfied:
\beq
||c(\vec{x}_k)||>\max\left[\eta_k,\eta^*\right]. 
\eeq
In this case we put more weight on the penalty term by reducing $\mu_{k+1}$ for the next iteration. 
At the same time, the Lagrange multiplier vector, ${\boldsymbol \lambda}_k$, remains unchanged. 
In the next iteration, the starting value of the minimizer, $\vec{x}_{k+1}^s$, is set to be the minimizer 
obtained in the previous iteration, $\vec{x}_{k}$. Finally, the feasibility tolerance, $\eta_{k+1}$, is also evolved. 
The detailed updating rules for Phase 1 are thus
\bea
\mu_{k+1}&=&\tau_{\mu}\cdot \mu_k, \label{tauupdate}\\ [1mm]
\lambda_a^{k+1}&=&\lambda_a^{k}, \\ [1mm]
\vec{x}_{k+1}^s&=&\vec{x}_{k}, \\ [1mm]
\eta_{k+1}&=& \bar{\eta}(\bar{\gamma}\cdot\mu_{k+1})^{\beta_{\eta}^{k+1}}.
\eea

\paragraph*{Phase 2.} 
The feasibility condition is converging, but insufficient to terminate the algorithm:
\beq
\eta^*<||c(\vec{x}_k)||<\eta_k.
\eeq
In this case, we do not reduce the penalty parameter, $\mu_k$, and instead 
adjust the values of the Lagrange multiplier vector by the rule in eq.~(\ref{eq:lmalm2}). 
The starting value of the minimizer, $\vec{x}_{k+1}^s$, for the next iteration is set as in Phase 1. 
The feasibility tolerance is also modified. The detailed updating rules are given by
\bea
\mu_{k+1}&=&\mu_k, \\ [1mm]
\lambda_a^{k+1}&=&\lambda_a^{k}-\frac{c_a(\vec{x}_k)}{\mu_k}, \\ [1mm]
\vec{x}_{k+1}^s&=&\vec{x}_{k}, \\ [1mm]
\eta_{k+1}&=& \eta_k\cdot \mu_{k+1}^{\beta_{\eta}^{k+1}}.
\eea
   
\paragraph*{Phase 3.} In this phase, we have achieved sufficient feasibility:
\beq
||c(\vec{x}_k)||<\eta^\ast.
\eeq
 We therefore break the sub-minimization loop and return $\vec{x}_k$ as the final value, $\vec{x}^\ast$, of the minimizer.

\subsection{Validation}
\label{sec:validation}

We now demonstrate the performance of the algorithm described in the
previous subsection 
with two simple examples, for which one can also obtain analytic
solutions for the minimizer, $\vec{x}^*$, and the 
Lagrange multiplier vector, ${\boldsymbol \lambda}^*$. The first example, considered in section~\ref{sec:example-1},
yields a well-defined solution at a unique global minimum. In the second example, treated in section~\ref{sec:example-2},
we find that the solution for the Lagrange multiplier vector is {\it not} well-defined because the 
relevant objective function is ``folded'' and is not differentiable at $\vec{x}^*$. 
The examples illustrate the evolution of the ALM parameters and demonstrate how the solution 
found in the $k^{\rm th}$ iteration converges to the true value in terms of feasibility and optimality. 

\subsubsection{Example one} 
\label{sec:example-1}

Our first example involves minimizing the objective function 
\begin{equation}
\label{eq:ex-1-objective}
f(x,y) = x+y,
\end{equation}
over the usual plane, $\vec{x}\equiv (x,y)$, subject to the constraint
\begin{equation}
\label{eq:ex-1-constraint}
x^2+y^2-1=0.
\end{equation}
This constraint implies that our solution must lie on a unit circle centered at the origin.  
Clearly, the function (\ref{eq:ex-1-objective}) is minimized along the circle at the point
\begin{equation}
\label{eq:ex-1-analytic-solution}
(x^\ast, y^\ast) = \bigg( -\frac{\sqrt{2}}{2},-\frac{\sqrt{2}}{2}\bigg), 
\end{equation}
which is the global minimizer in this example.

\begin{figure}[t]
\includegraphics[trim= 1cm 1cm 1.5cm 1.5cm, width=7cm]{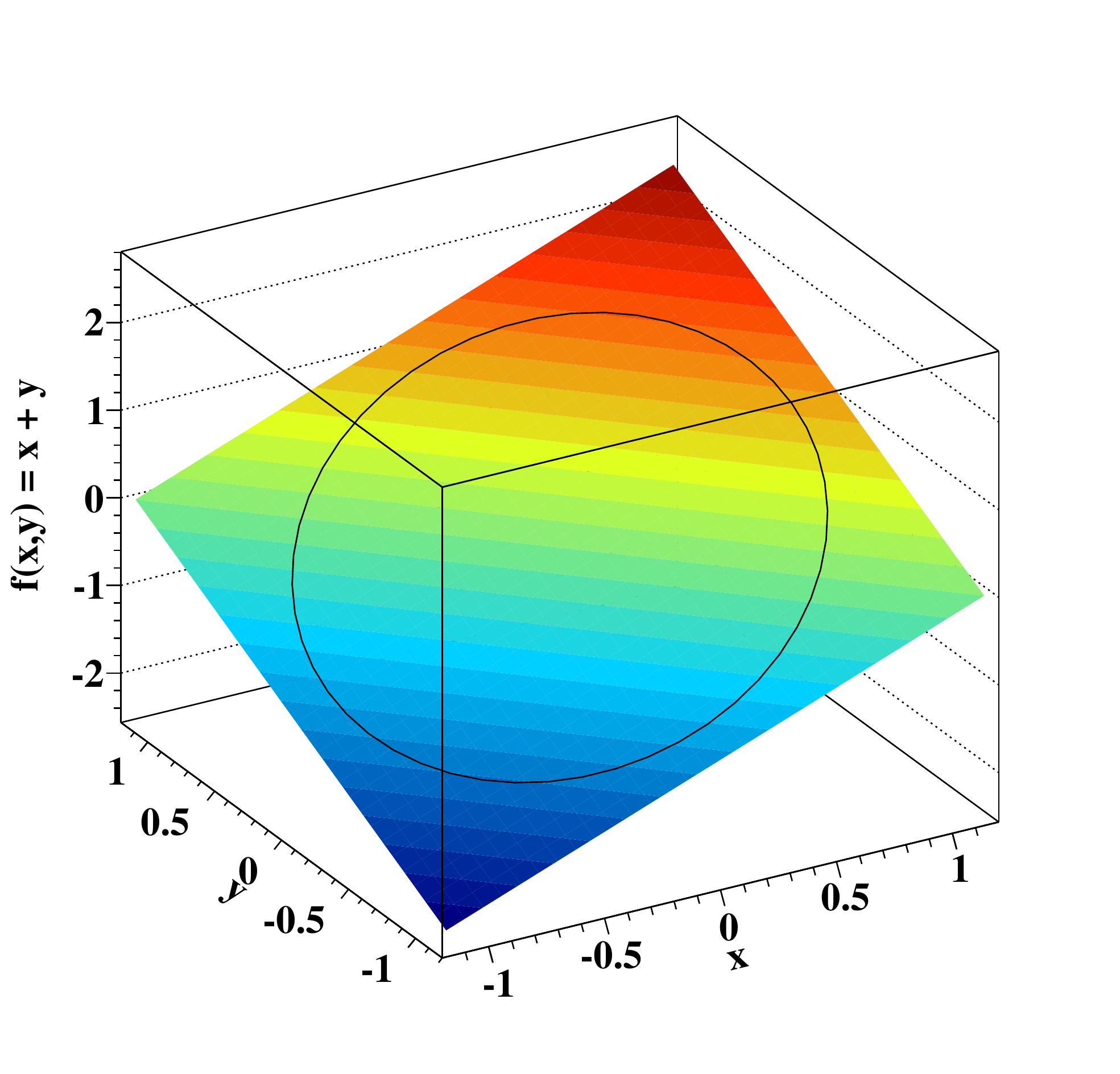}
\includegraphics[trim= 0.cm 1cm 3cm 1cm, width=7cm]{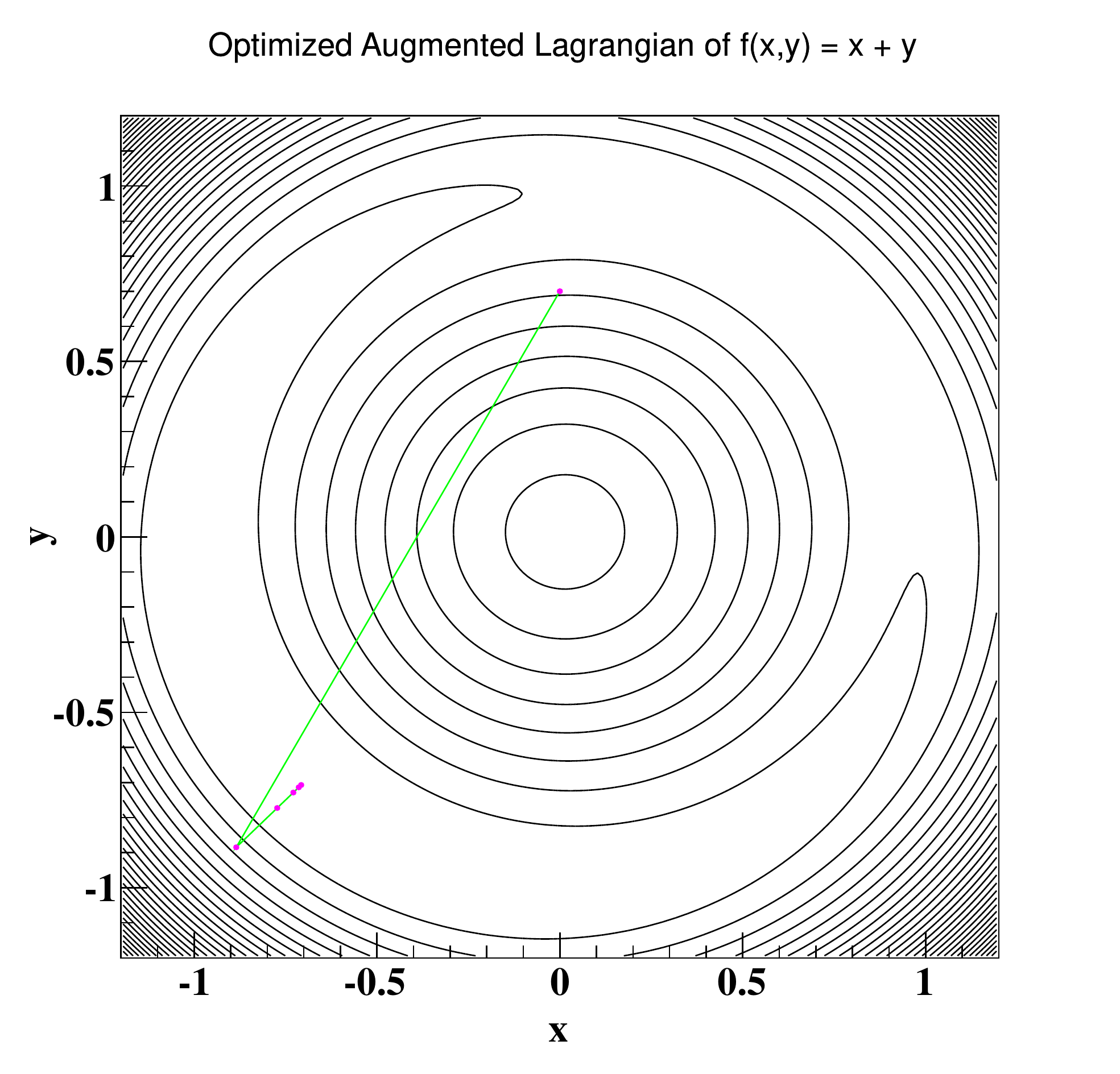}
\caption{Test of the ALM for the objective function $f(x,y)=x+y$
  subject to $x^2+y^2=1$. (a) Plot of the objective function (color coded) and
  the constraint curve (in black). (b) Contour plot of the augmented Lagragian (\ref{eq:ex-1-alm}) for
  the last (fifth) iteration. The magenta points denote the minimizers, $\vec{x}_k$, found at each iteration.}
\label{fig:example1}
\end{figure}

The objective function, (\ref{eq:ex-1-objective}), is plotted in the left panel of figure~\ref{fig:example1}.
The locus of feasible points (i.e., the unit circle about the origin,) is shown in black.
With the Lagrange multiplier method, one adds a Lagrange multiplier
term to the objective function as in (\ref{eq:lag function})
\begin{equation}
\label{eq:ex-1-lag}
\mathcal{L}(x,y,\lambda) = x+y - \lambda\, (x^2+y^2-1).
\end{equation}
There are two \emph{stationary points} given by
\begin{equation}
\label{eq:ex-1-stationary}
 (x^*,\,y^*\,,\,\lambda^*) = \pm \bigg(\frac{\sqrt{2}}{2}, \frac{\sqrt{2}}{2}\, ,\,\frac{\sqrt{2}}{2} \bigg).
\end{equation}
The Hessian corresponding to the function (\ref{eq:ex-1-lag}) is
\begin{eqnarray}
\label{eq:ex-1-hess}
H_{\mathcal{L}}=\begin{pmatrix}
-2\lambda & 0 \\
0 & - 2\lambda
\end{pmatrix},
\end{eqnarray}
and the condition for a minimum (i.e., that $H_{\mathcal{L}}$ should be positive definite) 
requires us to choose $\lambda < 0$, which selects the correct minimizer among the two
stationary points (\ref{eq:ex-1-stationary}):
\begin{equation}
\label{eq:ex-1-solution}
(x^*,\,y^*\,,\,\lambda^*) =
(-\frac{\sqrt{2}}{2},-\frac{\sqrt{2}}{2}\,,\,-\frac{\sqrt{2}}{2} ),
\end{equation}
confirming the earlier result (\ref{eq:ex-1-analytic-solution}).

We now wish to verify that the {\sc Optimass} algorithm reproduces this
solution.
After running the code, we obtain
\begin{equation}
\label{eq:example1sol}
(x^*,y^*;\lambda^*)=(-0.707106,-0.707106;-0.707180),
\end{equation}
which is consistent with (\ref{eq:ex-1-solution}) ($\sqrt{2}/2 = 0.707107...$).
We note that this convergence only required five steps, suggesting that
the minimum was found relatively easily. The right panel of
figure~\ref{fig:example1} shows a contour plot of the
augmented Lagrangian,
\begin{eqnarray}
\label{eq:ex-1-alm}
\AugLag(x,y;\mu_5,\lambda_5)&=&(x+y)-\lambda_5\,(x^2+y^2-1)+\frac{1}{2\mu_5}
(x^2+y^2-1)^2\quad\text{with} \\
(\mu_5,\lambda_5) &=& (0.027,-0.707180),\nonumber
\end{eqnarray}
for the final minimization step ($k=5$) at which the correct
solution, (\ref{eq:example1sol}), was obtained. The convexification
due to the penalty term
\begin{equation}
\frac{1}{2\mu_5} (x^2+y^2-1)^2 \nonumber
\end{equation}
 ensures that the solution of the constrained optimization problem
 is (at least) a local minimum. The magenta dots (connected by green lines) in
 the figure denote the solution, $\vec{x}_k$, of the $k^{\rm th}$ minimization,
 starting with a randomly chosen initial point, $\vec{x}_0^s=(0,0.7)$. 
\begin{figure}[t]
\begin{center}
\includegraphics[width=7.2cm]{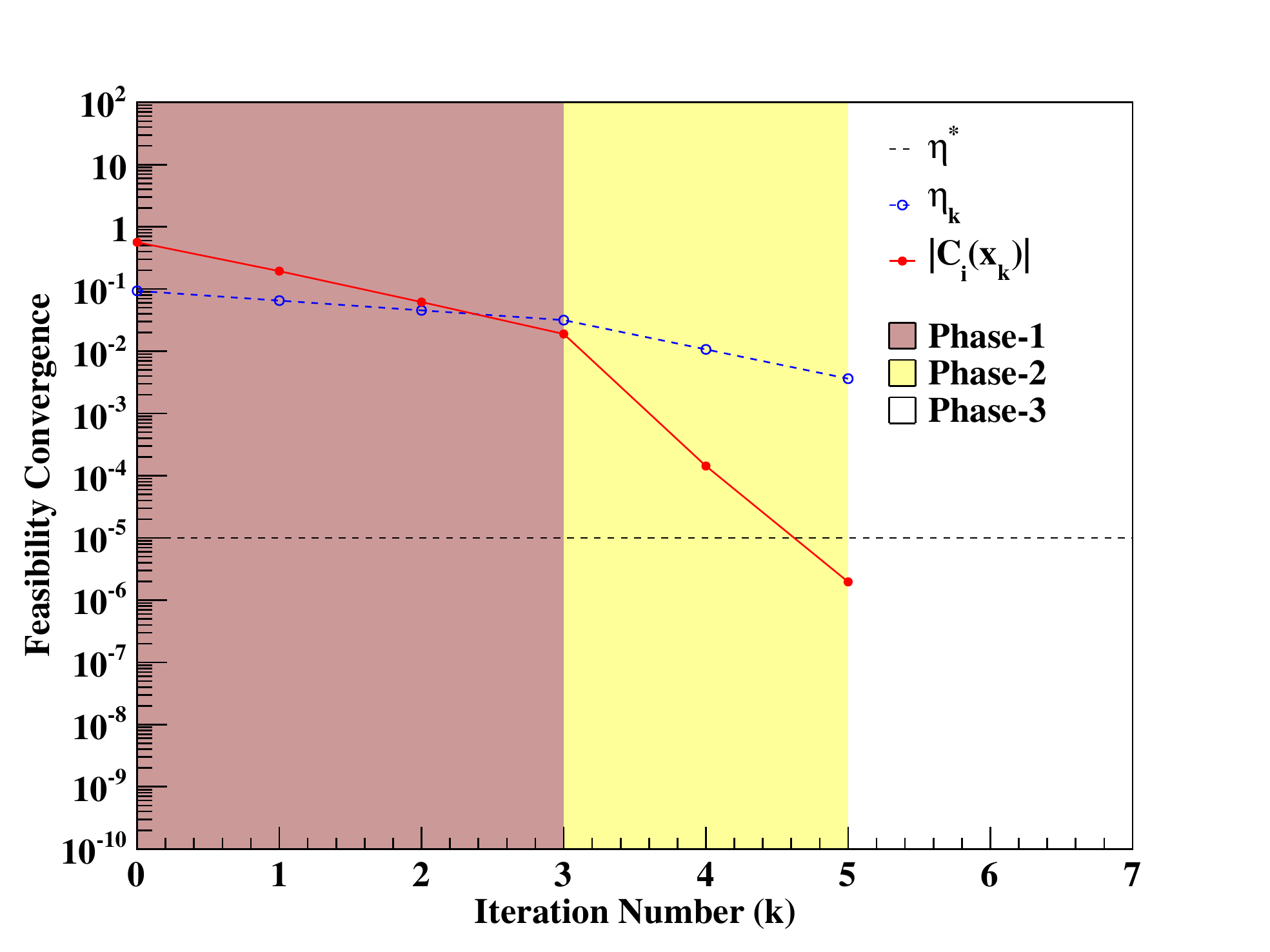}
\includegraphics[width=7.2cm]{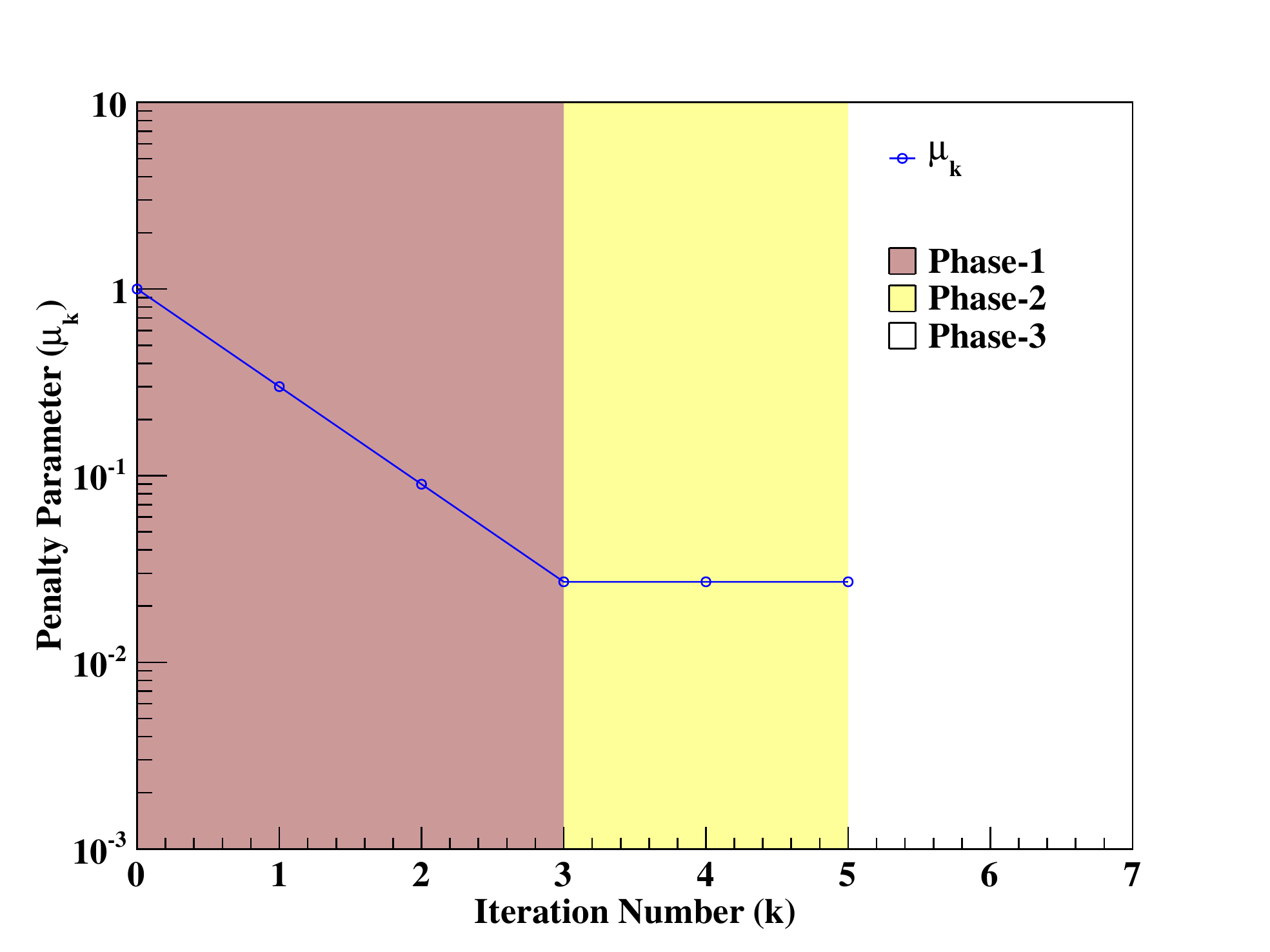}\\
\includegraphics[width=7.2cm]{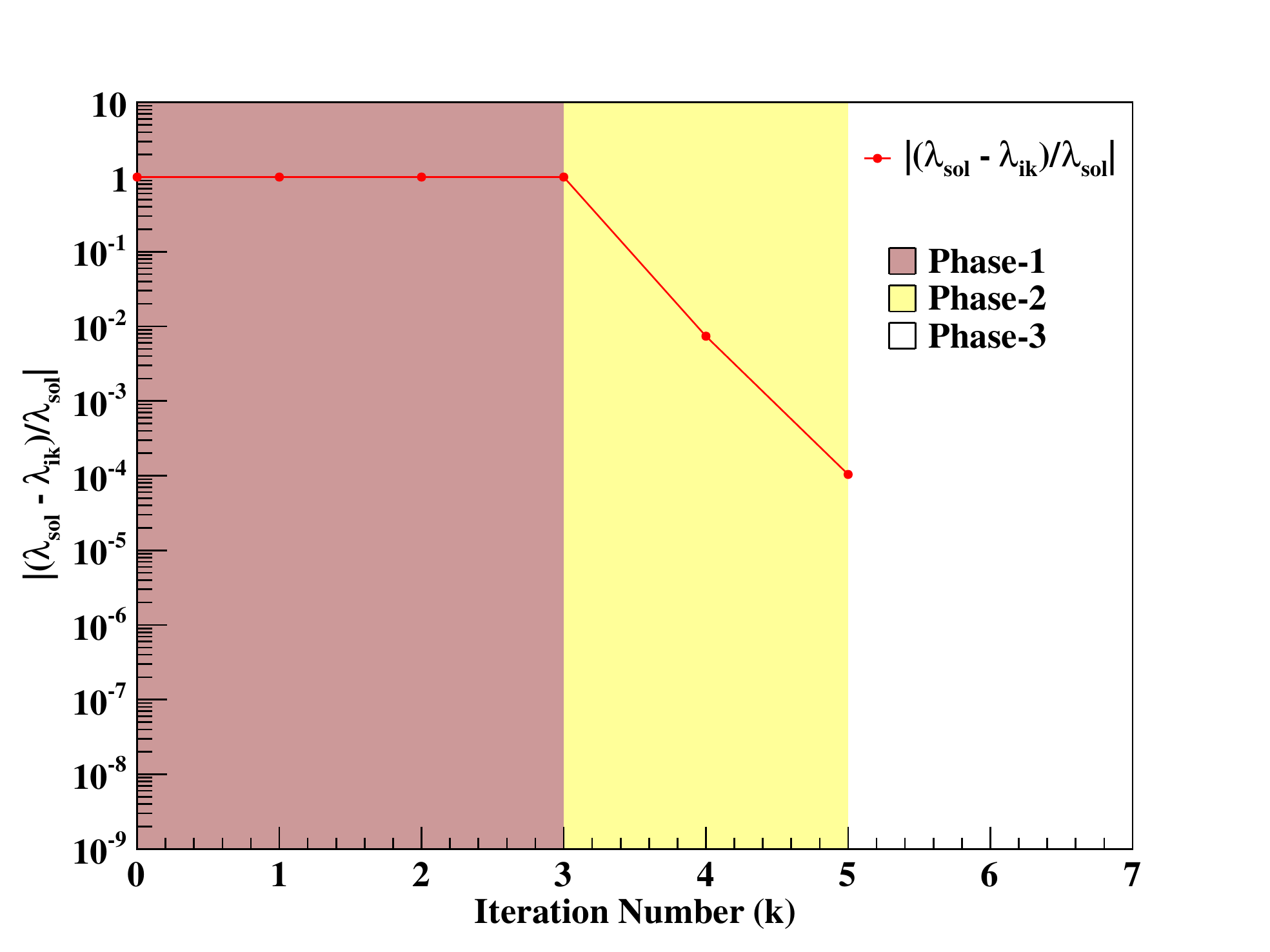}
\includegraphics[width=7.2cm]{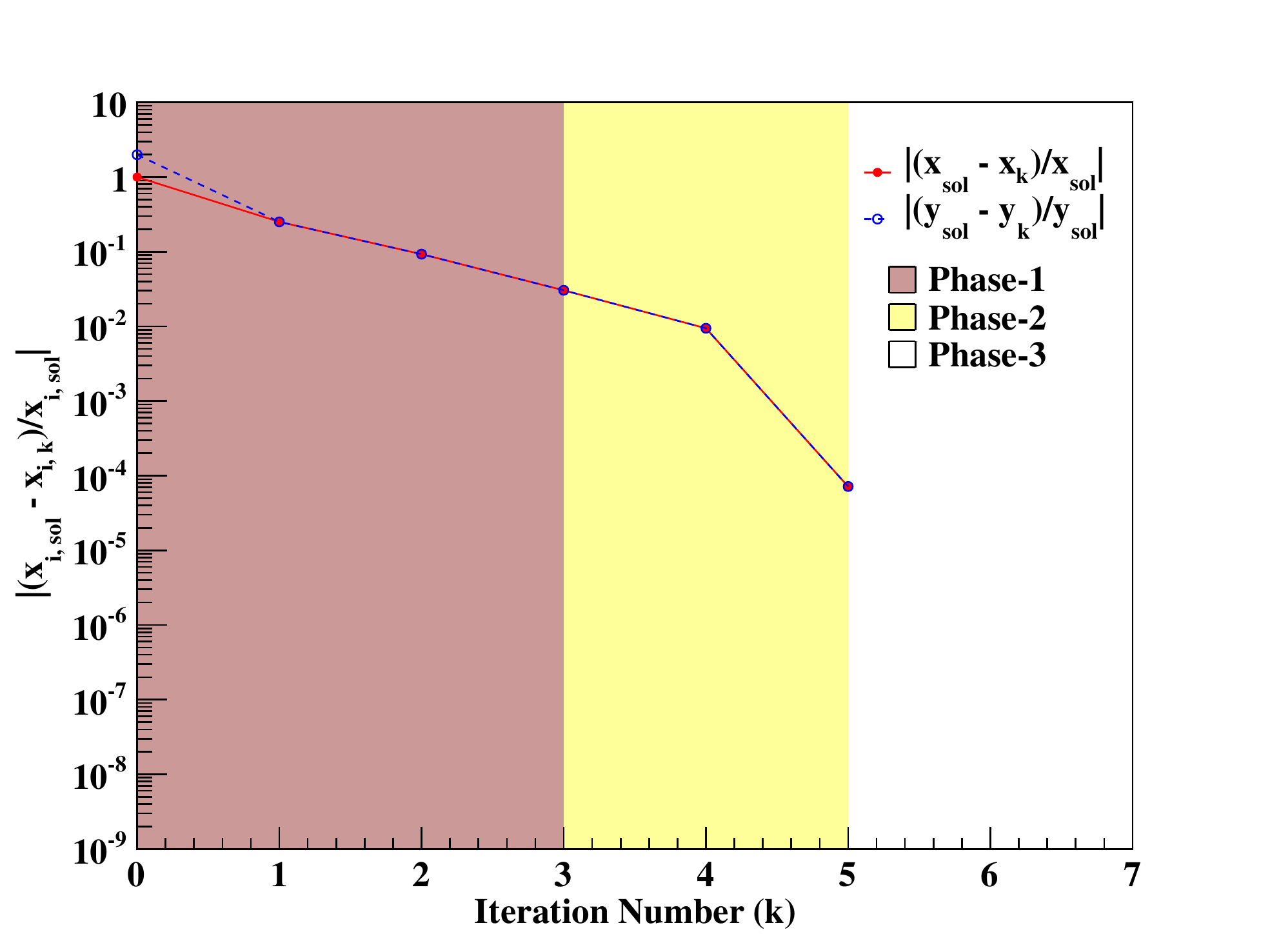}
\end{center}
\caption{Tracing the convergence state of the solution and the 
evolution of relevant parameters of the ALM algorithm for the minimization
problem depicted in figure~\ref{fig:example1}.}
\label{fig:example1trace}
\end{figure}

In the four panels of figure~\ref{fig:example1trace}, we trace
the evolution of several parameters of the algorithm as well as the
properties of the approximate solutions, $\vec{x}_k$.  
The upper left panel shows the evolution of the
intermediate feasibility tolerance, $\eta^k$, (blue dashed line) and the real
feasibility, $||c_i(x_k)||$, (red solid line) calculated at the end of each iteration, as well as
the scale set by the ultimate feasibility tolerance, $\eta^\ast$,
given by the horizontal black dashed line.
We see that in the first two iterations $||c_i(x_k)||>\eta^k$,
and so we are in Phase 1(brown shade). The onset of Phase 2 is marked
by the crossing of the red solid and blue dashed lines; in iterations 3 and 4 
we are in Phase 2. The terminal Phase 3 is entered when the red solid line 
dips below the horizontal black dashed line marking the value of $\eta^\ast$,
and this finally occurs during the 5th iteration.

The upper right and lower left panels in figure~\ref{fig:example1trace} respectively show 
the evolution of (the initial values of) the penalty parameter, $\mu_k$, and the Lagrange multiplier, $\lambda_k$,
in each iteration. We see that during Phase 1, $\mu_k$ was updated while
$\lambda_k$ was fixed, while in Phase 2, $\lambda_k$
was updated while $\mu_k$ was held fixed.  
Throughout this process, the solution, $\vec{x}_k$, from each
step gradually converges to the analytic solution, (\ref{eq:ex-1-analytic-solution}),
as shown in the lower right panel in figure~\ref{fig:example1trace}. 
Note that from the first iteration on, the solutions are on the
diagonal line $x=y$ (see also the right panel in figure~\ref{fig:example1}) --- except for their starting values, 
the red and blue lines in the lower right panel of figure~\ref{fig:example1trace}
essentially overlap.

\subsubsection{Example two}
\label{sec:example-2}

\begin{figure}[t]
\includegraphics[trim= 1cm 1cm 1.5cm 1.5cm, width=7cm]{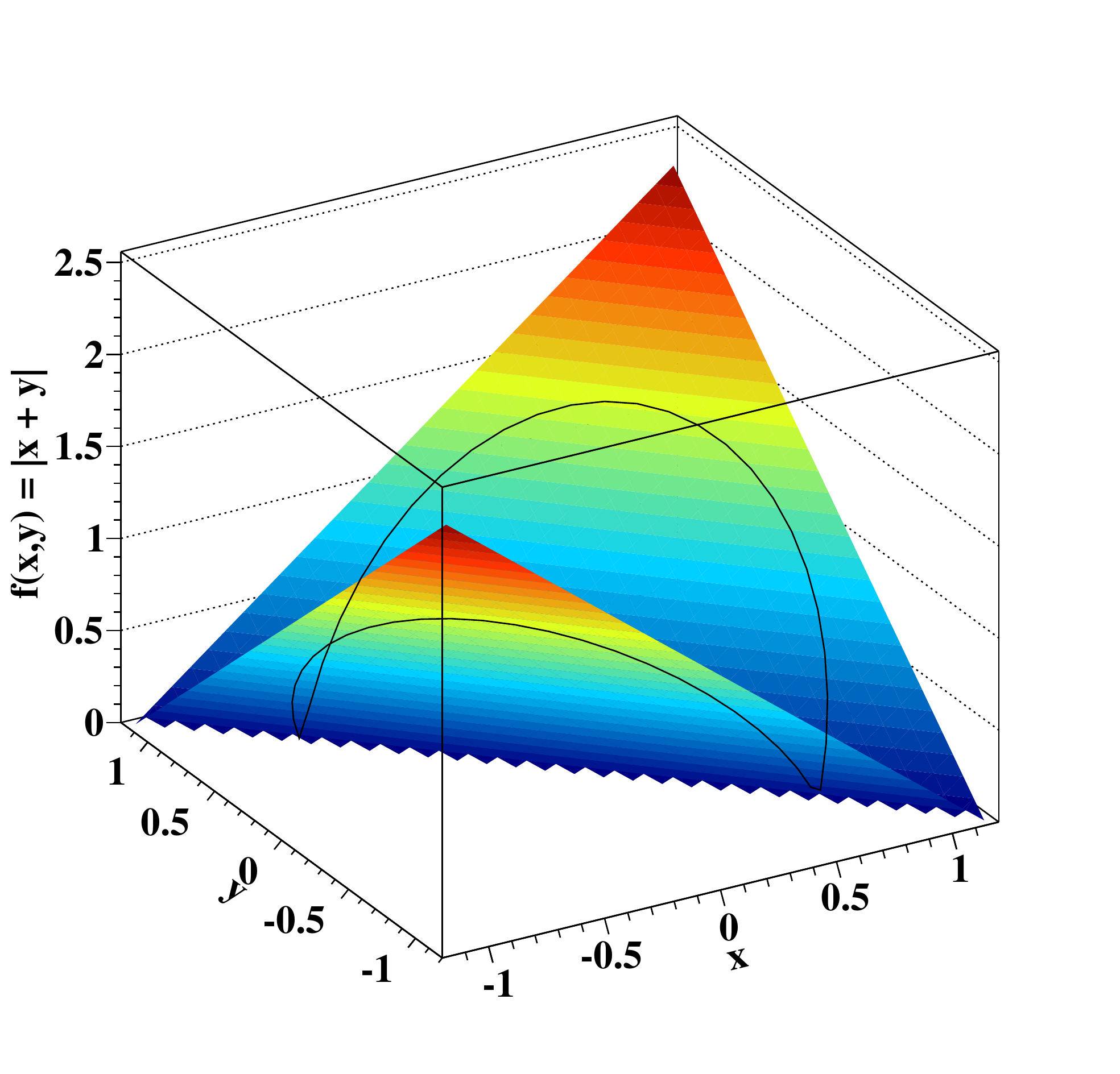}
\includegraphics[trim= 0.cm 1cm 3cm 1cm, width=7cm]{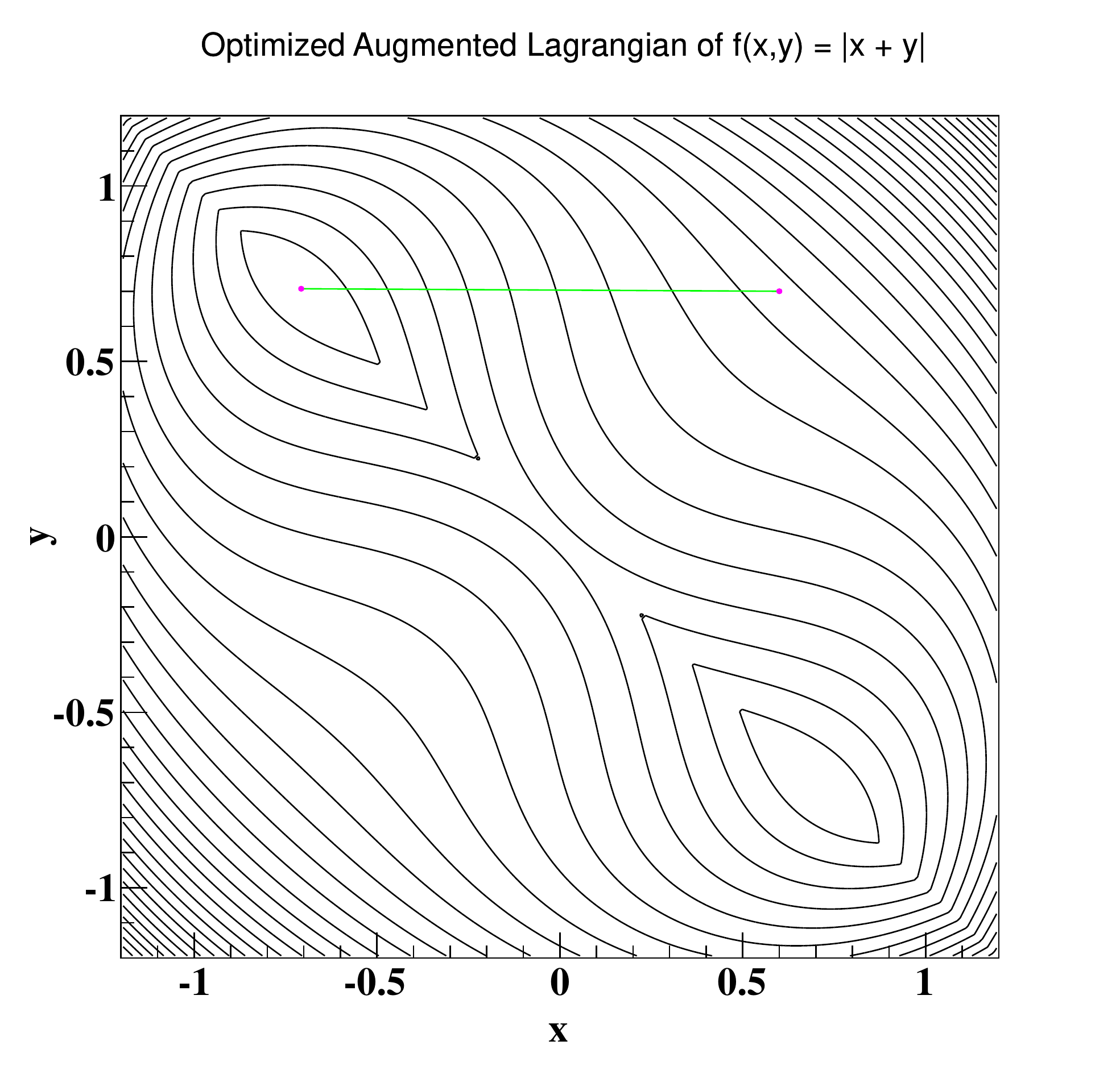}
\caption{The same as figure~\ref{fig:example2}, but using (\ref{eq:example2}) as an objective function instead.
  }
\label{fig:example2}
\end{figure}

Our second example is very similar to the one considered in the previous subsection, 
except now we change the objective function to 
\begin{equation}
f(x,y) = \left|x+y\right|,
\label{eq:example2}
\end{equation}
and we keep the same constraint as before:
\begin{equation}
x^2+y^2-1=0.
\label{eq:ex-2-constraint}
\end{equation}
The left panel in figure~\ref{fig:example2} plots the objective function,  $f(x,y)$, 
as well as the feasible set (the unit circle centered on the origin).  
Note the ``fold'' along the line $y=-x$.  On this line, the objective function 
is not a smooth function, hence one of the basic
assumptions generally employed in the theory of constrained
optimization (see section~\ref{sec:overview}) is not satisfied.  
Nevertheless, in such cases we typically find that the {\sc Optimass} 
algorithm still converges efficiently to the correct solution.

It is clear from figure~\ref{fig:example2} that the current problem has a
two-fold ambiguity, there are two equivalent solutions for the global minimizer:
\bea
(x^*,\,y^*) = \left(\frac{\sqrt{2}}{2}, -\frac{\sqrt{2}}{2}\right) \qquad {\rm or}\qquad 
(x^*,\,y^*) = \left(-\frac{\sqrt{2}}{2}, \frac{\sqrt{2}}{2}\right).
\label{eq:example2sol1}
\eea
Starting from the initial point $(x_0^s,y_0^s)=(0.6, 0.7)$, {\sc Optimass} converges to
\bea
(x^*,y^*)=(-0.707132, 0.707132)
\label{eq:example2sol}
\eea
in a single iteration (i.e., without taking $\mu \to 0$).
The right panel in figure~\ref{fig:example1} shows a contour plot 
of the augmented Lagrangian,
\bea
\AugLag(x,y;\mu_0,\lambda_0)=\left| x+y\right|-\lambda_0\,(x^2+y^2-1)+\frac{1}{2\mu_0}\left(x^2+y^2-1\right)^2,
\eea
for the only step that the algorithm needed, the initial step. 
The ``folded valley'' feature of the objective function
suggests that one does not need much additional convexification from the
penalty term (since we are already in a valley),  
and the algorithm converges very quickly.

In both of these two toy examples, as well as in numerous physics motivated studies described in the next section, 
we verified that the numerical solutions obtained with {\sc Optimass} are stable with respect to small variations of
the default initial values of the parameters, and in particular $\vec{x}^s_0$.

\section{Calculating $M_2$ variables with {\sc Optimass}}
\label{sec:application}

In this section, we describe the main intended use of {\sc Optimass},
the calculation of kinematic variables suitable for analyzing missing energy
events at hadron colliders. The calculation involves a minimization of a mass function 
over a number of invisible particle momenta, subject to certain constraints
(e.g., on-shell constraints, or the missing transverse momentum constraint (\ref{eq:mpt})).
In particular, we will show how the code can be used to calculate the recently proposed $M_2$
variables with non-linear constraints~\cite{Barr:2011xt,Mahbubani:2012kx,Cho:2014naa}.

We first provide a brief review of the $M_2$ variables;
then define the relevant objective function and identify the sorts
of constraints that may be imposed.  We then demonstrate the
performance of {\sc Optimass} in the calculation of $M_2$ in the
physically important case of top quark pair production, when
both tops decay leptonically.

\subsection{Introduction to $M_2$}
\label{m2-intro}

The $M_2$ variable~\cite{Barr:2011xt,Mahbubani:2012kx,Cho:2014naa} is a $(3+1)$-dimensional analogue of the
well-known $M_{T2}$ variable~\cite{Lester:1999tx}. Both are typically applied 
to final states that may result from (a) the pair\footnote{Hence the subscript ``2".} production of
``mother'' particles that (b) subsequently decay to both visible
and invisible particles. The best motivated scenarios typically have too many 
invisible particles in the final state, so that we cannot, in general, reconstruct the
masses or momenta of all of the intermediate particles in the event with certainty.
Both $M_2$ and $M_{T2}$ are thus constructed to provide an ansatz 
for the invisible particle momenta. This ansatz involves the minimization 
of a suitably defined kinematic mass function of the visible and invisible momenta in the event.
$M_{T2}$, by construction, is restricted to the transverse plane, and
does not involve the longitudinal momenta of the invisible particles.  
On the other hand, $M_2$ is not limited to the transverse plane, 
and thus provides an ansatz for the longitudinal invisible momenta as
well.  This ansatz
can be usefully applied to particle mass reconstruction and event topology disambiguation
\cite{Cho:2014naa,Cho:2014yma}:
once we obtain values for the three-momenta of the invisible particles, 
we can work backwards to reconstruct the masses of the
heavier particles produced in the intermediate steps of the relevant decay
in terms of the hypothesized masses of the invisible particles in the final state. 
While the mass of such intermediate resonances is {\it a priori}
unknown, in symmetric event topologies the two decay sides are
identical (by definition), and we can impose the condition that the
mass of an intermediate resonance of interest is the same in both decay chains.
Adding such on-shell constraints further restricts the
allowed solution space for the individual invisible momenta, 
leading in general to a different outcome from the procedure of minimization.
Thus the imposition of different on-shell constraints leads
to new, physically-motivated kinematic variables. 

\begin{figure}[t]
\centering
\includegraphics[scale=0.9]{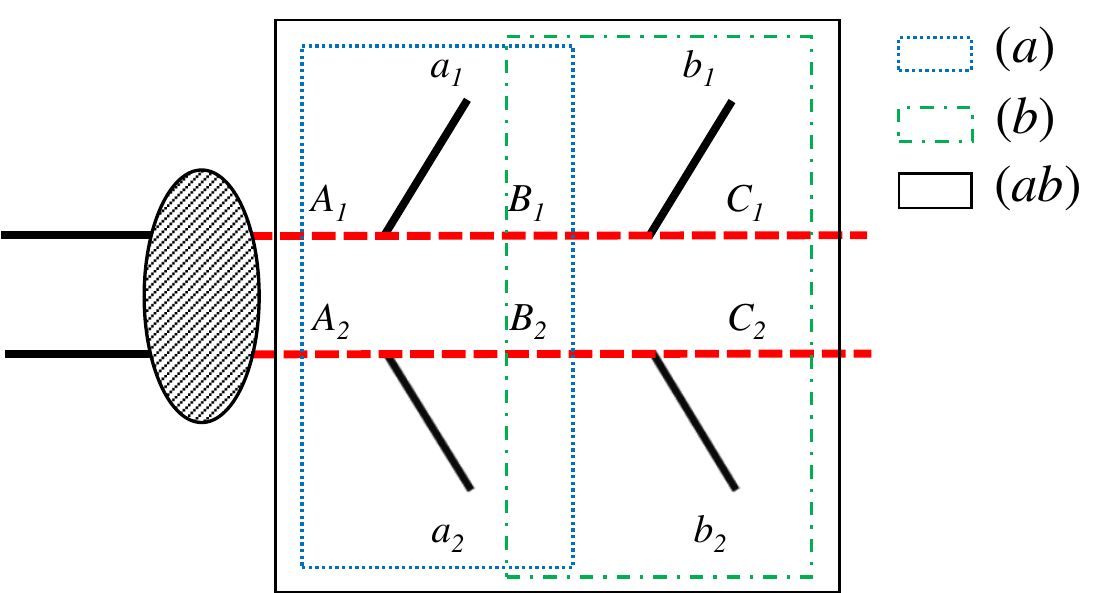}
\caption{The event topology for the decay process in~eq.~(\ref{eq:ModelProcess}), 
  together with the three possible subsystems. The blue dotted, the green dot-dashed, and
  the black solid lines indicate the subsystems $(a)$, $(b)$, and
  $(ab)$, respectively. \label{fig:decaysubsystem}}
\end{figure}

Let us consider for concreteness a process in which a pair of heavy particles,
$A$, undergo identical two-step, two-body, cascade decays, i.e., each
$A$ decays into two (massless) visible particles, $a$ and $b$, plus a
(massive) invisible particle, $C$, via an on-shell intermediate particle,
$B$, (see figure~\ref{fig:decaysubsystem})
\bea
A \rightarrow a\; B \rightarrow a\; b\; C.
\label{eq:ModelProcess}
\eea
For simplicity, we assume that all visible particles are fully
distinguishable and that particle $a$ is emitted before particle $b$,
i.e., we do not address the combinatorial issues since they are not relevant 
for the current discussion of computing the $M_2$ variables. 

Given the event topology of figure~\ref{fig:decaysubsystem},
one can consider three different subsystem topologies, depending on 
which of the particles along the red dashed lines are treated as
mothers and which are treated as daughters
\cite{Burns:2008va,Cho:2014naa}. 
For example, considering the event as a whole corresponds to subsystem $(ab)$,
in which $A_i$ are the mothers, $C_i$ are the daughters, while 
$B_i$ are intermediate resonances, dubbed ``relatives" in \cite{Cho:2014naa}. 
We are interested in placing the maximum possible lower bound on the mass 
of $A$, as a function of the hypothesized mass, $\tilde{m}$, of $C$.
The prescription for doing so is well-known (see, e.g., \cite{Barr:2011xt}):
we minimize of the heavier of the two parent masses, $M_{A_1}$ and $M_{A_2}$, 
subject to relevant kinematic constraints. In the simplest case, 
we only apply the missing transverse momentum constraint, (\ref{eq:mpt}),
and obtain
\bea
M_{2} (\tilde m) &\equiv&
\min_{\vec{q}_{1},\vec{q}_{2}}\left\{\max\left[M_{A_1}(\vec{q}_{1},\tilde
    m),\;M_{A_2} (\vec{q}_{2},\tilde m)\right] \right\},  
\label{eq:m2def}\\
\vec{q}_{1T}+\vec{q}_{2T} &=& \mpt   \nonumber
\eea 
where $\vec{q}_i$ denotes the three-momentum of the invisible particle, $C_i$.  
Note that the missing transverse momentum condition is linear and is easily solved, 
so that the minimization in (\ref{eq:m2def}) is unconstrained and can 
be performed over an unconstrained four-dimensional momentum space, 
e.g., $\{\vec{q}_{1T}, q_{1z}, q_{2z}\}$. 

The situation becomes much more interesting (and challenging) 
when we consider additional nonlinear constraints.
Given the process of figure~\ref{fig:decaysubsystem}, 
it is natural to consider additionally constrained versions of eq.~(\ref{eq:m2def}). 
Having already made the assumption that the two decay chains are identical\footnote{See 
refs.~\cite{Barr:2009jv,Konar:2009qr,Cho:2014yma} for relaxing this assumption.}, 
we can additionally impose that the particles $A_1$ and $A_2$ have the same mass,
\bea
M_{A_1}=M_{A_2},
\label{eq:parent}
\eea
that the particles $B_1$ and $B_2$ have the same mass, 
\bea
M_{B_1}=M_{B_2},
\label{eq:relative}
\eea
or both (\ref{eq:parent}) and (\ref{eq:relative}). 
Together with the case where neither (\ref{eq:parent}) or
(\ref{eq:relative}) is required to hold, a total of four variants are
therefore possible. Following the same notation as ref.~\cite{Cho:2014naa}, we introduce two more subscripts
on the $M_2$ variable to indicate whether the constraints in
eqs.~(\ref{eq:parent}) and~(\ref{eq:relative}) were applied during the
minimization or not. The first subscript will refer to the parent
constraint in eq.~(\ref{eq:parent}), while the second subscript 
will refer to the relative constraint in eq.~(\ref{eq:relative}). If
a constraint is imposed, the corresponding index is ``C'',
otherwise it is ``X''. Therefore, eq.~(\ref{eq:m2def}) can be
expressed as $M_{2XX}$ because no extra constraints are imposed: 
\bea
M_{2XX} (\tilde{m})&\equiv&
\min_{\vec{q}_{1},\vec{q}_{2}}\left\{\max\left[M_{A_1}(\vec{q}_{1},\tilde
    m),\;M_{A_2} (\vec{q}_{2},\tilde m)\right] \right\},  
\label{eq:m2XXdef}
\\
\vec{q}_{1T}+\vec{q}_{2T} &=& \mpt.   \nonumber
\eea 
The other three variables are formally defined as follows:
\bea
M_{2CX} (\tilde{m})&\equiv&
\min_{\vec{q}_{1},\vec{q}_{2}}\left\{\max\left[M_{A_1}(\vec{q}_{1},\tilde
    m),\;M_{A_2} (\vec{q}_{2},\tilde m)\right] \right\},  
\label{eq:m2CXdef}\\
\vec{q}_{1T}+\vec{q}_{2T} &=& \mpt   \nonumber \\
M_{A_1}&=& M_{A_2} \nonumber 
\eea 
\bea
M_{2XC} (\tilde{m})&\equiv&
\min_{\vec{q}_{1},\vec{q}_{2}}\left\{\max\left[M_{A_1}(\vec{q}_{1},\tilde
    m),\;M_{A_2} (\vec{q}_{2},\tilde m)\right] \right\},  
\label{eq:m2XCdef}\\
\vec{q}_{1T}+\vec{q}_{2T} &=& \mpt   \nonumber \\
M_{B_1}^2&=& M_{B_2}^2 \nonumber 
\eea 
\bea
M_{2CC} (\tilde{m})&\equiv&
\min_{\vec{q}_{1},\vec{q}_{2}}\left\{\max\left[M_{A_1}(\vec{q}_{1},\tilde
    m),\;M_{A_2} (\vec{q}_{2},\tilde m)\right] \right\}.
\label{eq:m2CCdef}\\
\vec{q}_{1T}+\vec{q}_{2T} &=& \mpt   \nonumber \\
M_{A_1}&=& M_{A_2} \nonumber  \\
M_{B_1}^2&=& M_{B_2}^2 \nonumber 
\eea 
Eqs.~(\ref{eq:m2XXdef}-\ref{eq:m2CCdef}) define the four possible $M_2$ variables for the $(ab)$
subsystem. One can similarly define four $M_2$ variables for each of the $(a)$
and $(b)$ subsystems, we refer the reader to \cite{Cho:2014naa} for the
exact definitions.

\subsection{Calculating $M_2$ for dilepton top events}

From the definitions (\ref{eq:m2CXdef}-\ref{eq:m2CCdef}) it is clear that the 
problem of computing the variables $M_{2CX}$,  $M_{2XC}$, and  $M_{2CC}$
falls into the general category of constrained minimization problems (\ref{eq:objective0},~\ref{eq:constraints0})
which {\sc Optimass} is designed to solve. In the remainder of this section, we shall therefore illustrate the 
functionality of {\sc Optimass} with the physics example of figure~\ref{fig:decaysubsystem}.
Specifically, we consider the case of pair-produced top quarks that decay fully leptonically:
\bea
pp \ra t\bar{t}, 
\qquad (t \ra bW^+ \ra bl^+\nu_l),
\qquad (\bar{t}\ra \bar{b}W^- \ra \bar{b}l^-\bar{\nu}_l).
\eea
Thus, in figure~\ref{fig:decaysubsystem}, particle $A$ is associated with the top quark,
particle $B$ with the $W$-boson, and particle $C$ with the neutrino.
For simplicity, since our major interest is not in the shape of the distributions but in 
the precision of the minimization procedure, we consider events where the top
quarks are produced at threshold and decay according to phase space.
We neglect initial and final state radiation, and also do not take into account
experimental efficiencies, cuts, combinatorics, and detector resolution. 
All those effects are important in a real physics analysis, but are irrelevant 
to the question of evaluating the performance of the {\sc Optimass} minimization 
algorithm, which is our goal here.

We use {\sc Optimass} to compute the values for the $M_{2CX}$,  $M_{2XC}$, and  $M_{2CC}$
variables for all three subsystems. In order to judge the precision of this numerical calculation, 
ideally we need to identify an alternative method for computing these answers, which would give us 
reference benchmarks. Fortunately, for the case of the $M_{2CX}$ variable, such a
benchmark is provided by the $M_{T2}$ variable itself --- we can use the result, 
proven in ref.~\cite{Cho:2014naa}, that $M_{2CX}$ and $M_{T2}$ are identical event by event.
This enables us to directly compare the values of $M_{T2}$ and $M_{2CX}$ for each of the
three possible subsystems. The Cambridge variable $M_{T2}$ can already be reliably computed with 
one of several publicly available codes; here we use the package from ref.~\cite{Cheng:2008hk}.
Furthermore, for the $(ab)$ and $(a)$ subsystems, analytical formulae for $M_{T2}$ 
are also available \cite{Cho:2007qv,Cho:2007dh}, facilitating the comparison.\footnote{In the case of 
the $(b)$ subsystem, the $b$-quarks simulate initial state radiation, in which case 
no analytical formula for $M_{T2}$ is known, so we need to rely on the computer code from ref.~\cite{Cheng:2008hk}.}

\begin{figure}[t]
\centering
\includegraphics[width=4.7cm]{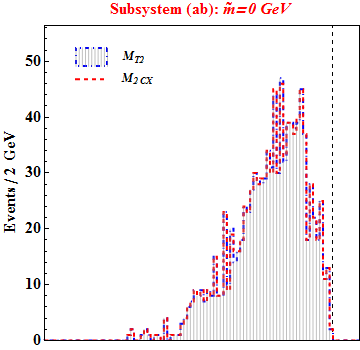}
\includegraphics[width=4.7cm]{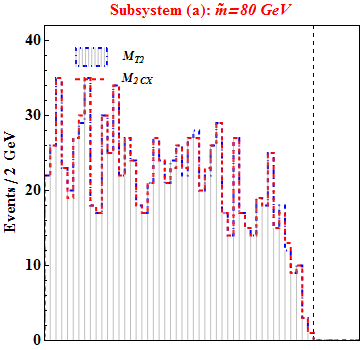}
\includegraphics[width=4.7cm]{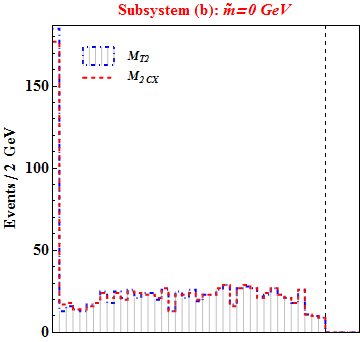}\\
\includegraphics[width=4.7cm]{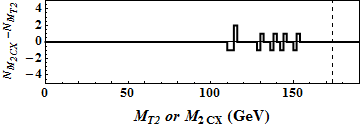}
\includegraphics[width=4.7cm]{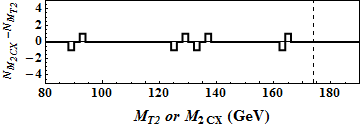}
\includegraphics[width=4.7cm]{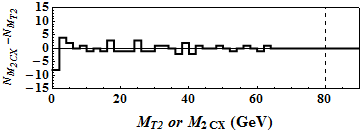}
\caption{\label{fig:compareMT2vsM2CX} Comparison between $M_{T2}$ (blue shaded histograms)
  and $M_{2CX}$ (red histograms) for each subsystem. For the (ab) and (a) subsystems,
  the relevant $M_{T2}$ values are evaluated by the well-known
  analytic formula, whereas those for the (b) subsystem are computed
  numerically with the package from ref.~\cite{Cheng:2008hk}. 
  The vertical black dashed lines indicate the expected endpoint
  of the $M_{T2}$ distribution for the given test mass.}
\end{figure}
\begin{figure}[t]
\includegraphics[width=4.7cm]{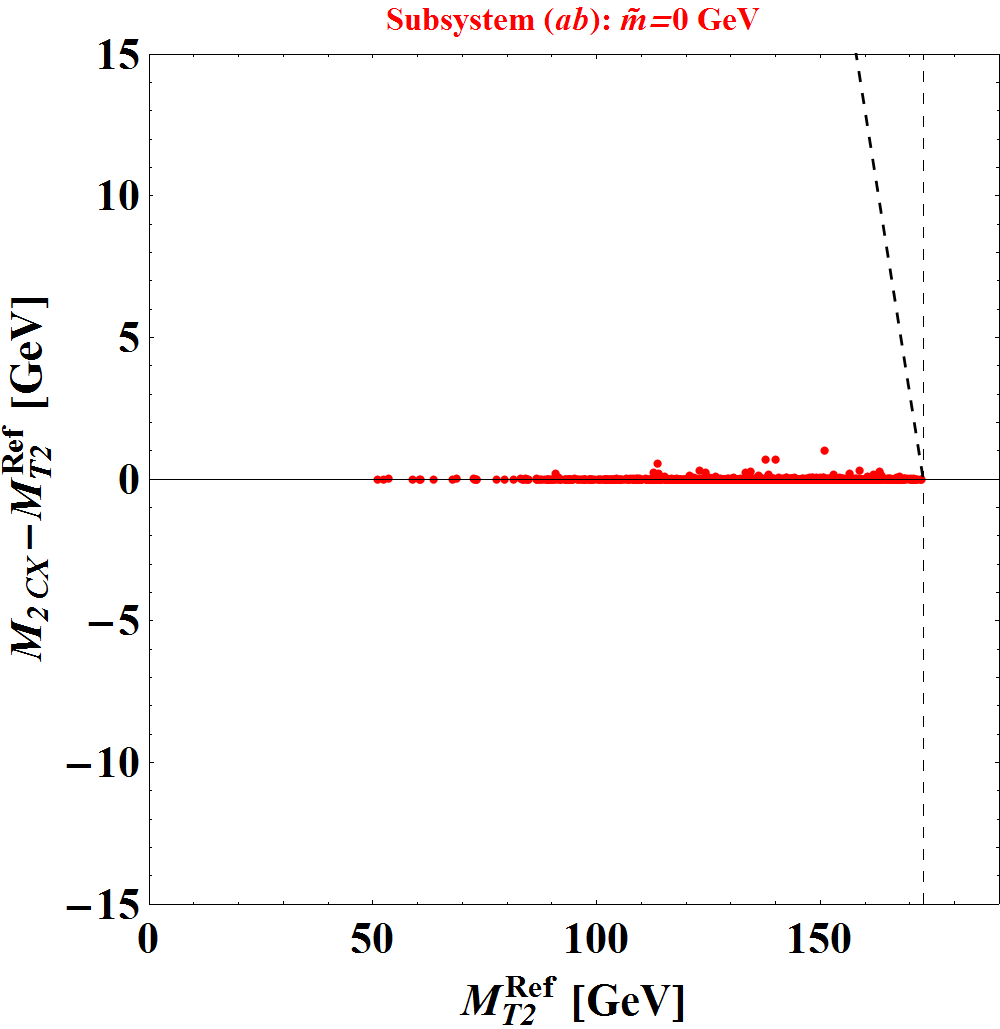}
\includegraphics[width=4.7cm]{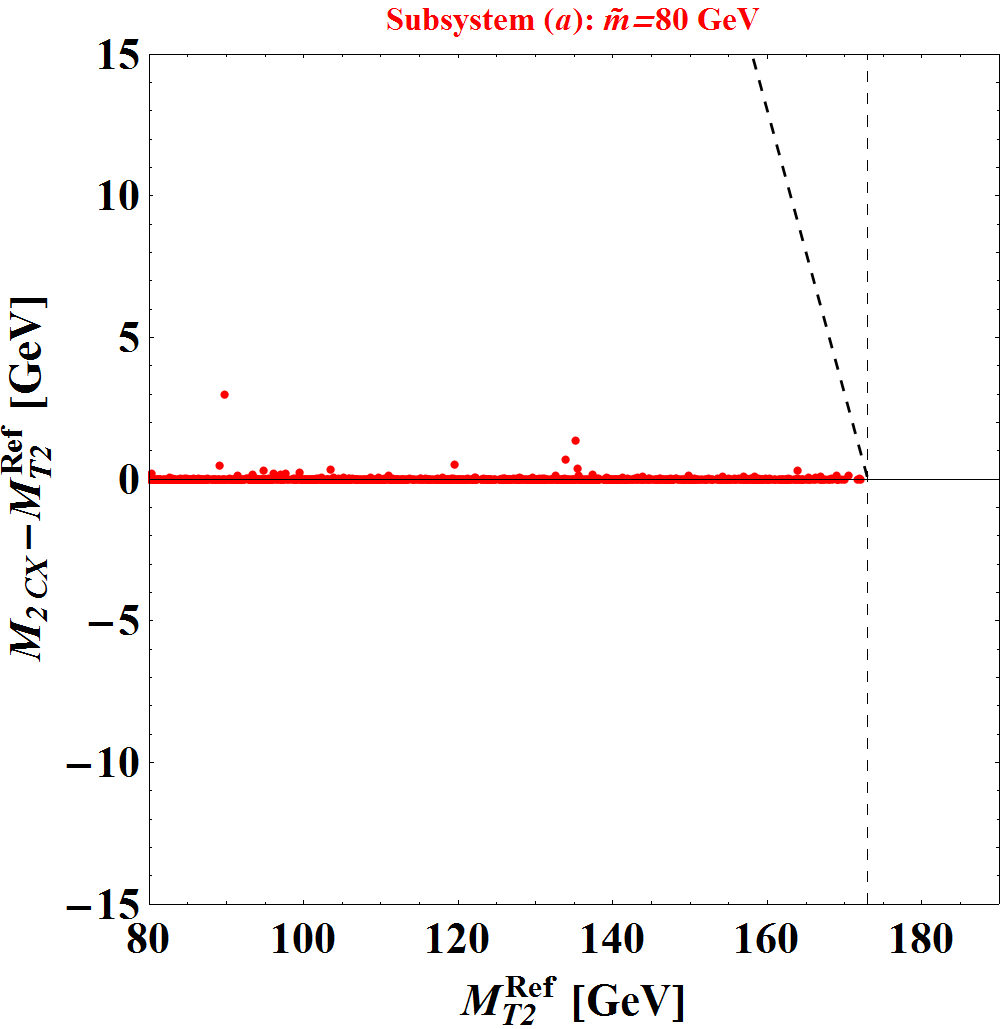}
\includegraphics[width=4.7cm]{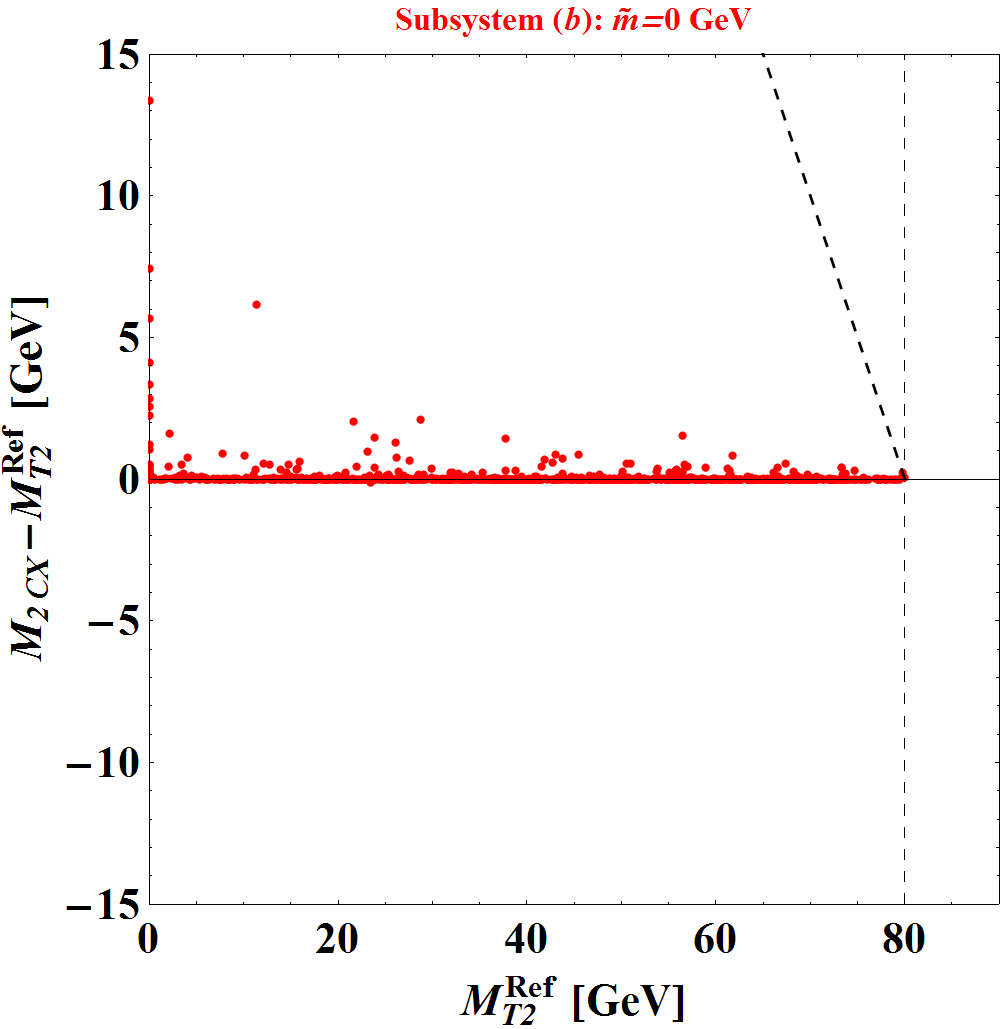}
\caption{\label{fig:scatterMT2vsM2CX} Scatter plots of the difference $M_{2CX}-M_{T2}^{\rm Ref}$ versus the 
reference value, $M_{T2}^{\rm Ref}$.}
\end{figure}

\begin{figure}[t]
\centering
\includegraphics[width=4.7cm]{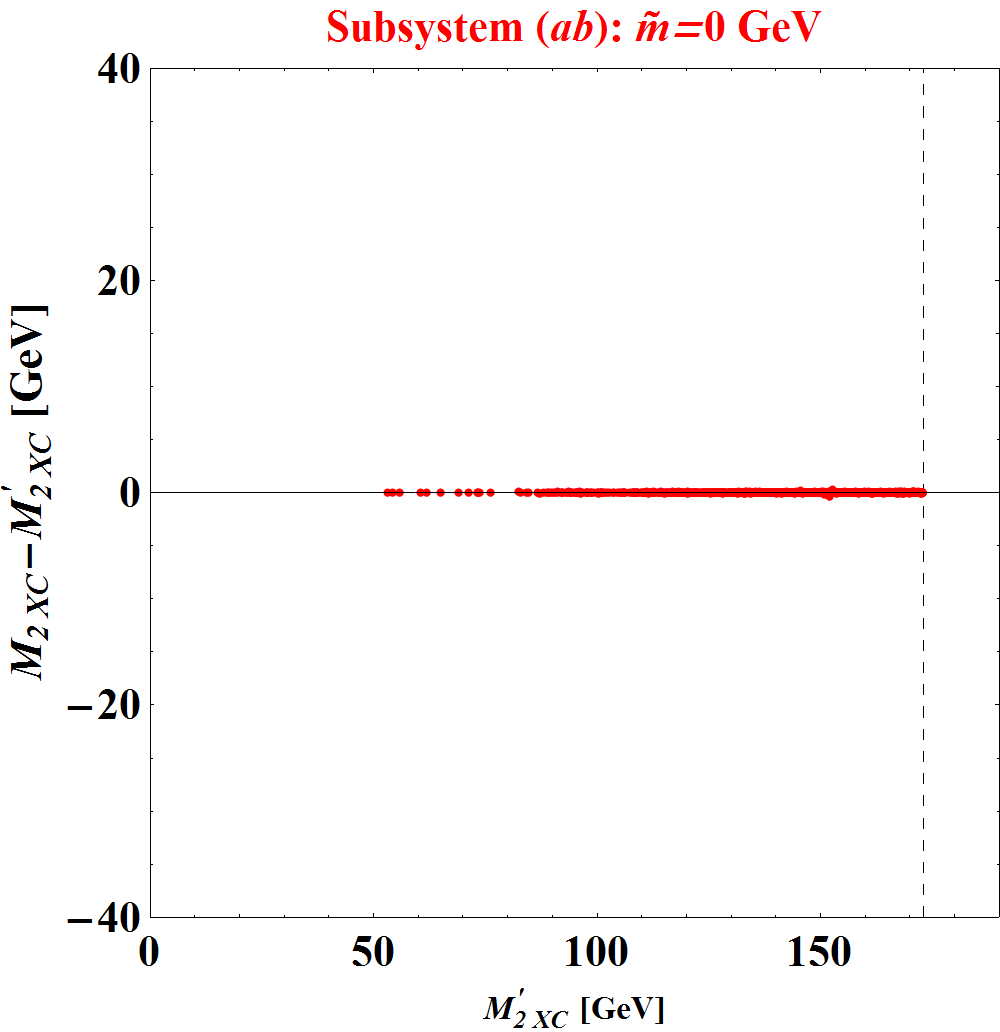}
\includegraphics[width=4.7cm]{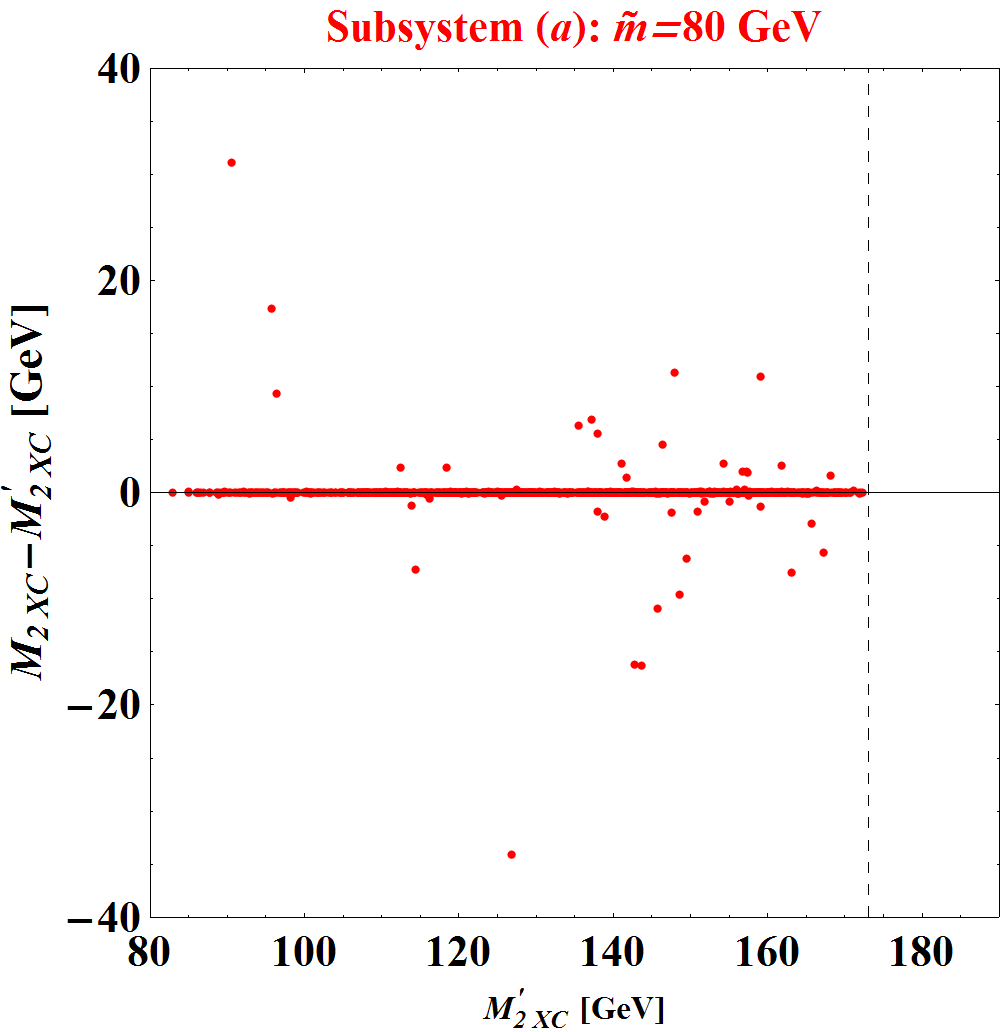}
\includegraphics[width=4.7cm]{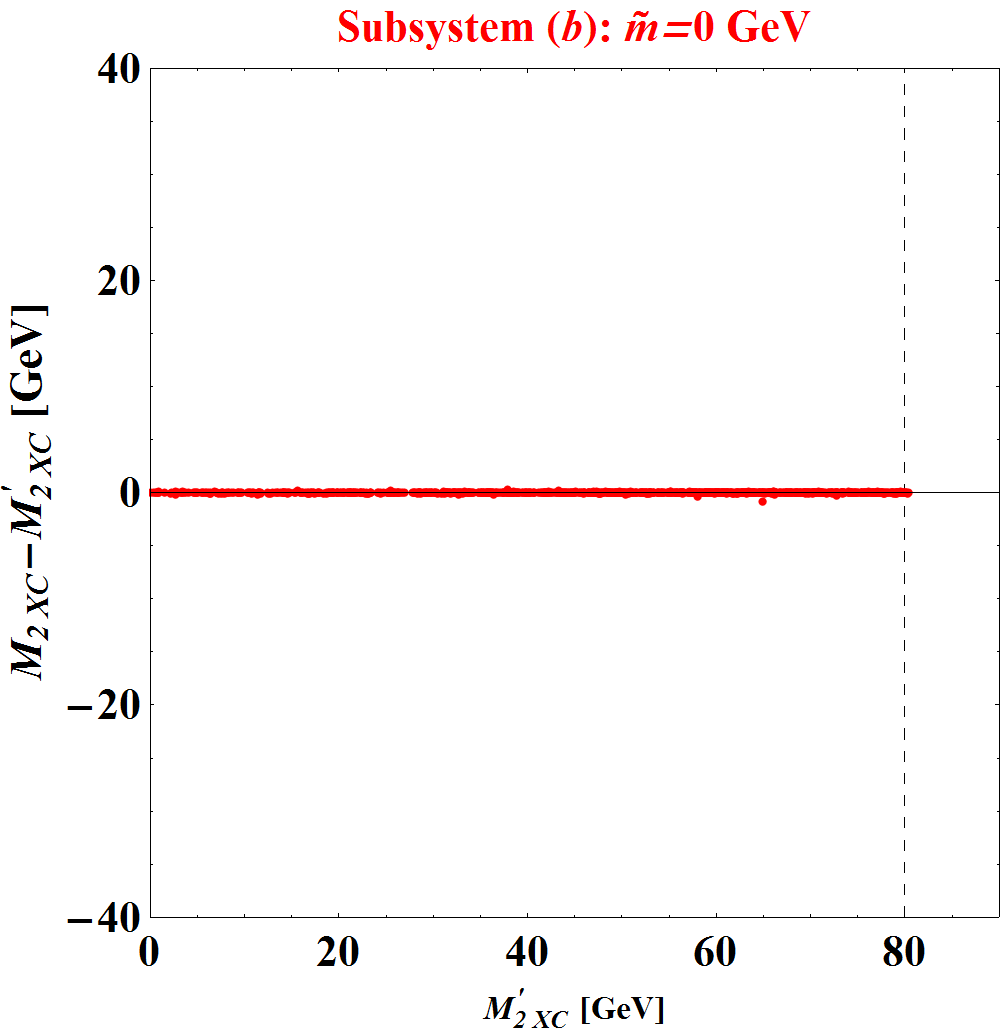}
\caption{\label{fig:compareXC} The comparison between the values of $M_{2XC}$ obtained by two different internal codes.}
\end{figure}
\begin{figure}[h]
\centering
\includegraphics[width=4.7cm]{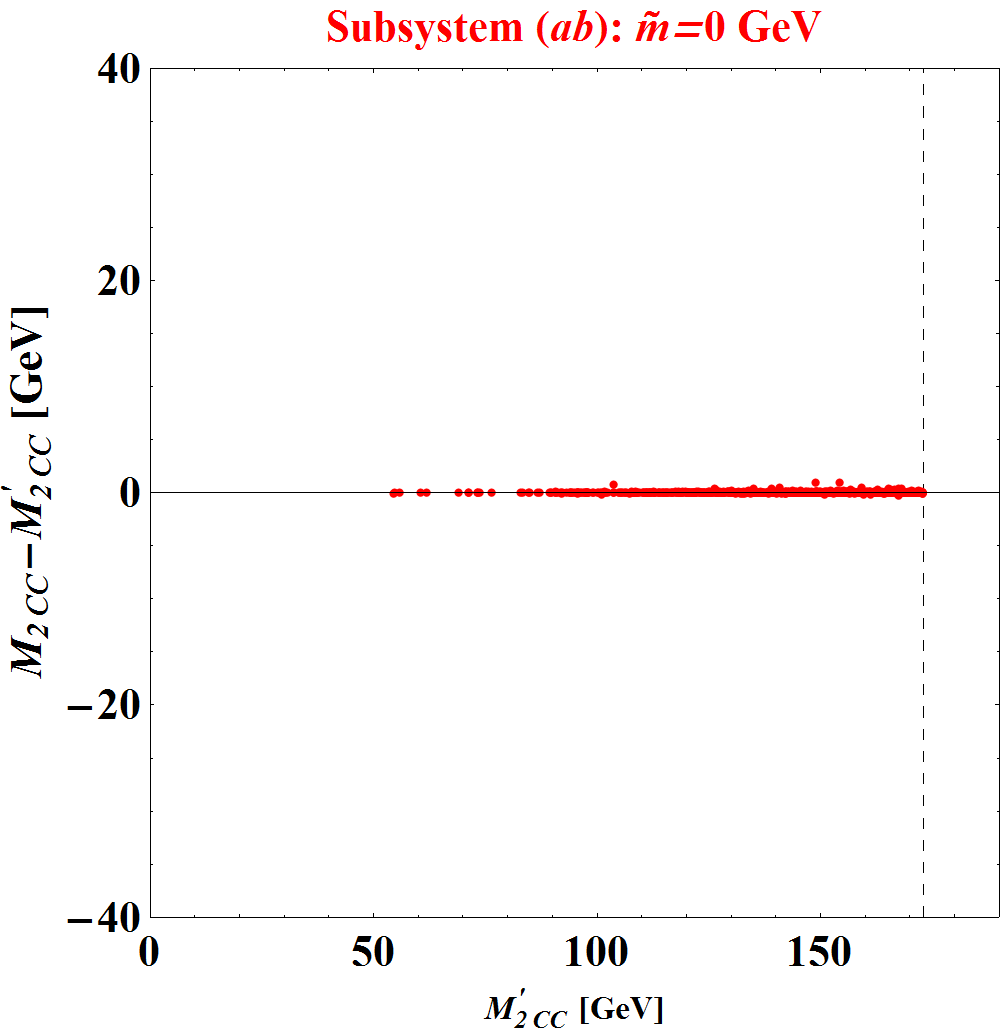}
\includegraphics[width=4.7cm]{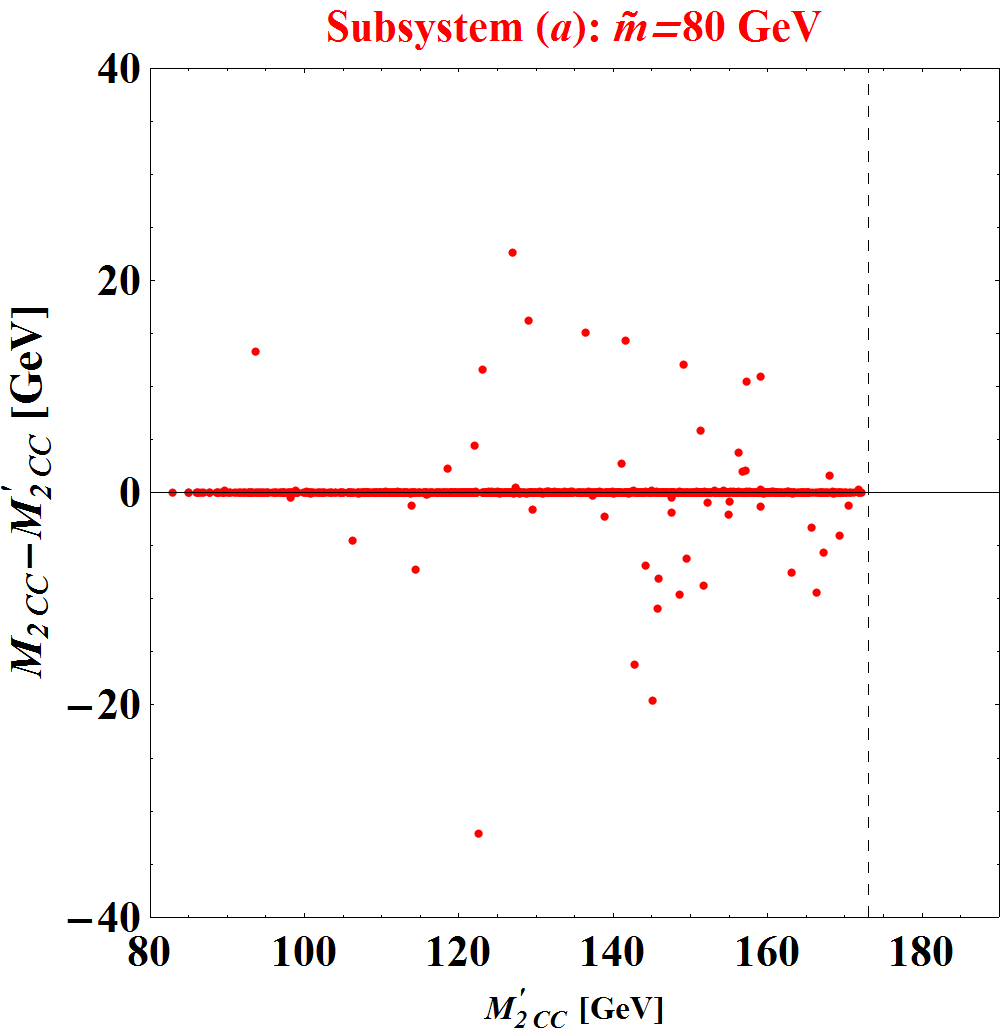}
\includegraphics[width=4.7cm]{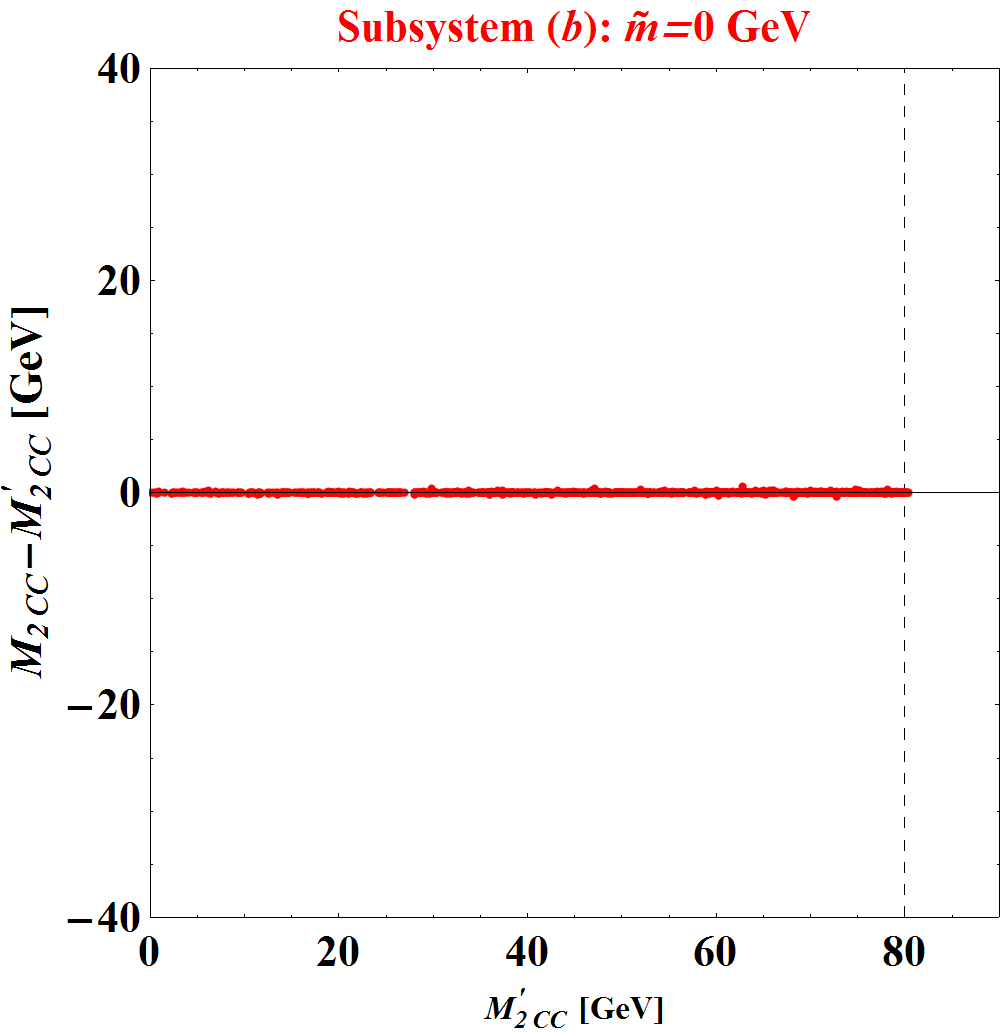}
\caption{\label{fig:compareCC} The same as figure~\ref{fig:compareXC}, but for $M_{2CC}$.}
\end{figure}

Figures~\ref{fig:compareMT2vsM2CX} and \ref{fig:scatterMT2vsM2CX}
show the results from the comparison between the value of $M_{2CX}$ obtained from
{\sc Optimass} and the corresponding reference value, $M_{T2}^{\rm Ref}$,
for all three subsystems. Since the exercise is performed with the $t\bar{t}$ decay
sample, we take the trial mass to be the true mass of the daughter particle:
$\tilde{m}=0$ GeV for the $(ab)$ and the $(b)$ subsystems and
$\tilde{m}=80$ GeV for the $(a)$ subsystem.  
We choose the $M$ parameter in eq.~(\ref{eq:eta_star}) to be
$200$ GeV for subsystems $(ab)$ and $(a)$, and $100$ GeV for subsystem $(b)$.
Figure~\ref{fig:compareMT2vsM2CX} 
reveals that the distributions of the on-shell constrained $M_2$ variable, $M_{2CX}$, (red dashed
histograms) are almost identical to the corresponding $M_{T2}$ distributions 
(blue dot-dashed shaded histograms). Only a handful of events show 
a difference on the order of $1-2$ GeV, as seen in the scatter plots of figure~\ref{fig:scatterMT2vsM2CX}.
The results shown in figures~\ref{fig:compareMT2vsM2CX} and \ref{fig:scatterMT2vsM2CX}
were obtained with the default values of the {\sc Optimass} parameters. Of course, the
precision can be further improved by tweaking the relevant tolerance parameters, 
increasing the maximum number of iterations, or improving on the initial guess of
$\vec{x}_0^s$. However, this will come at the cost of increased
computation time; we believe that the level of precision seen in these
figures should be sufficient for most practical analyses.

\begin{figure}[t]
\begin{center}
\includegraphics[width=7.2cm]{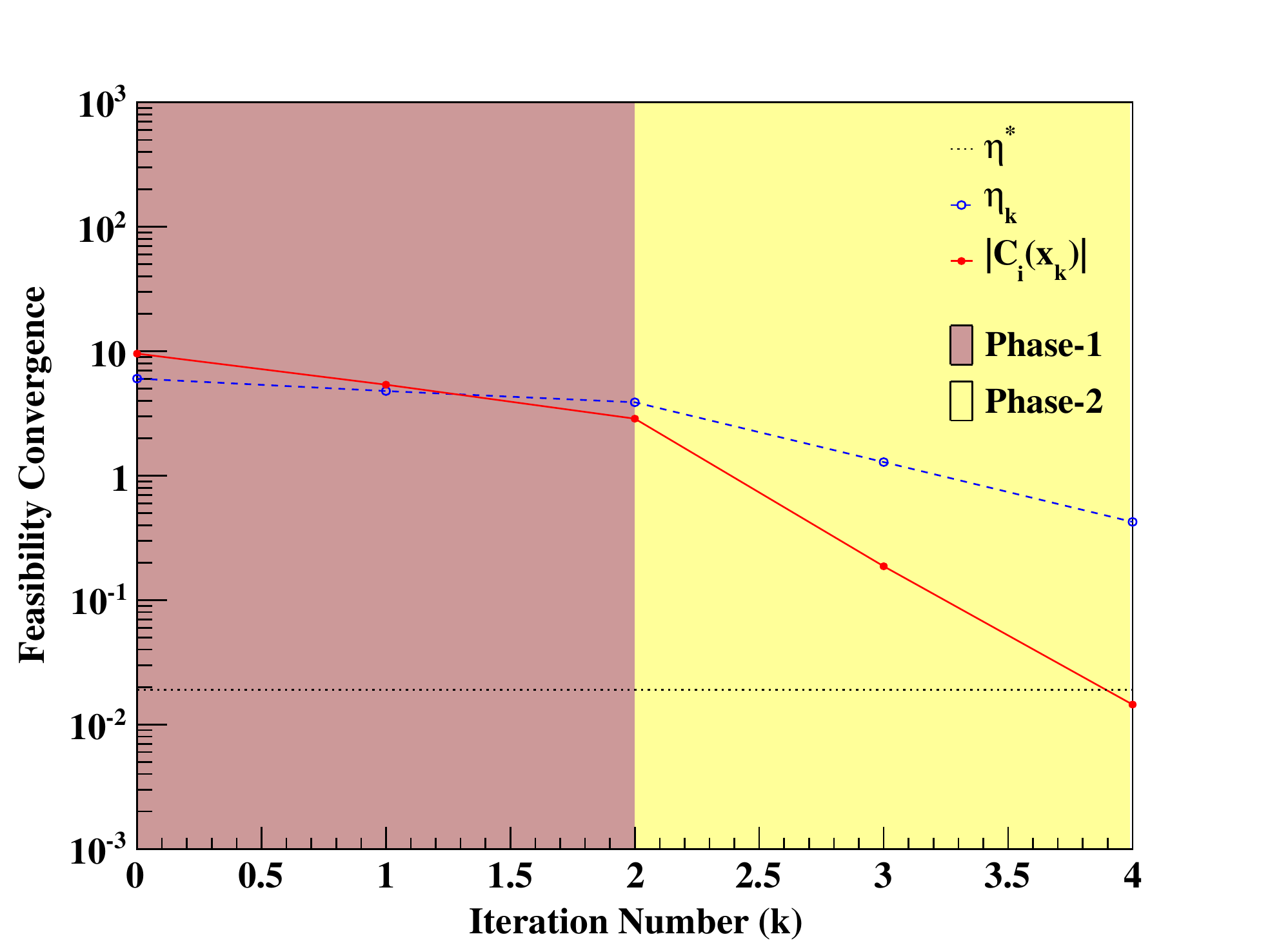}
\includegraphics[width=7.2cm]{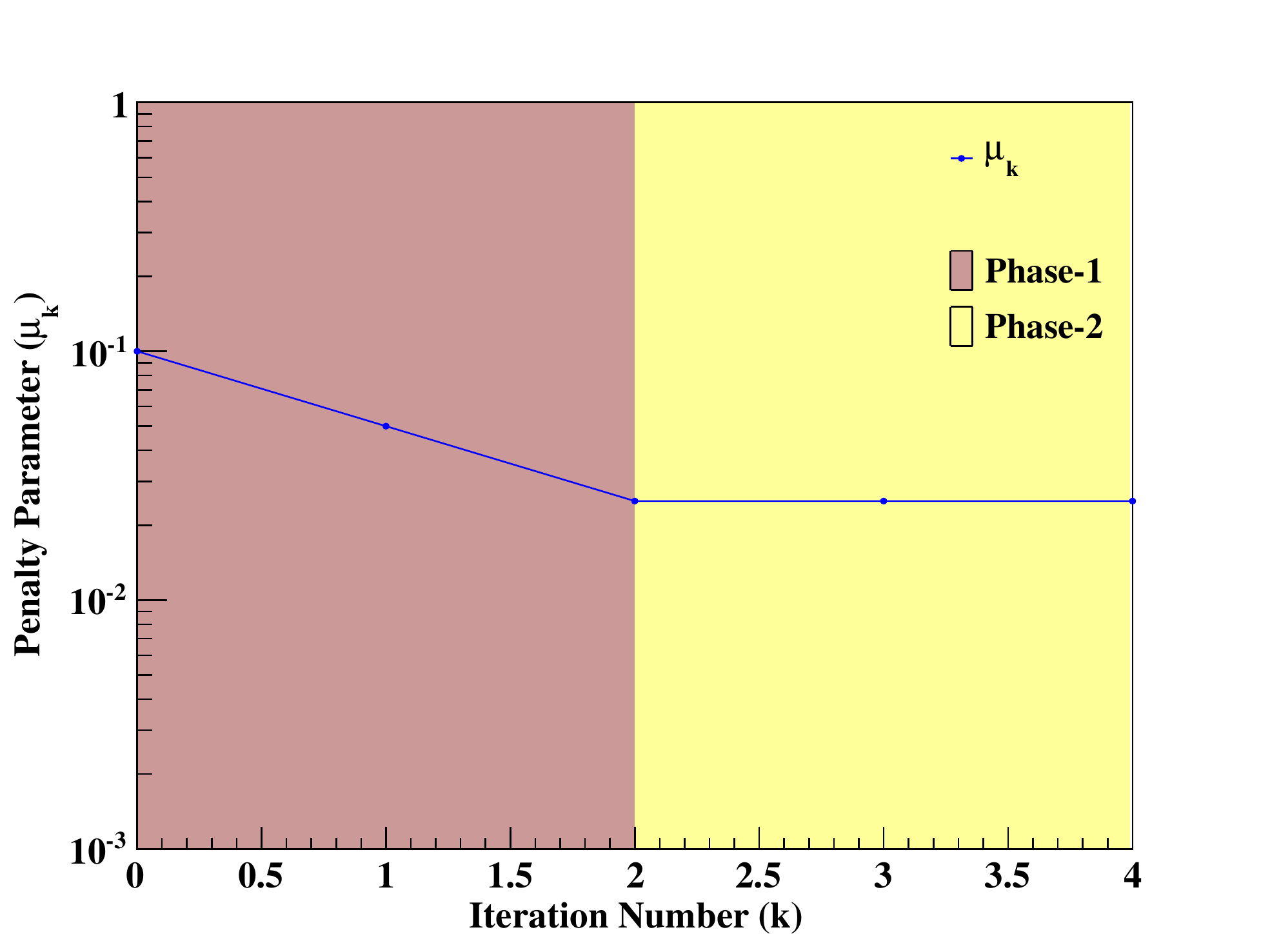}\\
\includegraphics[width=7.2cm]{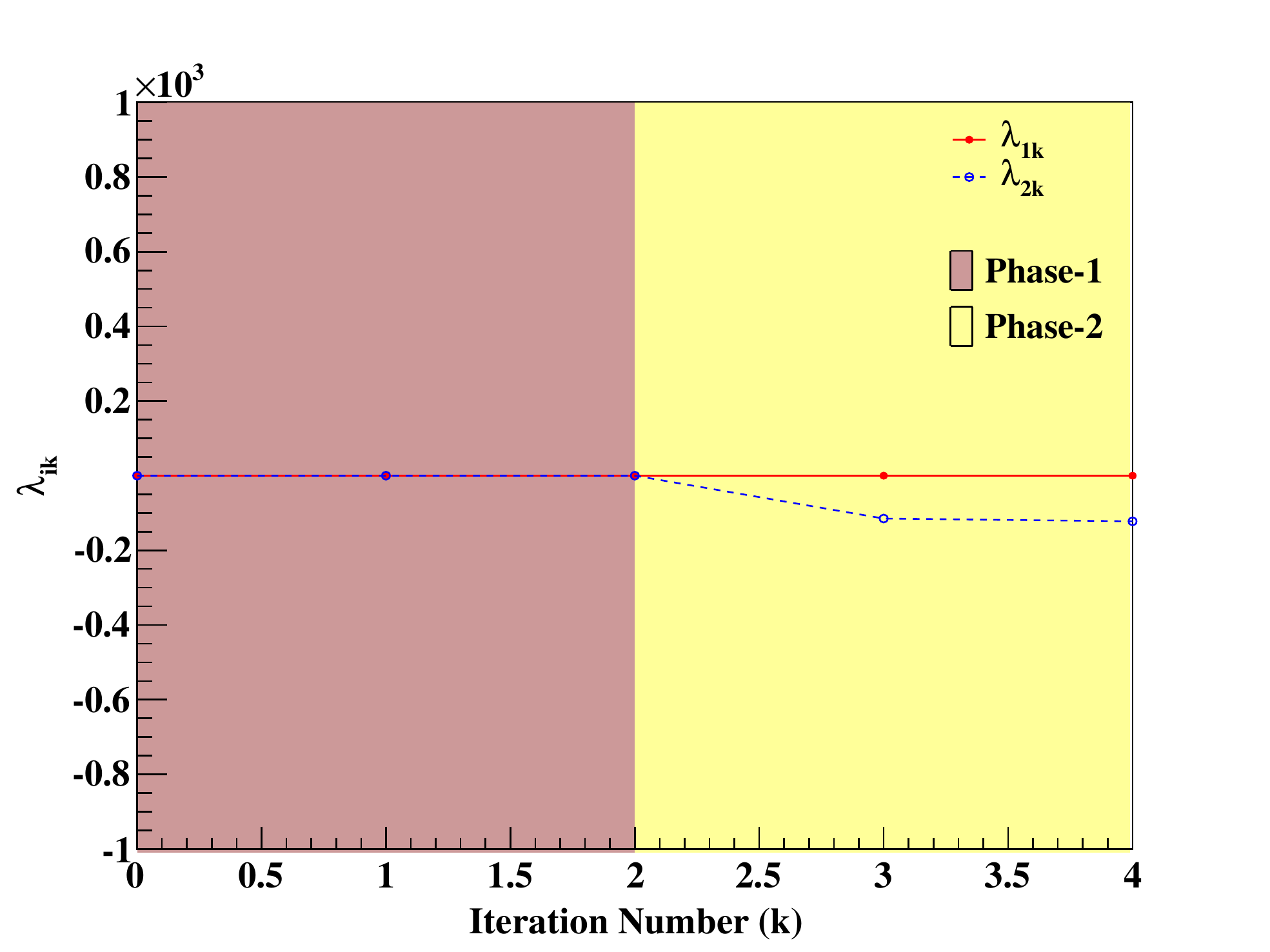}
\includegraphics[width=7.2cm]{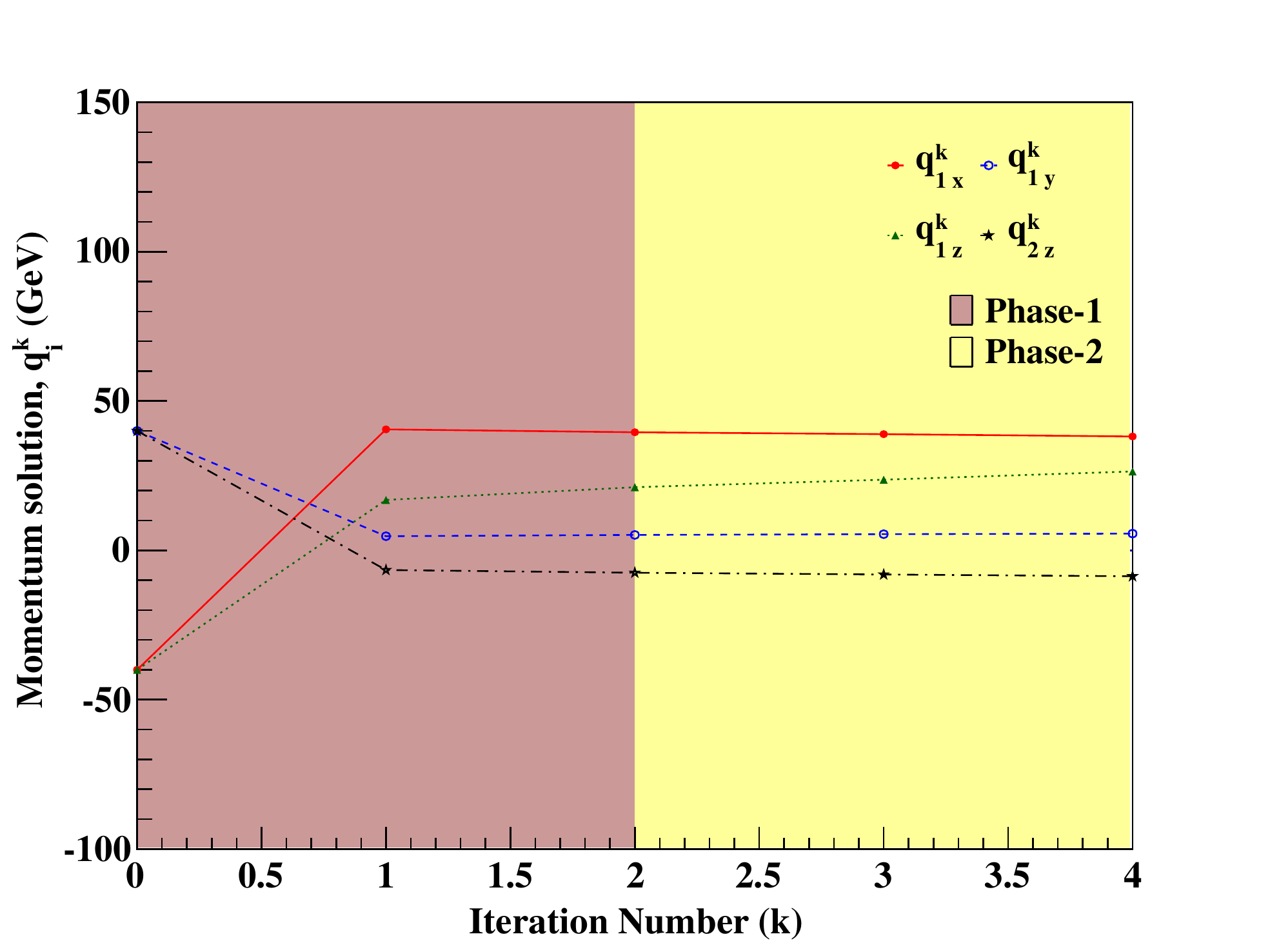}
\end{center}
\caption{The same as Fig.~\ref{fig:example1trace}, but for 
the $M_{2CC}$ calculation in the $(ab)$ subsystem of the single event considered in section~\ref{sec:1event}.}
\label{fig:M2cc_trace}
\end{figure}
\begin{figure}[h]
\begin{center}
\includegraphics[trim= 1cm 1cm 1.5cm 1.5cm, width=7cm]{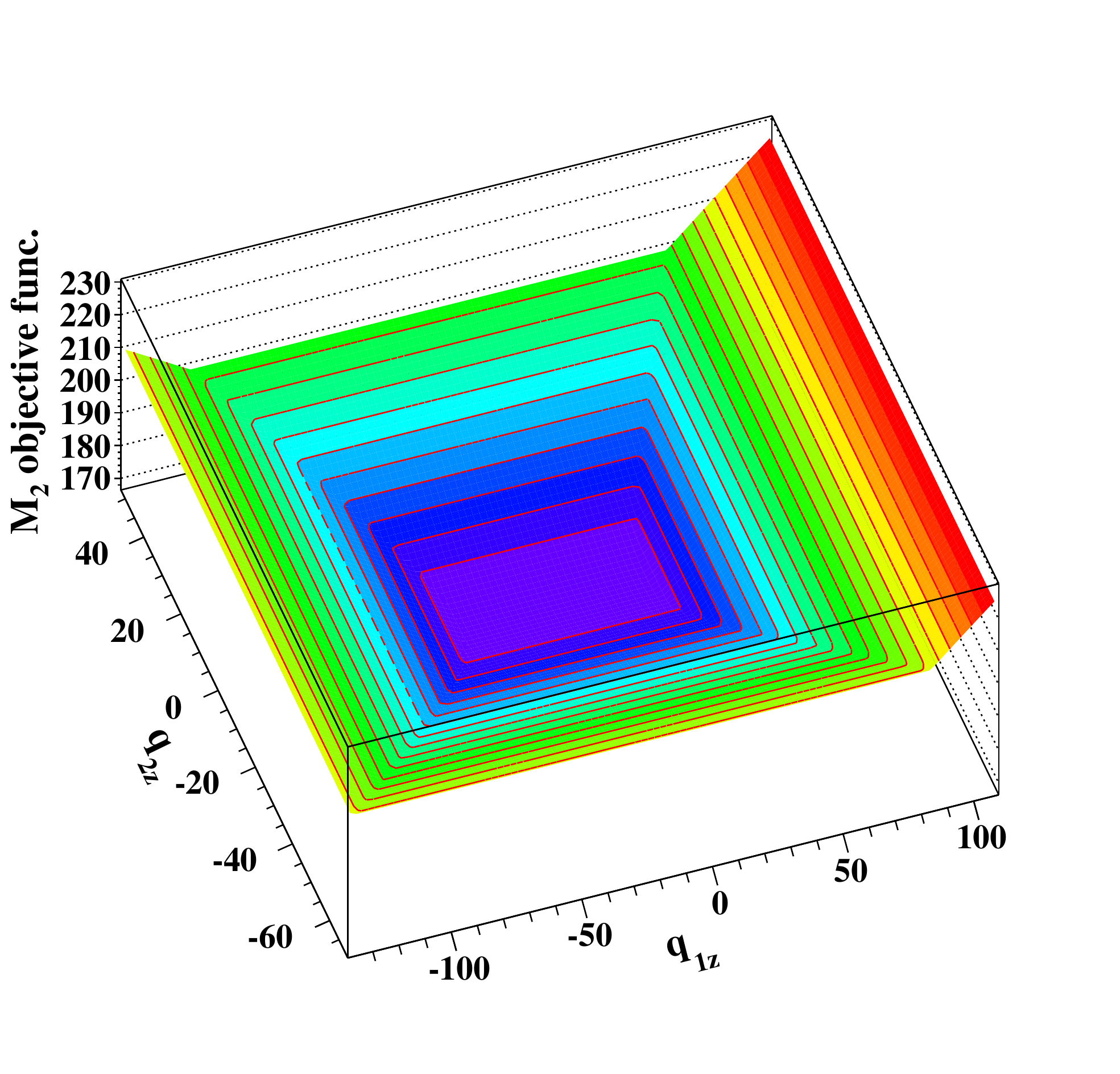}
\includegraphics[trim= 0.cm 0cm 1cm 2cm, clip, width=7cm]{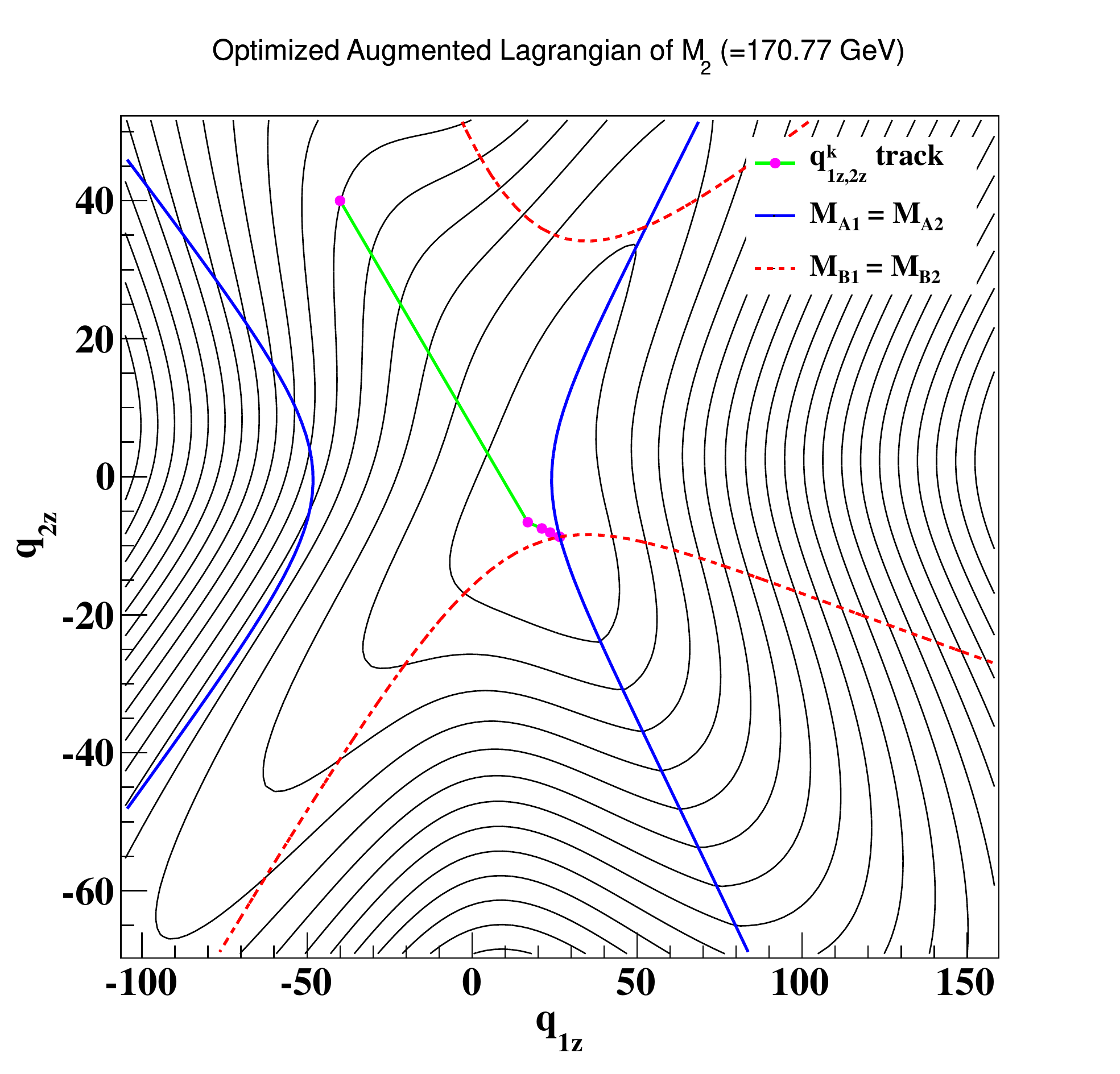}
\caption{The same as figures~\ref{fig:example1} and \ref{fig:example2}, but for the single event
considered in section~\ref{sec:1event}. Since the objective function has four independent arguments, 
in order to visualize the evolution of the minimizer, we plot $q_{1z}$ and $q_{2z}$,
having fixed the other two variables, $q_{1x}$ and $q_{1y}$, to the values which minimize the 
objective function for the given choice of $q_{1z}$ and $q_{2z}$. }
\label{fig:exampleM2}
\end{center}
\end{figure}

Figures~\ref{fig:compareXC} and \ref{fig:compareCC} provide a similar validation 
for the case of the $M_{2XC}$ and $M_{2CC}$ variables. In this case, however,
we do not have a readily available benchmark for comparison: first, because 
analytical formulas for those cases do not exist, and second, because there is 
no publicly available code which is able to handle $M_{2XC}$ and $M_{2CC}$.  
This is why we produced two different versions of our code (created independently
by different sets of the current authors), and proceeded to compare their results
for $M_{2XC}$ and $M_{2CC}$ in figures~\ref{fig:compareXC} and \ref{fig:compareCC},
respectively. The figures show that the two internal codes agree reasonably well,
with notable differences only in about 1\% of the events. The events with the 
largest deviations were scrutinized further, revealing that one of the codes typically 
found a local minimum, due to a different choice of starting values for $\vec{x}_0^s$.
When repeating the minimization with a range of possible choices for $\vec{x}_0^s$,
and taking the minimum of the obtained set of values of $M_2$, the two codes were 
shown to be in exact agreement.

\subsection{Demonstration of {\sc Optimass} for one event}
\label{sec:1event}

In conclusion, we supplement the toy examples from section~\ref{sec:validation} 
with one example of a real $t\bar{t}$ event. The 4-momenta $(E,p_x,p_y,p_z)$ 
of the four visible particles (in GeV) are 
\bea
 p_{a1}&=& (68.003, \,\, -8.404,\,\, 16.069,\,\, -65.541 )\nonumber\\
 p_{b1}&=& (56.168, \,\, -29.282,\,\, -29.683,\,\, 37.635)\\ 
 p_{a2}&=& (68.003, \,\, 6.881,\,\, -56.711,\,\, -36.890 )\nonumber\\
 p_{b2}&=& (81.160, \,\, -27.332,\,\, 68.553,\,\, 33.769), \nonumber \label{eq:exampleM2event}
\eea
thus the missing transverse momentum is 
\beq
\mpt = (58.137, 1.772).
\eeq
The initial parameters for the algorithm are given by
\bea
(q_{1x,0}^s,&q_{1y,0}^s&,q_{1z,0}^s,q_{2z,0}^s)=(-40,+40,-40,+40)\, \text{(GeV)},
\qquad (\lambda^{0}_1,\lambda^{0}_2)= (0,\,0).\nonumber   
\label{eq:exampleM2input}
\eea

Figure~\ref{fig:M2cc_trace} is the analogue of Fig.~\ref{fig:example1trace}, showing
the convergence to a satisfactory solution for $M_{2CC}$ in subsystem $(ab)$ after the $k=4$
step, giving 
\bea
(q_{1x}^*,q_{1y}^*,q_{1z}^*,q_{2z}^*)\,&=&\,(38.082,\,\, 5.612,\,\, 26.598,\,\, -8.717)\,\text{(GeV)}, \nonumber\\
(\lambda^{1*}_{4},\lambda^{2*}_{4}) \,&=&\, (0.000001, -122.329803),\\
\mu_{4}\,&=&\,0.025.\nonumber
\label{eq:exampleM2sol}
\eea
Figure~\ref{fig:exampleM2} is analogous to figures~\ref{fig:example1} and \ref{fig:example2}
for this case. The left panel plots the original objective function, while the right panel shows a contour plot of the 
augmented Lagrangian function as of the final ($k = 4$) iteration. In the right panel, the set of points
which satisfy the constraint eq.~(\ref{eq:parent}) (eq.~(\ref{eq:relative})) is shown in blue (red).
As before, the magenta points mark the locations of the minimizer,
$\vec{x}_k$, found in the $k^{\rm th}$ iteration.

\section{Conclusions}
\label{sec:conclusions}

With the restart of the LHC, the quest for new physics has resumed.
We believe that kinematic variables like $M_2$ will play an increasingly 
important role in searching for SUSY and related models; the gain 
in sensitivity that these variables provide (see, e.g.,~\cite{Cho:2014yma})
aids both in setting limits on and in discovering BSM physics.

Since the calculation of $M_2$, like many other kinematic variables, 
involves a constrained minimization that must be performed
numerically, it is important to ensure that this calculation is performed in an efficient and reliable way.
Thus we have introduced the public package, {\sc Optimass}, which
achieves these important goals.
Our algorithm utilizes the ALM
and interfaces with the popular unconstrained minimization package,
{\sc Minuit}. 

\begin{figure}
\centering
\includegraphics[width=8.0cm]{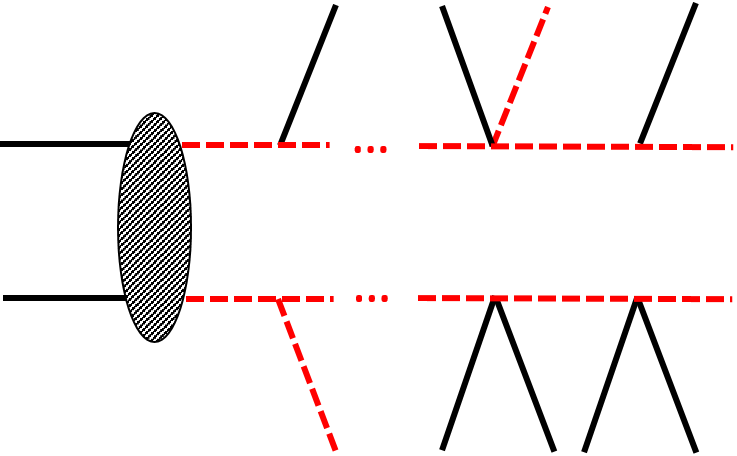}
\caption{\label{fig:GenProc} 
An example of the general event topology which can be handled by {\sc Optimass}.
We allow for two decay chains, involving a sequence of 2-body or 3-body decays.
In each decay, the final state particles can be visible, invisible, or both.}
\end{figure}

We have described the relevant issues in what we hope will be
sufficient depth to aid the physicist new to the challenges of
constrained optimization.  The {\sc Optimass} algorithm has been
described in detail and examples of its use have been provided.
We compared analytically calculated values of $M_{T2}$ to the $M_{2CX}$ variable
obtained using {\sc Optimass} and found excellent agreement.  
Other tests of {\sc Optimass} were performed and the results are encouraging.
We stress that while our physics example in section~\ref{sec:application}
was limited to the dilepton $t\bar{t}$ topology of figure~\ref{fig:decaysubsystem},
{\sc Optimass} has been designed to handle arbitrarily general event topologies,
as indicated in figure~\ref{fig:GenProc}:
\begin{itemize}
\item {\em Multiple invisible particles in each decay chain}. In many motivated scenarios, 
invisible particles may appear not only at the end of the decay chain (as is customary for dark matter particles),
but also at intermediate stages. The {\sc Optimass} code can handle such cases, 
since the total number of invisible particles is unrestricted. In figure~\ref{fig:GenProc},
the first decay in the lower chain and the second to last decay in the upper chain provide
examples of sources of such intermediate invisible particles.
\item {\em Multi-body decays.} The decay chains can be constructed of two-body decays, three-body decays, etc.  
Furthermore, a multi-body decay may result in a set of final state particles which can be
visible, invisible, or both. As an illustration, in figure~\ref{fig:GenProc} we show three two-body 
decays and three three-body decays. Two of the three-body decays result in visible particles only, 
while the remaining one produces both a visible and an invisible final state particle.
\end{itemize}
This more general functionality of {\sc Optimass} will be explored and demonstrated in a future publication \cite{us}.

In conclusion, we look forward to implementing improvements to and extensions of the
{\sc Optimass} framework, and to its use in the search for new physics
that lies ahead.

\appendix

\section{Unconstrained minimization with {\sc Minuit}}
\label{sec:minuit}

Using the methods of section \ref{sec:overview}, and in particular the ALM, we can convert 
a constrained optimization problem into an unconstrained minimization problem;  
the next step is to actually solve the resulting unconstrained 
minimization problem.  For this task we use {\sc  Migrad} and {\sc  Simplex}, 
two algorithms which are part of the {\sc  Minuit} library.  Thus, after
briefly introducing this ubiquitous library, we will discuss these algorithms
in sections~\ref{sec:Migrad} and~\ref{sec:Simplex}, respectively.

{\sc Minuit}~\cite{minuit} is a popular function minimization library.
It is often used for data analysis, as the minimization of $\chi^2$ functions and likelihoods
represents perhaps the main use of minimization in experimental particle physics.
{\sc Minuit} was initially written in {\sc  Fortran}, but has been
reimplemented (as {\sc Minuit2}~\cite{minuit2}) in {\sc  C++}, taking
advantage of its object-oriented features;
{\sc Minuit2} is included in the math library of the omnipresent
data analysis package {\sc Root}~\cite{root}. 
{\sc Minuit} and {\sc Minuit2} (which we will henceforth refer
to simply as ``{\sc Minuit}'') contain various minimization
algorithms, offering the user a choice. Among the several main algorithms 
({\sc Migrad}, {\sc Seek}, {\sc Scan}, {\sc Simplex}), we choose to use 
{\sc Migrad} and {\sc Simplex} which are briefly described in the next two subsections.

\subsection{{\sc  Migrad}}
\label{sec:Migrad}

{\sc Migrad} utilizes a variation of the Newton's method called a Variable Metric Method (VMM)~\cite{vmm}. 
We remind the reader that Newton's method is an iterative method for
finding the root of a function $f(x)$ in which
\begin{equation}
\label{eq:newton}
x_{n+1}-x_n = -\frac{f(x_n)}{f^\prime(x_n)}.
\end{equation}
Finding a minimum of $f(x)$ rather than a root corresponds to finding
a zero of $f^\prime(x)$.  
In this case the sequence of approximate solutions obtained by
Newton's method are described by
\begin{equation}
\label{eq:newton-min}
x_{n+1}-x_n = -\frac{f^\prime(x_n)}{f^{\prime\prime}(x_n)}.
\end{equation}
The analogous expression in the multidimensional case is
\begin{equation}
\label{eq:newton-min-multi}
\vec{x}_{n+1}-\vec{x}_n = -H^{-1}(\vec{x}_n) \nabla f(\vec{x}_n),
\end{equation}
where $H^{-1}(\vec{x}_n)$ is the inverse of the Hessian matrix.
 The name ``Variable Metric Method" is due to an interesting parallel with General Relativity.
 Namely, in the limit where the objective function, $f(\vec{x})$, is a quadratic form with minimum 
 at $\vec{x}={\vec 0}$, then, for small ${\vec{x}}$, 
 \begin{equation}
 f({\vec x}) - f({\vec 0}) \approx {\vec x}^{\, T}\left( \frac{1}{2} H({\vec x}) \right) {\vec x}. 
 \label{eq:metric}
 \end{equation}
The expression on the right hand side is a bilinear form with
$(1/2) H({\vec x})$ playing the role of the metric tensor.  
If one chooses
\begin{equation}
f(\vec{x}) = \sum_{i=1}^n x_i^2
\label{eq:euclid}
\end{equation}
 in $n$ dimensions, then $(1/2) H({\vec x})$ is precisely the $n$-dimensional Euclidean metric.  
The quantity on the right hand side of eq.~(\ref{eq:metric}) is known as the ``estimated vertical distance
to minimum" (EDM), i.e., 
\bea
{\rm EDM}\equiv  {\vec x}^{\, T}\left(\frac{1}{2} H({\vec x})\right) {\vec x}. \label{eq:edm}
\eea
Generally, when using a VMM, the optimality condition will check whether the calculated 
value for the EDM exceeds a certain tolerance parameter.

Eq.~(\ref{eq:newton-min-multi}) describes the essential idea of the VMM.  However, the VMM is a 
``quasi-Newton method" (as opposed to Newton's method itself) because
instead of calculating the Hessian (or ``metric") exactly, 
it approximates it iteratively.
The main differences among different VMM algorithms lie 
in the precise form of this iterative approximation procedure.
The first VMMs used the so-called DFP updating
formula~\cite{vmm,vmmFP} (after Davidon, Fletcher, and
Powell). Currently, the most common algorithms, including {\sc
  Migrad}, use the BFGS method~\cite{vmmBFGS1,vmmBFGS2,vmmBFGS3,vmmBFGS4}.

A very useful property of VMMs is that subsequent steps are in
``conjugate" directions, i.e., in orthogonal directions with respect to
the metric provided by the Hessian; additionally, convergence to the
minimum is efficient.  Thus, the algorithm rarely crosses the folded
region of the $M_2$-objective function where the gradient and Hessian
are not defined.  This fact motivates the use of the {\sc Migrad}
implementation of the VMM for our constrained $M_2$ calculations.

\subsection{Simplex}
\label{sec:Simplex}
 
 Unlike VMMs, the downhill simplex (or Nelder-Mead) method \cite{simplex}, does not require the 
 calculation of gradients.  Instead, one calculates the values of $f(\vec{x})$ at the $n+1$ 
 vertices of a simplex, a non-degenerate solid in $n$ dimensions with $n+1$ sides and $n+1$ vertices. 
 A new vertex for the simplex is generated in each iteration of the method.  
 If the value of the objective function at the new point is 
 lower than the value at one of the existing vertices, the worst vertex is replaced by the new point.
 In this way, the volume of the simplex becomes smaller; the algorithm stops when the simplex,
 now enclosing the minimum, has shrunk to a specified size.

To be more concrete, let us first consider a (large) simplex of $n+1$ points in $n$ dimensions,
with vertices
\bea
p_1,p_2,...,p_n,p_{n+1}.
\eea
These points are ordered so that 
\bea
f_1 \leq f_2 \leq ... \leq f_n \leq f_{n+1},
\label{forder}
\eea
where $f_i\equiv f(p_i)$.
We define the ``center of mass", $\bar{p}$, using all points except
$p_{n+1}$ as follows,
\beq
\bar{p} = 
\frac{1}{n} 
\sum_{i=1}^{n} p_i.
\eeq
In each step, the algorithm tries to replace the worst point,
$p_{n+1}$.  First, a new test point, $p_r$, is obtained by reflection of
the worst point about the center of mass,
\bea
p_r = \bar{p} + \alpha(\bar{p}-p_{n+1}),
\eea
for some typical value of the expansion factor (generally $\alpha =
1$; this is the value we shall use in our use of {\sc Simplex}). 
A new point is then determined using the value of $f(p_r)$ as follows: 
\begin{description}
\item[ 1. $f_1\leq f_r\leq f_{n}$: ] The previous worst point, $p_{n+1}$, is replaced by $p_r$, and the points are relabeled 
in accordance with (\ref{forder}).
\item[ 2. $f_r\leq f_1$: ] The test point, $p_r$, is the best point, so the current search direction 
is considered to be effective. 
We therefore shift the first $n$ points
\begin{equation}
\label{eq:shift}
p_1 \to p_2, p_2 \to p_3, ...., p_n \to p_{n+1}.
\end{equation}
To determine the new value for $p_1$, we try one additional point,
$p_{s1}=\bar{p}+\beta(p_r-\bar{p})$ (typically $\beta=2$), and
evaluate its functional value, $f_{s1}$.
If $f_{s1}<f_r$, we set $p_1 = p_{s1}$, otherwise $p_1 = p_r$. 
\item[ 3. $f_r>f_n$: ] The simplex may be too big and therefore its size must be reduced. 
  If $f_r>f_{n+1}$, then a new contracted simplex is
  defined around the best point, $p_1$, by replacing all points except
  $p_1$ by $p_i=p_1+\delta(p_i-p_1)$ with $0<\delta<1$ (typically
  $\delta=0.5$). If $f_n<f_r<f_{n+1}$, then $p_{n+1}$ is replaced by
  $p_r$.  A test of the new inner point,
  $p_{s2}=\bar{p}-\gamma(\bar{p}-p_{n+1})$ (typically $\gamma=0.5$), is
  then performed, and $p_{n+1}$ is replaced by $p_{s2}$ if $f_{s2} <
  f_{n+1}$.
\end{description}
Since the simplex method is always designed to take as big a step as possible, it is rather 
{\em insensitive} to shallow local minima and other small-scale structures in the objective 
function.  Thus, we use the method to identify promising candidates for global minima.
Once the location of a possible global minimum has been identified, the {\sc Migrad} 
algorithm described above is used to obtain a more precise value of any local 
minimum in this area, hopefully obtaining an accurate value for the
location of the global minimum.  The downhill simplex method is
implemented in {\sc Minuit} using the {\sc Simplex} algorithm.

\section{Installation and user instructions}
\label{app:optimass}
The latest version of the {\sc Optimass} has been developed and designed to achieve the automation 
of kinematic mass function minimization with constraints for general particle decay system. In particular, 
\begin{enumerate}
\item{It has generality to treat various decay topologies from multiple parent particles where in general the multiplicity can be larger than two.}
\item{It also has generality to include decay vertices where in general the multiplicity of branch legs can be larger than three.}
\item{It has flexibility to easily define a specific sub-system of intermediate parent particles with effective invisibles, in the full decay system.}
\item{It also has flexibility to define kinematic constraint functions of user's own interest, in terms of visible and invisible particles' momentum degrees of freedom.}
\item{All these generality and flexibility can be initiated from user's simple model card file which defines 1) full decay process with user's own particle label scheme, 2) parent node particle in each decay chain, 3) effective invisible nodes in each decay chain, and 4) constraint functions of particle momenta which can be interactively expressed by the user's particle label scheme.}
\end{enumerate}
The {\sc Optimass} is free software written in {\sc C++} and {\sc Python} under the copyleft of the \href{http://www.gnu.org/copyleft/gpl.html}{GNU General Public License}. The latest version of the {\sc Optimass} can be downloaded from the following web page :
\begin{center}
\href{http://hep-pulgrim.ibs.re.kr/optimass}{\texttt{http://hep-pulgrim.ibs.re.kr/optimass}}
\end{center}
More detailed {\sc Optimass} installation guide and the tutorial with examples on how to run the code implementing user's own decay topologies, can be found on the webpage as well.

\acknowledgments
We would like to thank K.~Kong for useful discussions. 
JG, KM, FM, and LP thank their CMS colleagues for useful discussions. 
WC thanks C. B. Park for useful discussions.
DK acknowledges support by LHC-TI postdoctoral fellowship under grant NSF-PHY-0969510.
MP is supported by the Korea Ministry of Science, ICT and Future Planning, Gyeongsangbuk-Do and Pohang City for 
Independent Junior Research Groups at the Asia Pacific Center for Theoretical Physics.
MP is also supported by World Premier International Research Center Initiative (WPI Initiative), MEXT, Japan.
Work also supported by DOE Grant No.~DE-SC0010296 and by IBS under the project code, IBS-R018-D1.


\begin{thebibliography}{99}

\bibitem{Aad:2012tfa} 
  G.~Aad {\it et al.}  [ATLAS Collaboration],
  {\it Observation of a new particle in the search for the Standard Model Higgs boson with the ATLAS detector at the LHC,}
  Phys.\ Lett.\ B {\bf 716}, 1 (2012)
  [arXiv:1207.7214].

\bibitem{Chatrchyan:2012ufa} 
  S.~Chatrchyan {\it et al.}  [CMS Collaboration],
  {\it Observation of a new boson at a mass of 125 GeV with the CMS experiment at the LHC,}
  Phys.\ Lett.\ B {\bf 716}, 30 (2012)
  [arXiv:1207.7235].


\bibitem{Martin:1997ns} 
  S.~P.~Martin,
  {\it A Supersymmetry primer,}
  [hep-ph/9709356].

\bibitem{Drees:2004jm} 
  M.~Drees, R.~Godbole and P.~Roy,
  {\it Theory and phenomenology of sparticles: An account of four-dimensional N=1 supersymmetry in high energy physics,}
  Hackensack, USA: World Scientific (2004)

\bibitem{Baer:2006rs} 
  H.~Baer and X.~Tata,
  {\it Weak scale supersymmetry: From superfields to scattering events,}
  Cambridge, UK: Univ. Pr. (2006)

\bibitem{Dine:2007zp} 
  M.~Dine,
  {\it Supersymmetry and string theory: Beyond the standard model,}
  Cambridge, UK: Cambridge Univ. Pr. (2007)


\bibitem{Appelquist:2000nn} 
  T.~Appelquist, H.~-C.~Cheng and B.~A.~Dobrescu,
  {\it Bounds on universal extra dimensions,}
  Phys.\ Rev.\ D {\bf 64}, 035002 (2001)
  [hep-ph/0012100].
  
\bibitem{Cheng:2002ab} 
  H.-C.~Cheng, K.~T.~Matchev and M.~Schmaltz,
  {\it Bosonic supersymmetry? Getting fooled at the CERN LHC,}
  Phys.\ Rev.\ D {\bf 66}, 056006 (2002)
  [hep-ph/0205314].

\bibitem{Barr:2010zj} 
  A.~J.~Barr and C.~G.~Lester,
  {\it A Review of the Mass Measurement Techniques proposed for the Large Hadron Collider,}
  J.\ Phys.\ G {\bf 37}, 123001 (2010)
  [arXiv:1004.2732].


\bibitem{Wang:2008sw} 
  L.~-T.~Wang and I.~Yavin,
  {\it A Review of Spin Determination at the LHC,}
  Int.\ J.\ Mod.\ Phys.\ A {\bf 23}, 4647 (2008)
  [arXiv:0802.2726].

\bibitem{Burns:2008cp} 
  M.~Burns, K.~Kong, K.~T.~Matchev and M.~Park,
  {\it A General Method for Model-Independent Measurements of Particle Spins, Couplings and Mixing Angles in Cascade Decays with Missing Energy at Hadron Colliders,}
  JHEP {\bf 0810}, 081 (2008)
  [arXiv:0808.2472].
    
\bibitem{Abbott:2000gx} 
  V.~M.~Abazov {\it et al.}  [D0 Collaboration],
  {\it A Quasi model independent search for new physics at large transverse momentum,}
  Phys.\ Rev.\ D {\bf 64}, 012004 (2001)
  [hep-ex/0011067].

\bibitem{Aaltonen:2011wt} 
  T.~Aaltonen {\it et al.}  [CDF Collaboration],
  {\it Measurement of the Top Quark Mass in the Lepton+Jets Channel Using the Lepton Transverse Momentum,}
  Phys.\ Lett.\ B {\bf 698}, 371 (2011)
  [arXiv:1101.4926].

\bibitem{Tovey:2000wk} 
  D.~R.~Tovey,
  {\it Measuring the SUSY mass scale at the LHC,}
  Phys.\ Lett.\ B {\bf 498}, 1 (2001)
  [hep-ph/0006276].

\bibitem{Bramante:2011xd} 
  J.~Bramante, J.~Kumar and B.~Thomas,
  {\it Large Jet Multiplicities and New Physics at the LHC,}
  Phys.\ Rev.\ D {\bf 86}, 015014 (2012)
  [arXiv:1109.6014].

\bibitem{Hedri:2013pvl} 
  S.~El Hedri, A.~Hook, M.~Jankowiak and J.~G.~Wacker,
  {\it Learning How to Count: A High Multiplicity Search for the LHC,}
  JHEP {\bf 1308}, 136 (2013)
  [arXiv:1302.1870].
  
\bibitem{Hook:2012fd} 
  A.~Hook, E.~Izaguirre, M.~Lisanti and J.~G.~Wacker,
  {\it High Multiplicity Searches at the LHC Using Jet Masses,}
  Phys.\ Rev.\ D {\bf 85}, 055029 (2012)
  [arXiv:1202.0558].


\bibitem{Tovey:2008ui} 
  D.~R.~Tovey,
  {\it On measuring the masses of pair-produced semi-invisibly decaying particles at hadron colliders,}
  JHEP {\bf 0804}, 034 (2008)
  [arXiv:0802.2879].

\bibitem{Polesello:2009rn} 
  G.~Polesello and D.~R.~Tovey,
  {\it Supersymmetric particle mass measurement with the boost-corrected contransverse mass,}
  JHEP {\bf 1003}, 030 (2010)
  [arXiv:0910.0174].

\bibitem{Matchev:2009ad} 
  K.~T.~Matchev and M.~Park,
  {\it A General method for determining the masses of semi-invisibly decaying particles at hadron colliders,}
  Phys.\ Rev.\ Lett.\  {\bf 107}, 061801 (2011)
  [arXiv:0910.1584].
  
\bibitem{Agashe:2012bn} 
  K.~Agashe, R.~Franceschini and D.~Kim,
  {\it Simple invariance of two-body decay kinematics,}
  Phys.\ Rev.\ D {\bf 88}, no. 5, 057701 (2013)
  [arXiv:1209.0772].

\bibitem{Agashe:2012fs} 
  K.~Agashe, R.~Franceschini, D.~Kim and K.~Wardlow,
  {\it Using Energy Peaks to Count Dark Matter Particles in Decays,}
  Phys.\ Dark Univ.\  {\bf 2}, 72 (2013)
  [arXiv:1212.5230].

\bibitem{Agashe:2013eba} 
  K.~Agashe, R.~Franceschini and D.~Kim,
  {\it Using Energy Peaks to Measure New Particle Masses,}
  JHEP {\bf 1411}, 059 (2014)
  [arXiv:1309.4776 ].
  
\bibitem{Nojiri:2000wq} 
  M.~M.~Nojiri, D.~Toya and T.~Kobayashi,
  {\it Lepton energy asymmetry and precision SUSY study at hadron colliders,}
  Phys.\ Rev.\ D {\bf 62}, 075009 (2000)
  [hep-ph/0001267].

\bibitem{Cheng:2011ya} 
  H.~C.~Cheng and J.~Gu,
  {\it Measuring Invisible Particle Masses Using a Single Short Decay Chain,}
  JHEP {\bf 1110}, 094 (2011)
  [arXiv:1109.3471 ].
    
\bibitem{Barr:2011xt} 
  A.~J.~Barr, T.~J.~Khoo, P.~Konar, K.~Kong, C.~G.~Lester, K.~T.~Matchev and M.~Park,
  {\it Guide to transverse projections and mass-constraining variables,}
  Phys.\ Rev.\ D {\bf 84}, 095031 (2011)
  [arXiv:1105.2977 ].

\bibitem{CL_tasi}
 C.~G.~Lester, {\it Mass and Spin Measurement Techniques (for the Large Hadron Collider),} 
 lectures given at TASI 2011, in 
 {\it The Dark Secrets of the Terascale} : Proceedings, TASI 2011, Boulder, Colorado, USA, Jun 6 - Jul 11, 2011,
 Eds.~T.~M.~P.~Tait and K.~T.~Matchev,
 World Scientific, 2013.

\bibitem{Smith:1983aa} 
  J.~Smith, W.~L.~van Neerven and J.~A.~M.~Vermaseren,
  {\it The Transverse Mass and Width of the $W$ Boson,}
  Phys.\ Rev.\ Lett.\  {\bf 50}, 1738 (1983).

\bibitem{Barger:1983wf} 
  V.~D.~Barger, A.~D.~Martin and R.~J.~N.~Phillips,
  {\it Perpendicular $\nu_e$ Mass From $W$ Decay,}
  Z.\ Phys.\ C {\bf 21}, 99 (1983).


\bibitem{Hinchliffe:1996iu} 
  I.~Hinchliffe, F.~E.~Paige, M.~D.~Shapiro, J.~Soderqvist and W.~Yao,
  {\it Precision SUSY measurements at CERN LHC,}
  Phys.\ Rev.\ D {\bf 55}, 5520 (1997)
  [hep-ph/9610544].


\bibitem{Konar:2008ei} 
  P.~Konar, K.~Kong and K.~T.~Matchev,
  {\it $\sqrt{\hat{s}}_{min}$ : A Global inclusive variable for determining the mass scale of new physics in events with missing energy at hadron colliders,}
  JHEP {\bf 0903}, 085 (2009)
  [arXiv:0812.1042 ].
  
\bibitem{Konar:2010ma} 
  P.~Konar, K.~Kong, K.~T.~Matchev and M.~Park,
  {\it RECO level $\sqrt{s}_{min}$ and subsystem $\sqrt{s}_{min}$: Improved global inclusive variables for measuring the new physics mass scale in $\met$ events at hadron colliders,}
  JHEP {\bf 1106}, 041 (2011)
  [arXiv:1006.0653 ].

\bibitem{Robens:2011zm} 
  T.~Robens,
  {\it $\sqrt{\hat{s}}_{\rm min}$ resurrected,}
  JHEP {\bf 1202}, 051 (2012)
  [arXiv:1109.1018 ].

\bibitem{Rogan:2010kb} 
  C.~Rogan,
  {\it Kinematical variables towards new dynamics at the LHC,}
  arXiv:1006.2727 .

\bibitem{Buckley:2013kua} 
  M.~R.~Buckley, J.~D.~Lykken, C.~Rogan and M.~Spiropulu,
  {\it Super-Razor and Searches for Sleptons and Charginos at the LHC,}
  Phys.\ Rev.\ D {\bf 89}, no. 5, 055020 (2014)
  [arXiv:1310.4827 ].
  
\bibitem{CMS:2014ets} 
  CMS Collaboration [CMS Collaboration],
  {\it Search for supersymmetry in two-photons+jet events with razor variables in pp collisions at sqrt(s) = 8 TeV,}
  CMS-PAS-SUS-14-008.

\bibitem{Khachatryan:2015pwa} 
  V.~Khachatryan {\it et al.} [CMS Collaboration],
  {\it Search for supersymmetry using razor variables in events with $b$-tagged jets in $pp$ collisions at $\sqrt{s} =$ 8 TeV,}
  Phys.\ Rev.\ D {\bf 91}, 052018 (2015)
  [arXiv:1502.00300 ].



\bibitem{Bhat:2010zz} 
  P.~C.~Bhat,
  {\it Multivariate Analysis Methods in Particle Physics,}
  Ann.\ Rev.\ Nucl.\ Part.\ Sci.\  {\bf 61}, 281 (2011).

\bibitem{Hubisz:2008gg} 
  J.~Hubisz, J.~Lykken, M.~Pierini and M.~Spiropulu,
  {\it Missing energy look-alikes with 100 pb$^{-1}$ at the LHC,}
  Phys.\ Rev.\ D {\bf 78}, 075008 (2008)
  [arXiv:0805.2398 ].
  
\bibitem{Alves:2012ft} 
  D.~S.~M.~Alves, M.~R.~Buckley, P.~J.~Fox, J.~D.~Lykken and C.~T.~Yu,
  {\it Stops and $\not E_T$: The shape of things to come,}
  Phys.\ Rev.\ D {\bf 87}, no. 3, 035016 (2013)
  [arXiv:1205.5805].


\bibitem{kondo1} 
  K.~Kondo,
  {\it Dynamical Likelihood Method for Reconstruction of Events With Missing Momentum. 1: Method and Toy Models,}
  J.\ Phys.\ Soc.\ Jap.\  {\bf 57}, 4126 (1988).

\bibitem{kondo2} 
  K.~Kondo,
  {\it Dynamical likelihood method for reconstruction of events with missing momentum. 2: Mass spectra for $2 \to 2$ processes,}  J.\ Phys.\ Soc.\ Jap.\  {\bf 60}, 836 (1991).

\bibitem{Gainer:2013iya} 
  J.~S.~Gainer, J.~Lykken, K.~T.~Matchev, S.~Mrenna and M.~Park,
  {\it The Matrix Element Method: Past, Present, and Future,}
  arXiv:1307.3546 .

\bibitem{kondo3} 
  K.~Kondo, T.~Chikamatsu and S.~H.~Kim,
  {\it Dynamical likelihood method for reconstruction of events with missing momentum. 3: Analysis of a CDF high p(T) e mu event as t anti-t production,}
  J.\ Phys.\ Soc.\ Jap.\  {\bf 62}, 1177 (1993).

\bibitem{dalitz} 
  R.~H.~Dalitz and G.~R.~Goldstein,
  {\it The Decay and polarization properties of the top quark,}
  Phys.\ Rev.\ D {\bf 45}, 1531 (1992).

\bibitem{oai:arXiv.org:hep-ex/9808029} 
  B.~Abbott {\it et al.}  [D0 Collaboration],
  {\it Measurement of the top quark mass in the dilepton channel,}
  Phys.\ Rev.\ D {\bf 60}, 052001 (1999)
  [hep-ex/9808029].

\bibitem{vigil} 
  J.~C.~Estrada Vigil,
  {\it Maximal use of kinematic information for the extraction of the mass of the top quark in single-lepton t anti-t events at D0,}
  FERMILAB-THESIS-2001-07.

\bibitem{canelli} 
  M.~F.~Canelli,
  {\it Helicity of the $W$ boson in single - lepton $t \bar{t}$ events,}
  UMI-31-14921.

\bibitem{abazov} 
  V.~M.~Abazov {\it et al.}  [D0 Collaboration],
  {\it A precision measurement of the mass of the top quark,}
  Nature {\bf 429}, 638 (2004)
  [hep-ex/0406031].

\bibitem{Aaltonen:2010yz} 
  T.~Aaltonen {\it et al.}  [CDF Collaboration],
  {\it Top Quark Mass Measurement in the Lepton + Jets Channel Using a Matrix Element Method and \textit{in situ} Jet Energy Calibration,}
  Phys.\ Rev.\ Lett.\  {\bf 105}, 252001 (2010)
  [arXiv:1010.4582].
  
\bibitem{Volobouev:2011vb} 
  I.~Volobouev,
  {\it Matrix Element Method in HEP: Transfer Functions, Efficiencies, and Likelihood Normalization,}
  arXiv:1101.2259 [physics.data-an].

\bibitem{Demilly:2014baa} 
  A.~Demilly,
  {\it Mesure de la masse du quark top dans le canal dileptonique \'electron-muon avec la m\'ethode des \'el\'ements de matrice dans l'exp\'erience ATLAS aupr\'es du LHC,}
  CERN-THESIS-2014-140.
  
\bibitem{Khachatryan:2015ila} 
  V.~Khachatryan {\it et al.}  [CMS Collaboration],
  {\it Search for a standard model Higgs boson produced in association with a top-quark pair and decaying to bottom quarks using a matrix element method,}
  Eur.\ Phys.\ J.\ C {\bf 75}, no. 6, 251 (2015)
  [arXiv:1502.02485].

\bibitem{Gainer:2014bta} 
  J.~S.~Gainer, J.~Lykken, K.~T.~Matchev, S.~Mrenna and M.~Park,
  {\it Exploring Theory Space with Monte Carlo Reweighting,}
  JHEP {\bf 1410}, 78 (2014)
  [arXiv:1404.7129.

\bibitem{Alwall:2014hca} 
  J.~Alwall, R.~Frederix, S.~Frixione, V.~Hirschi, F.~Maltoni, O.~Mattelaer, H.-S.~Shao and T.~Stelzer {\it et al.},
  {\it The automated computation of tree-level and next-to-leading order differential cross sections, and their matching to parton shower simulations,}
  JHEP {\bf 1407}, 079 (2014)
  [arXiv:1405.0301.
    
\bibitem{Schouten:2014yza} 
  D.~Schouten, A.~DeAbreu and B.~Stelzer,
  {\it Accelerated Matrix Element Method with Parallel Computing,}
  Comput.\ Phys.\ Commun.\  {\bf 192}, 54 (2015)
  [arXiv:1407.7595 [physics.comp-ph]].
    

\bibitem{Lester:1999tx} 
  C.~G.~Lester and D.~J.~Summers,
  {\it Measuring masses of semiinvisibly decaying particles pair produced at hadron colliders,}
  Phys.\ Lett.\ B {\bf 463}, 99 (1999)
  [hep-ph/9906349].

\bibitem{Barr:2003rg} 
  A.~Barr, C.~Lester and P.~Stephens,
  {\it m(T2): The Truth behind the glamour,}
  J.\ Phys.\ G {\bf 29}, 2343 (2003)
  [hep-ph/0304226].


\bibitem{Konar:2009wn} 
  P.~Konar, K.~Kong, K.~T.~Matchev and M.~Park,
  {\it Superpartner Mass Measurement Technique using 1D Orthogonal Decompositions of the Cambridge Transverse Mass Variable $M_{T2}$,}
  Phys.\ Rev.\ Lett.\  {\bf 105}, 051802 (2010)
  [arXiv:0910.3679.

\bibitem{Barr:2009jv} 
  A.~J.~Barr, B.~Gripaios and C.~G.~Lester,
  {\it Transverse masses and kinematic constraints: from the boundary to the crease,}
  JHEP {\bf 0911}, 096 (2009)
  [arXiv:0908.3779.

\bibitem{Konar:2009qr} 
  P.~Konar, K.~Kong, K.~T.~Matchev and M.~Park,
  {\it Dark Matter Particle Spectroscopy at the LHC: Generalizing M(T2) to Asymmetric Event Topologies,}
  JHEP {\bf 1004}, 086 (2010)
  [arXiv:0911.4126].



\bibitem{Ross:2007rm} 
  G.~G.~Ross and M.~Serna,
  {\it Mass determination of new states at hadron colliders,}
  Phys.\ Lett.\ B {\bf 665}, 212 (2008)
  [arXiv:0712.0943].
  
\bibitem{Barr:2008ba} 
  A.~J.~Barr, G.~G.~Ross and M.~Serna,
  {\it The Precision Determination of Invisible-Particle Masses at the LHC,}
  Phys.\ Rev.\ D {\bf 78}, 056006 (2008)
  [arXiv:0806.3224].

\bibitem{Cho:2009ve} 
  W.~S.~Cho, J.~E.~Kim and J.~-H.~Kim,
  {\it Amplification of endpoint structure for new particle mass measurement at the LHC,}
  Phys.\ Rev.\ D {\bf 81}, 095010 (2010)
  [arXiv:0912.2354].

\bibitem{Cho:2010vz} 
  W.~S.~Cho, W.~Klemm and M.~M.~Nojiri,
  {\it Mass measurement in boosted decay systems at hadron colliders,}
  Phys.\ Rev.\ D {\bf 84}, 035018 (2011)
  [arXiv:1008.0391].

\bibitem{Barr:2011ux} 
  A.~J.~Barr, B.~Gripaios and C.~G.~Lester,
  {\it Re-weighing the evidence for a Higgs boson in dileptonic W-boson decays,}
  Phys.\ Rev.\ Lett.\  {\bf 108}, 041803 (2012)
  [Erratum-ibid.\  {\bf 108}, 109902 (2012)]
  [arXiv:1108.3468].


\bibitem{Bai:2012gs} 
  Y.~Bai, H.~C.~Cheng, J.~Gallicchio and J.~Gu,
  {\it Stop the Top Background of the Stop Search,}
  JHEP {\bf 1207}, 110 (2012)
  [arXiv:1203.4813].

\bibitem{Papaefstathiou:2014oja} 
  A.~Papaefstathiou, K.~Sakurai and M.~Takeuchi,
  {\it Higgs boson to di-tau channel in Chargino-Neutralino searches at the LHC,}
  JHEP {\bf 1408}, 176 (2014)
  [arXiv:1404.1077].



\bibitem{Cho:2007qv} 
  W.~S.~Cho, K.~Choi, Y.~G.~Kim and C.~B.~Park,
  {\it Gluino Stransverse Mass,}
  Phys.\ Rev.\ Lett.\  {\bf 100}, 171801 (2008)
  [arXiv:0709.0288].

\bibitem{Gripaios:2007is} 
  B.~Gripaios,
  {\it Transverse observables and mass determination at hadron colliders,}
  JHEP {\bf 0802}, 053 (2008)
  [arXiv:0709.2740].
  
\bibitem{Barr:2007hy} 
  A.~J.~Barr, B.~Gripaios and C.~G.~Lester,
  {\it Weighing Wimps with Kinks at Colliders: Invisible Particle Mass Measurements from Endpoints,}
  JHEP {\bf 0802}, 014 (2008)
  [arXiv:0711.4008].

\bibitem{Cho:2007dh} 
  W.~S.~Cho, K.~Choi, Y.~G.~Kim and C.~B.~Park,
  {\it Measuring superparticle masses at hadron collider using the transverse mass kink,}
  JHEP {\bf 0802}, 035 (2008)
  [arXiv:0711.4526].

\bibitem{Burns:2008va} 
  M.~Burns, K.~Kong, K.~T.~Matchev and M.~Park,
  {\it Using Subsystem $M_{T2}$ for Complete Mass Determinations in Decay Chains with Missing Energy at Hadron Colliders,}
  JHEP {\bf 0903}, 143 (2009)
  [arXiv:0810.5576].

\bibitem{Lester:2011nj} 
  C.~G.~Lester,
  {\it The stransverse mass, MT2, in special cases,}
  JHEP {\bf 1105}, 076 (2011)
  [arXiv:1103.5682].

\bibitem{Lally:2012uj} 
  C.~H.~Lally and C.~G.~Lester,
  {\it Properties of MT2 in the massless limit,}
  arXiv:1211.1542 .
  
\bibitem{Mahbubani:2012kx} 
  R.~Mahbubani, K.~T.~Matchev and M.~Park,
  {\it Re-interpreting the Oxbridge stransverse mass variable MT2 in general cases,}
  JHEP {\bf 1303}, 134 (2013)
  [arXiv:1212.1720].

\bibitem{Cho:2014naa} 
  W.~S.~Cho, J.~S.~Gainer, D.~Kim, K.~T.~Matchev, F.~Moortgat, L.~Pape and M.~Park,
  {\it On-shell constrained $M_2$ variables with applications to mass measurements and topology disambiguation,}
  JHEP {\bf 1408}, 070 (2014)
  [arXiv:1401.1449].

\bibitem{Cho:2014yma} 
  W.~S.~Cho, J.~S.~Gainer, D.~Kim, K.~T.~Matchev, F.~Moortgat, L.~Pape and M.~Park,
  {\it Improving the sensitivity of stop searches with on-shell constrained invariant mass variables,}
  JHEP {\bf 1505}, 040 (2015)
  [arXiv:1411.0664].


\bibitem{Cheng:2008hk} 
  H.~C.~Cheng and Z.~Han,
  {\it Minimal Kinematic Constraints and m(T2),}
  JHEP {\bf 0812}, 063 (2008)
  [arXiv:0810.5178].

\bibitem{Walker:2013uxa} 
  J.~W.~Walker,
  {\it A complete solution classification and unified algorithmic treatment for the one- and two-step asymmetric S-transverse mass 
  $ {\tilde{M}}_{\mathrm{T}2} $ event scale statistic,}
  JHEP {\bf 1408}, 155 (2014)
  [arXiv:1311.6219].
  
\bibitem{Lester:2014yga} 
  C.~G.~Lester and B.~Nachman,
  {\it Bisection-based asymmetric M$_{T2}$ computation: a higher precision calculator than existing symmetric methods,}
  JHEP {\bf 1503}, 100 (2015)
  [arXiv:1411.4312].
          

\bibitem{ALM1} 
M.~R.~Hestenes,
{\it Multiplier and gradient methods,}
Journal of Optimization Theory and Applications {\bf 4}, 303-320 (1969).

\bibitem{ALM2}
 M.~J.~D.~Powell, 
{\it A method for nonlinear constraints in minimization problems,} 
in Optimization ed. by R. Fletcher, Academic Press, New York, NY, pp. 283-298 (1969).

\bibitem{minuit}
   F.~James and M.~Roos,
  {\it Minuit: A System for Function Minimization and Analysis of the Parameter Errors and Correlations,}
  Comput.\ Phys.\ Commun.\  {\bf 10}, 343 (1975),
  \href{http://seal.web.cern.ch/seal/snapshot/work-packages/mathlibs/minuit/
 }{http://seal.web.cern.ch/seal/snapshot/work-packages/mathlibs/minuit/}.
 

\bibitem{nocedal}
{\it Numerical optimization} - Jorge Nocedal, and Stephen
J. Wright. Springer series in operations research and financial
engineering Springer, New York, NY, 2. ed. edition, (2006)

\bibitem{alm}
D. P. Bertsekas. Constrained Optimization and Lagrange Multiplier
Methods. Academic Press, New York, 1982.

\bibitem{Press:1992zz} 
  W.~H.~Press, S.~A.~Teukolsky, W.~T.~Vetterling and B.~P.~Flannery,
  {\it Numerical Recipes in {\sc Fortran}: The Art of Scientific Computing,}
  ISBN-9780521430647.


\bibitem{Cho:2008tj} 
  W.~S.~Cho, K.~Choi, Y.~G.~Kim and C.~B.~Park,
  {\it M(T2)-assisted on-shell reconstruction of missing momenta and its application to spin measurement at the LHC,}
  Phys.\ Rev.\ D {\bf 79}, 031701 (2009)
  [arXiv:0810.4853].

\bibitem{Park:2011uz} 
  C.~B.~Park,
  {\it Reconstructing the heavy resonance at hadron colliders,}
  Phys.\ Rev.\ D {\bf 84}, 096001 (2011)
  [arXiv:1106.6087].

\bibitem{convexification_alm} 
See Ref.~\cite{alm}, Proposition 4.2.3, Theorem 17.6, {\it Numerical optimization}.

\bibitem{lancelot}
A. R. Conn, N. I. M. Gould, and Ph. L. Toint, {\it A Globally convergent augmented Lagrangian algorithm for optimization with general constraints and simple bounds}. SIAM J. Numer. Anal., 28:545-572, (1991).\\
%
A. R. Conn, N. I. M. Gould, and Ph. L. Toint, {\it LANCELOT:  A Fortran Package for Large-scale Nonlinear Optimization (Release A)}. Lecture Notes in Computation Mathematics 17. Springer Verlag, Berlin, Heidelberg, New York, London, Paris and Tokyo, (1992).\\
%
{\it LANCELOT optimization software,} \href{http://www.numerical.rl.ac.uk/lancelot/blurb.html}{http://www.numerical.rl.ac.uk/lancelot/blurb.html}.


\bibitem{minuit2}
{\sc Minuit2}, \href{http://seal.web.cern.ch/seal/MathLibs/Minuit2/html/index.html}{http://seal.web.cern.ch/seal/MathLibs/Minuit2/html/index.html}.
  
\bibitem{root}
{\sc Root}-'A Data Analysis Framework`, \href{http://root.cern.ch/drupal/}{http://root.cern.ch/drupal}.\\

\bibitem{vmm}
W. C. Davidon, {\it Variable metric method for minimization,} Technical Report ANL-5990 (revised), Argonne National Laboratory, Argonne, Il, (1959)\\
W. C. Davidon, SIAM J. OPTIMIZATION Vol. 1, No. 1, pp. 1-17, February (1991).

\bibitem{vmmFP}
R. Fletcher and M. Powell, {\it A rapidly converging descent method for minimization,} Comput. J., 6:163, 1963.

\bibitem{vmmBFGS1}
C. G. Boyden, {\it The convergence of a class of double-rank minimization algorithms,} Journal of the Institute of Mathematics and Its Applications 6: 76-90 (1970).

\bibitem{vmmBFGS2}
R. Fletcher, {\it A New Approach to Variable Metric Algorithms,} Computer Journal 13 (3): 317-322 (1970).

\bibitem{vmmBFGS3}
D. Goldfarb, {\it A Family of Variable Metric Updates Derived by Variational Means,} Mathematics of Computation 24 (109): 23-26 (1970).

\bibitem{vmmBFGS4}
David F. Shanno, {\it Conditioning of quasi-Newton methods for function minimization,} Math. Comput. 24 (111): 647-656 (1970).

\bibitem{simplex}
J. A. Nelder and R. Mead, {\it A simplex method for function minimization,} Comput. J., 7:308, (1965)

\bibitem{us}
W.~S.~Cho, J.~S.~Gainer, D.~Kim, S.~H.~Lim, K.~T.~Matchev, F.~Moortgat, L.~Pape and M.~Park, {\it Scrutinizing missing energy events at the LHC with OPTIMASS,} in preparation.


 

\end{thebibliography}
\end{document}